\documentclass[numberedappendix,iop]{emulateapj}
\usepackage{apjfonts}
\usepackage{array}
\usepackage[usenames]{color}
\begin{document}
\newcommand{\comment}[1]{}
\newcommand{\risa}[1]{\textcolor{red}{(\bf #1)}}
\newcommand{\michael}[1]{\textcolor{blue}{(\bf #1)}}
\definecolor{purple}{RGB}{160,32,240}
\newcommand{\peter}[1]{\textcolor{purple}{(\bf #1)}}
\newcommand{\macc}{M_\mathrm{acc}}
\newcommand{\mpeak}{M_\mathrm{peak}}
\newcommand{\mnow}{M_\mathrm{now}}
\newcommand{\vacc}{v_\mathrm{acc}} 
\newcommand{\vpeak}{v_\mathrm{peak}} 
\newcommand{\vnow}{v^\mathrm{now}_\mathrm{max}}

\newcommand{\Mnfw}{M_\mathrm{NFW}}
\newcommand{\Msun}{M_{\odot}}
\newcommand{\mvir}{M_\mathrm{vir}}
\newcommand{\rvir}{R_\mathrm{vir}}
\newcommand{\vmax}{v_\mathrm{max}}
\newcommand{\vmac}{v_\mathrm{max}^\mathrm{acc}}
\newcommand{\mvac}{M_\mathrm{vir}^\mathrm{acc}}
\newcommand{\sfr}{\mathrm{SFR}}
\newcommand{\plotgrace}[1]{\includegraphics[angle=-90,width=\columnwidth,type=eps,ext=.eps,read=.eps]{#1}}
\newcommand{\plotgraceflip}[1]{\includegraphics[angle=-90,width=\columnwidth,type=eps,ext=.eps,read=.eps]{#1}}
\newcommand{\plotlargegrace}[1]{\includegraphics[angle=-90,width=2\columnwidth,type=eps,ext=.eps,read=.eps]{#1}}
\newcommand{\plotlargegraceflip}[1]{\includegraphics[angle=-90,width=2\columnwidth,type=eps,ext=.eps,read=.eps]{#1}}
\newcommand{\plotminigrace}[1]{\includegraphics[angle=-90,width=0.5\columnwidth,type=eps,ext=.eps,read=.eps]{#1}}
\newcommand{\plotmicrograce}[1]{\includegraphics[angle=-90,width=0.25\columnwidth,type=eps,ext=.eps,read=.eps]{#1}}
\newcommand{\plotsmallgrace}[1]{\includegraphics[angle=-90,width=0.66\columnwidth,type=eps,ext=.eps,read=.eps]{#1}}

\newcommand{\hinv}{h^{-1}}
\newcommand{\mpc}{\rm{Mpc}}
\newcommand{\hmpc}{$\hinv\mpc$}

\shortauthors{BEHROOZI \& SILK}
\shorttitle{Predicting High-Redshift Galaxy Evolution}

\title{A Simple Technique for Predicting High-Redshift Galaxy Evolution}

\author{Peter S. Behroozi\altaffilmark{1}, Joseph Silk\altaffilmark{2,3,4}}

\altaffiltext{1}{Space Telescope Science Institute, Baltimore, MD 21218 USA} 
\altaffiltext{2}{Institut d'Astrophysique, UMR 7095 CNRS, Universit\'e Pierre et Marie Curie, 98bis Blvd Arago, 75014 Paris, France}
\altaffiltext{3}{Department of Physics and Astronomy, The Johns Hopkins University, Baltimore MD 21218, USA}
\altaffiltext{4}{Beecroft Institute of Particle Astrophysics and Cosmology, Department of Physics, University of Oxford, Oxford OX1 3RH, UK}

\begin{abstract}
We show that the ratio of galaxies' specific star formation rates (SSFRs) to their host halos' specific mass accretion rates (SMARs) strongly constrains how the galaxies' stellar masses, specific star formation rates, and host halo masses evolve over cosmic time.  This evolutionary constraint provides a simple way to probe $z>8$ galaxy populations without direct observations.    Tests of the method with galaxy properties at $z=4$ successfully reproduce the known evolution of the stellar mass--halo mass (SMHM) relation, galaxy SSFRs, and the cosmic star formation rate (CSFR) for $5<z<8$.  We then predict the continued evolution of these properties for $8<z<15$.  In contrast to the non-evolution in the SMHM relation at $z<4$, the median galaxy mass at fixed halo mass increases strongly at $z>4$.  We show that this result is closely linked to the flattening in galaxy SSFRs at $z>2$ compared to halo specific mass accretion rates; we expect that average galaxy SSFRs at fixed stellar mass will continue their mild evolution to $z\sim 15$.  The expected CSFR shows no breaks or features at $z>8.5$; this constrains both reionization and the possibility of a steep falloff in the CSFR at $z=9$--10.  Finally, we make predictions for stellar mass and luminosity functions for the \textit{James Webb Space Telescope}, which should be able to observe one galaxy with $M_\ast \gtrsim 10^{8}\Msun$ per $10^{3}$ Mpc$^{3}$ at $z=9.6$ and one such galaxy per $10^{4}$ Mpc$^{3}$ at $z=15$.  
\end{abstract}

\section{Introduction}
\label{s:intro}

The \textit{James Webb Space Telescope} (JWST; launch: 2018) will strongly constrain galaxy formation from $z=20$ to $z=8$ \citep{JWST}.  Population III stars, globular clusters, supermassive black hole seeds, and the first galaxies are all expected to form during this epoch \citep[see][for a review]{Bromm11}.  Early galaxies are also expected to reionize the intergalactic medium \citep[see][for reviews]{Loeb01,Haardt12}.  Yet, constraints on how many $z>8$ galaxies JWST will observe have been contradictory.  Some observations with the \textit{Hubble Space Telescope} (HST) show a deficiency of star-forming galaxies at $z=9$ and above \citep{Ellis13,Oesch13}.  However, HST imaging of clusters has returned evidence for both steeply declining and mildly evolving cosmic star formation rates \citep{Bouwens12b,Coe12,Shu14,Oesch14}.  And to date, long gamma-ray bursts detected with \textit{Swift} (which measure the total cosmic star formation rate, as opposed to only star formation above a threshold) suggest only mild evolution in the total cosmic star formation rate at $z>8$ \citep{Robertson12,Kistler13,Wang13GRB};

Resolving this discrepancy by observing more $z>8$ galaxies is difficult with current instruments.  Indirect approaches based on existing measurements of $z=7$ and $z=8$ galaxies \citep{BORG12,Schmidt14,Bouwens14} represent an alternative.  Stars formed at $z=15$ or $z=10$ are still present in $z=7$--8 galaxies, so at the most basic level, a measurement of the current specific star formation rate can be used to guess how many stars were in place at earlier times \citep[see, e.g.,][]{Leitner11}.  Yet, the strong connection between galaxy growth and the growth of collapsed, self-bound, virialized dark matter structures (i.e., ``halos'') in the Lambda Cold Dark Matter paradigm \citep{more-09,yang-09,Leauthaud12,Reddick12,Tinker13,Behroozi13} suggests that a more physical extrapolation can be made.

Previous theoretical work on the growth of galaxies in dark matter halos at high redshift has included hydrodynamical simulations \citep[e.g.,][]{Finlator11,Wise12,Wise12b,Wise14,Jaacks12,Jaacks12b,Dayal13}, semi-analytical \citep[e.g.,][]{Lacey11}, and semi-empirical models \citep[e.g.,][]{Trenti10,Tacchella13,Wyithe13}.  These models have tended to require significant assumptions for how galaxy luminosity is connected to dark matter halo properties, imposed either indirectly (through sub-grid models in hydrodynamical simulations) or directly (as in a constant halo mass---luminosity relation in \citealt{Trenti10} or as in the semi-analytical models in \citealt{Lacey11}).  Although the accuracy of existing luminosity functions at $z>8$ is disputed (as noted above), some of these models predict a sharp decline in the cosmic star formation rate above $z=8$ due to the lack of galaxies above the HST observational threshold \citep{Trenti10,Finlator11,Jaacks12,Dayal13}.

We adopt an approach with somewhat weaker assumptions, which constrains how the relative growth of galaxies (i.e., their specific star formation rates) relates to the relative growth of their dark matter halos (i.e., their specific halo mass accretion rates).  We show that the ratio of these two quantities predicts the trajectory along which high-redshift galaxies evolve in the stellar mass---halo mass plane---i.e., it constrains the historical stellar mass---halo mass relationship.  As we will discuss, this relationship is valuable not only for predicting high-redshift and faint galaxy properties, but also for understanding the SSFR ``plateau''---that is, why galaxy SSFRs at fixed stellar mass are relatively flat at $z>2$ \citep{Weinmann11}.  In addition, the connection to SSFRs also offers a better understanding of why the stellar mass---halo mass relationship appears to evolve at redshifts $z>4$ \citep{BWC12}, but remains relatively constant for $z<4$ \citep{Behroozi13}.

We describe the details of our approach in \S \ref{s:theory}, discuss existing constraints from observations and simulations in \S \ref{s:constraints}, present tests and error analyses of the approach in \S \ref{s:tests}, and present results for $z>8$ galaxy populations in \S \ref{s:results}.  We discuss how these results affect the SSFR ``plateau,'' dwarf galaxy star formation efficiency, abundance matching (including the ``too big to fail'' problem), 
reionization, and expectations for JWST in \S \ref{s:discussion}; finally, we summarize conclusions in \S \ref{s:conclusions}.  Our adopted cosmology is a flat, $\Lambda$CDM universe with $\Omega_M=0.27$, $\Omega_b = 0.045$, $h=0.7$, $\sigma_8 = 0.82$, and $n_s = 0.95$; these are very similar to the WMAP9 best-fit parameters \citep{WMAP9}.  Throughout this paper, we use the virial spherical overdensity definition in \cite{mvir_conv} for halo masses.

\begin{figure}
\includegraphics[width=\columnwidth]{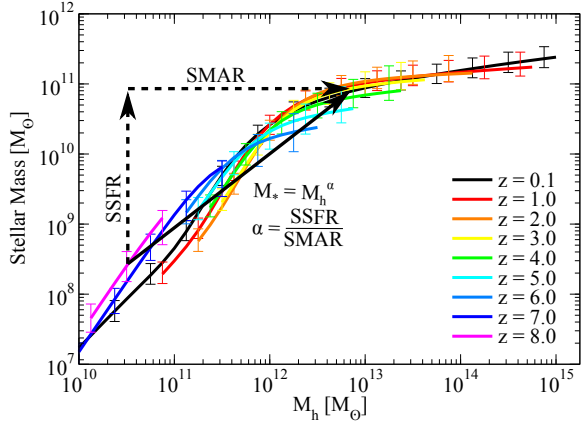}
\caption{Galaxy evolution in the stellar mass---halo mass plane.  The background figure shows the stellar mass--halo mass relations from $z=8$ to $z=0$ from \cite{BWC12}.  Evolution in this plane is set by two factors.  Galaxies' specific star formation rates (SSFRs) set the rate at which they increase in log stellar mass.  Similarly, their specific halo mass accretion rates (SMARs) set the rate at which they increase in log halo mass.  The power-law slope of their future stellar mass---halo mass trajectory is therefore set by the ratio of their SSFRs to their SMARs, recalling that this slope equals rise (SSFR) over run (SMAR).  Combining measured specific star formation rates with halo mass accretion histories from simulations therefore lets one predict galaxies' previous evolution in the stellar mass--halo mass plane.}
\label{f:schematic}
\end{figure}

\section{Theory}

\label{s:theory}

\subsection{Basic Principles}

\label{s:basic}

Galaxy star formation histories are typically parametrized as a function of time, i.e., $SFR(t)$.  Yet, physically interpreting $SFR(t)$ is not straightforward, because changes in $SFR$ with time come both from large-scale environment (e.g., the redshift dependence of mass accretion rates) and local properties (e.g., host halo mass and gravitational potential well depth).  We therefore consider a more physical parametrization which separates these two effects.

\begin{figure*}
\plotgrace{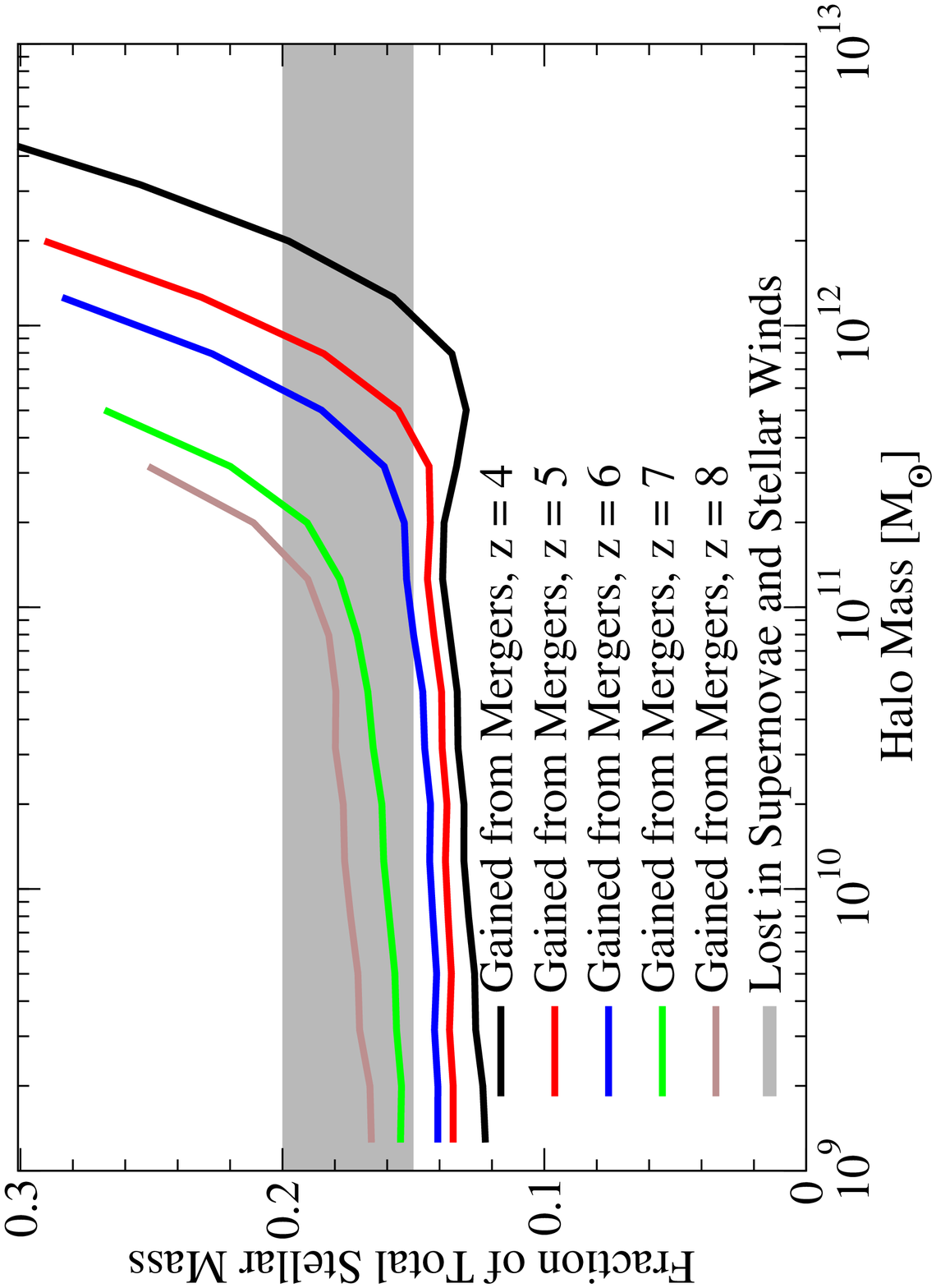}\plotgrace{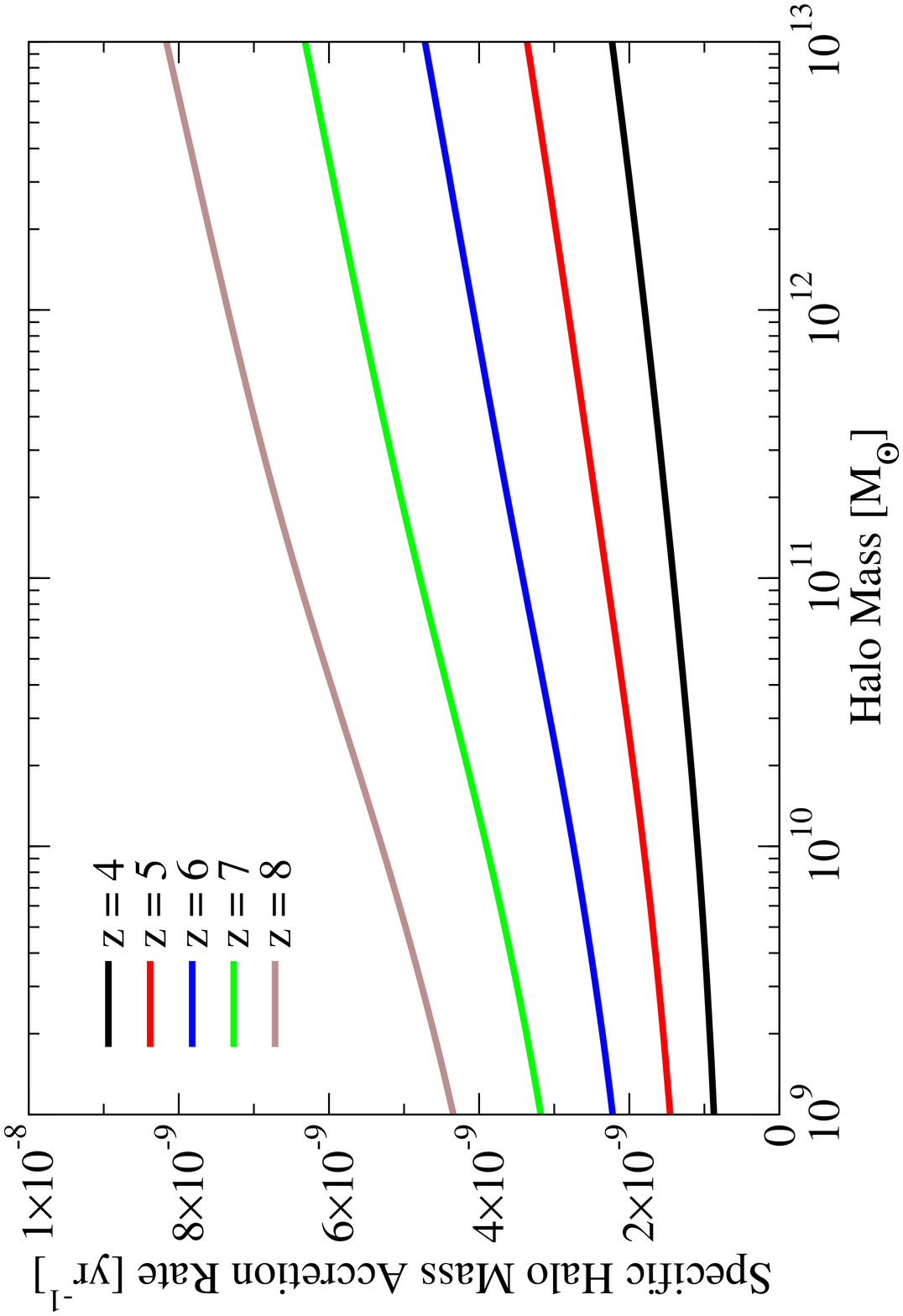}
\caption{Systematic error considerations.  \textbf{Left} panel: A comparison of the stellar mass lost through supernovae and stellar winds to the stellar mass gained in mergers for galaxies at $4<z<8$, calculated from \cite{BWC12}.  In all cases, the stellar mass lost is very similar to the stellar mass gained. \textbf{Right} panel: average specific halo mass accretion rates, as a function of halo mass and redshift.  The specific halo mass accretion rate depends only weakly on halo mass.}
\label{f:gain_loss}
\end{figure*}

Galaxies at time $t_f$ in a given stellar mass bin will have a well-defined average stellar mass, $M_{\ast,f}$, and a well-defined average halo mass, $M_{h,f}$.  The halo mass is determinable through abundance matching/modeling if it is not known through other means ($\S \ref{s:amodel}$).  In this paper, we parametrize the average growth of the galaxies in terms of the average growth of their host dark matter halos, i.e., as $M_\ast(M_h(t))$.  The changing cosmological accretion rate is captured in the average halo mass growth history, $M_h(t)$.  The historical efficiency with which the galaxies converted infalling baryons into stars is captured in how $M_{\ast}$ depends on $M_h$. We can write the galaxies' average star formation histories in terms of $M_\ast(M_h)$ as
\begin{equation}
SFR(M_h(t)) = \frac{dM_\ast}{dt} = \frac{dM_\ast}{dM_h} \frac{dM_h}{dt}.
\label{e:sfr}
\end{equation}
Just as for $SFR(t)$-parametrized star formation histories, the form of $M_\ast(M_h)$ will depend on the redshift and stellar mass of the galaxies in question.\footnote{We note that stellar mass loss (through supernovae and stellar winds) and stellar mass gain through mergers will also change $M_\ast$.  These effects cancel to within 10 percent (see \S \ref{s:effects} and Fig.\ \ref{f:gain_loss}), so we continue to write $SFR(M_h(t)) = \frac{dM_\ast}{dt}$ in this section for pedagogical clarity.}

As with fitting $SFR(t)$ directly, we have to make an assumption about the functional form of $M_\ast(M_h)$.  Regardless of its true form, we can approximate the recent stellar mass---halo mass history as a power law.  This form is especially reasonable for galaxies at $z>4$, which are typically not massive enough to experience AGN feedback, and so only experience stellar feedback (supernovae and reionization) over much of their growth histories.  We also test this assumption directly in \S \ref{s:tests} for galaxies from $z=4$ to $z=8$.  So, we let
\begin{equation}
M_\ast(M_h) = M_{\ast,f} \left(\frac{M_h}{M_{h,f}}\right)^\alpha
\label{e:smhm}
\end{equation}
where $\alpha$ is the unknown power-law dependence.  We re-emphasize that Eq.\ \ref{e:smhm} describes the historical evolution of a galaxy population, so $\alpha$ differs from the local slope of the stellar mass---halo mass (SMHM) relation at time $t_f$ if (and only if) the SMHM relation is evolving with redshift.

Combining Eq.\ \ref{e:smhm} with Eq.\ \ref{e:sfr}, the star formation rate history becomes a simple expression:
\begin{equation}
SFR(t) = \frac{\alpha M_{\ast}}{M_h} \frac{dM_h}{dt}
\label{e:sfr_a}
\end{equation}
Letting $SSFR(t)$ be the specific star formation rate ($M_\ast^{-1} \frac{dM_\ast}{dt}$) as a function of time, and letting $SMAR(t)$ be the halo specific mass accretion rate ($M_h^{-1} \frac{dM_h}{dt}$), the beautiful symmetry in Eq.\ \ref{e:sfr_a} is apparent:
\begin{equation}
\frac{SSFR(t)}{SMAR(t)} = \alpha.
\label{e:ssfr}
\end{equation}
That is, the ratio of a galaxy population's average SSFR to its average specific host halo mass accretion rate will be constant, under the weak assumption that the recent historical stellar mass---halo mass relation for the population's progenitors has a power-law form.

If we remove the assumption of a power-law growth history, we note that this ratio still describes the recent growth of the galaxies:
\begin{equation}
\label{e:ssfr_smar}
\frac{SSFR(t)}{SMAR(t)} = \frac{\frac{d\log M_\ast}{dt}}{\frac{d\log M_h}{dt}} = \frac{d\log M_\ast}{d\log M_h}
\end{equation}
which is valid as long as $SSFR(t)$ and $SMAR(t)$ are both positive.  As shown in Fig.\ \ref{f:schematic}, Eq.\ \ref{e:ssfr_smar} is really a geometric statement about the logarithmic stellar mass---halo mass plane.  The SSFR sets the galaxies' velocity along the logarithmic stellar mass axis, and the SMAR sets the velocity along the logarithmic halo mass axis.  The ratio of these two velocities (``rise over run'') corresponds to the slope of the galaxies' trajectory in this plane; this slope will be constant as long as a power-law relationship between the galaxies stellar masses and halo masses holds (Eq.\ \ref{e:smhm}).

Eqs.\ \ref{e:ssfr} and \ref{e:smhm} therefore yield a simple way to constrain galaxy progenitors.  For this method, two galaxy stellar mass functions at nearby redshifts would suffice for observational constraints.  The remaining inputs include host halo masses for the galaxies, which may be found by abundance matching/modeling (\S \ref{s:amodel}); halo mass accretion histories, determined via dark matter simulations (\S \ref{s:sims}); and galaxy specific star formation rates, which are set by the growth of the stellar mass function between the two redshifts (\S \ref{s:ssfrs}, \ref{s:amodel}).  The ratio of the galaxies' specific star formation rates to their host halos' specific halo mass accretion rates (Eq.\ \ref{e:ssfr}) then yields the power-law slope of the progenitors' historical stellar mass---halo mass relation.  This can be used to infer basic information about the galaxies' progenitors, including the progenitors' star formation histories, specific star formation rates, stellar masses, and halo masses.  As no observations of the progenitors are required, this method can yield constraints on galaxy populations which are otherwise too faint or at too high of a redshift to be observed directly.

We note in passing that Eq.\ \ref{e:ssfr_smar} applies for any pair of coupled variables.  E.g., one may also model the growth of black holes in galaxies using the ratio of specific black hole mass accretion rates to specific star formation rates.  Current research in the coevolution of black holes in galaxies has focused on the correlation between black hole accretion rates and star formation rates \citep[see, e.g.][]{Silverman08,Aird10,Mullaney12,Conroy13b,Chen13,Hickox14}.  Considering the relationship between the specific rates instead may be an interesting avenue for future research.

\subsection{Stellar Mass Loss, Mergers, Scatter, and the Initial Mass Function}

\label{s:effects}

A galaxy's stellar mass can change not only because of new star formation (Eq.\ \ref{e:sfr}), but also through stellar mass loss (stellar winds and supernovae) and mergers of smaller galaxies.  We adopt the stellar mass loss fraction as a function of time from \cite{BWC12}, which is based on the FSPS population synthesis model \citep{Conroy09,Conroy10} and the initial mass function of \cite{Chabrier03}:
\begin{equation}
\label{e:sm_loss}
f_\mathrm{loss}(t) = 0.05 \ln\left(1+ \frac{t}{1.4\;\mathrm{Myr}}\right).
\end{equation}
The logarithmic form results in a very rapid 15\% loss over the first 25 Myrs, followed by another more gradual loss of 15\% over the next 500 Myrs.  For the high-redshift ($z>4$) galaxies we consider, typical specific star formation rates are on the order of 5--10 $\times 10^{-9}$ yr$^{-1}$ \citep{BWC12}, which lead to characteristic formation timescales of 100--200 Myrs.  Since these timescales fall in the gradual loss regime, the overall stellar mass loss is quite insensitive to the exact star formation histories, and is in the range 15--20\%.

We next calculate the expected gain in stellar mass from mergers.  This is a convolution between halo merger rates and the stellar mass---halo mass relation, both of which we take from constraints in \cite{BWC12}.  As shown in the left panel of Fig.\ \ref{f:gain_loss}, merging halos contribute about 12--18\% of the total stellar mass in most galaxies for $4<z<8$; very massive galaxies may receive up to a 30\% contribution.  This small fraction can be understood from the fact that the stellar mass to halo mass ratio declines with halo mass, so most of the incoming stellar mass will come from major mergers \citep{BehrooziRpeak}.  The aforementioned star formation timescales ($\sim$150 Myrs) are shorter than the expected major merger timescales ($\sim$1 per 300 Myr from $z=8$ to $z=6$; \citealt{Fakhouri08,Fakhouri10,BWC12}), meaning that mergers contribute a minority of the stellar mass growth.

As a result, mergers and stellar mass loss nearly cancel each other's effects (Fig.\ \ref{f:gain_loss}, left panel), representing in only a 5--10\% total correction to $\alpha$ in Eq.\ \ref{e:ssfr}.  From $z=8$ to $z=4$, halos grow by a factor of $\sim 1$ dex, so this correction would result in a $0.1$--$0.2$ dex error in the inferred stellar masses over this range (see also Eq.\ \ref{e:errs}).  This is well within typical 0.3 dex systematic errors for stellar masses \citep{Conroy09,Behroozi10,BWC12}.  All the same, we correct the stellar mass in Eq.\ \ref{e:ssfr} to remove stellar mass from mergers and to include stellar mass lost due to passive evolution.  This process is described in \S \ref{s:amodel}.

Another potential source of error is scatter in halo mass at fixed stellar mass.  The average specific halo mass accretion rate is an extremely weak function of halo mass (Fig.\ \ref{f:gain_loss}, right panel).  E.g., a change of $\pm$0.3 dex in halo mass corresponds to a change of $\pm$ 5.5\% in the specific mass accretion rate for $10^{12}\Msun$ halos.  However, galaxy mass may correlate with a halo's recent accretion rate, which can bias the galaxies' host halo specific accretion rates.  We avoid these kinds of selection biases by directly constraining the average stellar mass and the average star formation rate as a function of halo mass (as in \citealt{BWC12}), as discussed in \S \ref{s:amodel}.  This allows us to select galaxies at fixed halo mass; the corresponding average specific halo mass accretion rates can then be calculated in an unbiased way from a dark matter simulation (\S \ref{s:sims} and Fig.\ \ref{f:gain_loss}, right panel).

Finally, the choice of initial mass function (IMF) has a minimal effect.  The primary effect of switching, e.g., to a \cite{Salpeter55} IMF would be to identically rescale the SFR and stellar mass, leaving their ratio unchanged \citep{Conroy09}.  Although the stellar mass normalization would be affected, the predicted evolution (from Eq.\ \ref{e:ssfr}) would remain the same.  A secondary effect would be to change the inferred stellar mass loss by 5--10\%, depending on galaxy metallicity \citep{Shimizu13}, but this would not change the ratio of the star formation rate to the total stellar mass ever formed, which is what is relevant for Eq.\ \ref{e:ssfr}.  However, we note the possibility that the IMF may change to be more top-heavy at high redshifts for reasons including lower metallicities, high gas opacity, ultracompact star clusters, tidal shear or rising CMB temperatures \citep[see][for reviews]{Scalo05,Bonnell07}.  Some specific recent examples are given for starbursts in \cite{Weidner11}, for low metallicity at high redshift in \cite{Narayanan13}, for ultracompact star clusters and low metallicity in \cite{Marks12}, and disturbed galaxies in \cite{Habergham10}.

\section{Constraints from Observations and Simulations}

We discuss direct specific star formation rate measurements in \S \ref{s:ssfrs}, our method for constraining stellar masses and specific star formation rates as a function of halo mass and redshift in \S \ref{s:amodel}, and the dark matter simulations we use in \S \ref{s:sims}.

\label{s:constraints}

\subsection{Measuring Specific Star Formation Rates}

\label{s:ssfrs}

\begin{figure}
\plotgrace{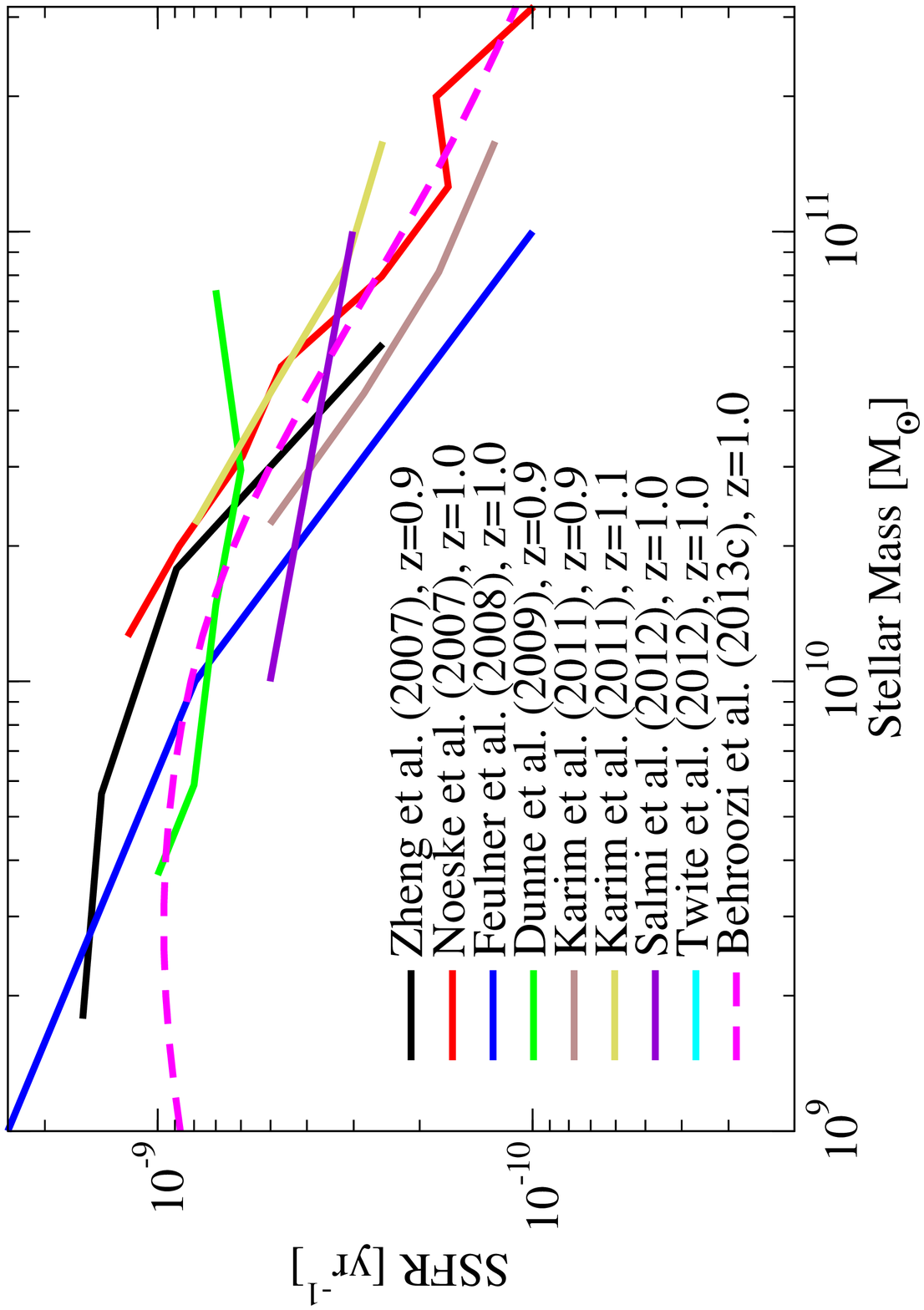}
\caption{Directly measured SSFRs at $0.9 < z < 1.1$ from recent literature \citep{Zheng07,Noeske07,Feulner08,Dunne09,Karim11,Salmi12,Twite12} show significant scatter in both  amplitudes and slopes.  Except where specified, we adopt the specific star formation rates from the meta-analysis technique in \cite{BWC12}, which incorporates these and other published constraints on galaxy SSFRs as well as constraints from the growth of the stellar mass function from $z=8$ to $z=0$.}
\label{f:ssfr_z1}
\end{figure}

Specific star formation rates are difficult to measure directly.  This is most clear at low redshifts, where many independent determinations are available; e.g., Fig.\ \ref{f:ssfr_z1} compares several direct measurements of the $z=1$ SSFR from recent literature.  These have all been normalized to a \cite{Chabrier03} Initial Mass Function (IMF); however, significant disagreement is present both in the amplitudes and slopes of the measurements.  \cite{BWC12} found that this level of scatter ($\pm0.3$ dex) between different publications' direct measurements is present at all redshifts---even for those now considered nearby ($z<0.5$).

The magnitude of these uncertainties is important, because they directly translate into the uncertainty in the power-law slope of the star formation efficiency inferred from Eq.\ \ref{e:ssfr} (see \S \ref{s:tests}).  For that reason, it is not appropriate to rely on specific SFR measurements from a single source.  We instead rely on the meta-analysis technique described in \cite{BWC12} and \S \ref{s:amodel}, which combines constraints from many different direct SSFR measurements with constraints from the growth rate of the stellar mass function.  For our methodology tests (\S \ref{s:tests}), we restrict the meta-analysis to data obtained at $z \le 5$; for our predictions (\S \ref{s:results}), we use the full data for $z\le 8$.

As shown in Fig.\ \ref{f:ssfr_z1}, this meta-analysis process is largely equivalent to taking the median of the published direct SSFR measurements.  However, for completeness, we discuss the results from using single-source direct measurements of the SSFR in Appendix \ref{a:single_ssfr}.

\subsection{Constraining the Stellar Mass---Halo Mass Relation and Specific Star Formation Rates through Abundance Modeling}

\label{s:amodel}

We here distinguish abundance \textit{modeling} from abundance \textit{matching}.  The latter nonparametrically constrains the stellar mass---halo mass (SMHM) relation by assigning galaxies ranked by decreasing stellar mass to halos ranked by decreasing halo mass within equal volumes.   Abundance modeling instead assumes a parametric form (usually informed by abundance matching) for the SMHM relation and uses a Markov Chain Monte Carlo (MCMC) method to constrain the posterior distribution for the SMHM parameters.  This approach benefits because observational constraints besides the stellar mass function (SMF) at a single redshift may be readily incorporated into the MCMC likelihood function.  Examples include SMFs at multiple redshifts \citep{moster-09,Moster12,Behroozi10,BWC12,Yang11,Bethermin12,Wang12,Lu14}, star formation rates \citep{BWC12}, correlation functions \citep{Yang11,Leauthaud12,Leauthaud11,Tinker13}, conditional stellar mass functions \citep{Yang11,Lu14}, and weak lensing constraints \citep{Leauthaud12,Leauthaud11,Tinker13}.  The ability to include systematic uncertainties and covariances in the likelihood function is also advantageous \citep{Behroozi10,BWC12,Yang11}.

We follow the same technique used in \cite{BWC12}, and we refer readers to that paper for full details.  Briefly, we adopt a six-parameter functional form to describe the SMHM relation at a single redshift.  This form includes parameters for a characteristic halo mass and stellar mass, a faint-end power-law slope, a massive-end turnoff, a shape for the transition between faint- and massive-end behavior, and the scatter in stellar mass at fixed halo mass.  For each parameter, three variables control the evolution at low ($z<0.5$), medium ($0.5<z<2$) and high ($z>2$) redshifts, giving a total of 18 variables to describe the evolution of the SMHM relation.  We additionally include nuisance parameters for systematic effects, including systematic offsets in recovered stellar masses for active and passive galaxies, random errors in stellar masses, incompleteness at high redshifts due to dusty and/or bursty star formation, and the fraction of stellar growth due to \textit{in situ} star formation vs.\ \textit{ex situ} mergers.

Every location in this parameter space corresponds to a specific choice for the SMHM relation, $SM(M_h,z)$, as a function of halo mass and redshift.  For the halo mass definition, we use the peak historical virial \citep{mvir_conv} halo mass along halos' growth trajectories ($M_\mathrm{peak}$), which was found by \cite{Reddick12} to be the best halo mass proxy for reproducing galaxy clustering and luminosity functions.  
Each choice of $SM(M_h,z)$ then represents a unique way to assign stellar masses for every halo at every timestep in a dark matter simulation. The expected stellar mass function at any redshift is therefore set by the number density of galaxies in dark matter halos; expected specific star formation rates and cosmic star formation rates are set by the growth of galaxies along dark matter merger trees.  The closeness of these expectations to observed constraints gives the relative likelihood for the chosen parameter set; an MCMC algorithm then determines the posterior distribution of allowable SMHM relations, as well as the implied specific star formation rates as a function of halo mass and redshift (required by \S \ref{s:effects} and \S \ref{s:ssfrs}).  As the dark matter halo merger trees allow full bookkeeping for the amount of stellar mass accreted in mergers and lost through passive stellar evolution, these corrections to galaxy stellar masses can be readily subtracted, as discussed in \S \ref{s:effects}.

For observational constraints, we use the compilation in \cite{BWC12}.  This includes stellar mass functions,\footnote{Stellar mass functions: \cite{Baldry08,perezgonzalez-2008,marchesini-2008,Marchesini10,Stark09,Mortlock11,Lee11b,Bouwens11,BORG12,Moustakas12}.} specific star formation rates,\footnote{Specific star formation rates: \cite{Yoshida06,Salim07,Zheng07,Smolcic09,Shim09,LeBorgne09,Dunne09,vdBurg10,Kajisawa10,Ly10,Ly11,Rujopakarn10,Cucciati11,Robotham11,Tadaki11,Magnelli11,Karim11,Bouwens11b,Sobral12}.} and cosmic star formation rates\footnote{Cosmic star formation rates: \cite{Salim07,Noeske07,Zheng07,Daddi07,Feulner08,Kajisawa10,Schaerer10,Lee11,Tadaki11,Karim11,McLure11,Gonzalez12,Reddy12,Twite12,Whitaker12,Salmi12,Labbe12}.} from $z=8$ to $z=0$.  For the tests in \S \ref{s:tests}, we have also performed a restricted analysis excluding all verification data---i.e., all data at $z>5$.

The main output of this model is the posterior distribution for $M_\ast(M_h, z)$---i.e., the stellar mass as a function of halo mass and redshift.  As averaged star formation rates as a function of halo mass and redshift are available from the intermediate steps in the calculation, we can straightforwardly calculate the posterior distribution of specific star formation rates as a function of redshift and halo mass.

\subsection{Dark Matter Simulations and Halo Mass Accretion Rates}
\label{s:sims}

Halo properties are derived primarily from the \textit{Bolshoi} simulation \citep{Bolshoi}.  Bolshoi follows $2048^3$ ($8.6\times 10^9$) dark matter particles in a 250 Mpc $h^{-1}$ comoving, periodic box from $z=80$ to $z=0$ using the \textsc{art} code \citep{kravtsov_etal:97,kravtsov_klypin:99}.  Its mass and force resolution ($1.9\times 10^8\Msun$ and $1$ kpc $h^{-1}$, respectively) allow it to resolve dark matter halos down to $10^{10}\Msun$.  The assumed cosmology is a flat, $\Lambda$CDM cosmology with parameters $\Omega_m = 0.27$, $\Omega_\Lambda = 0.73$, $h=0.7$, $\sigma_8 = 0.82$, and $n_s = 0.95$; these are very close to the WMAP9+BAO+$H_0$ best-fit values \citep{WMAP9}.  Dark matter halos were found using the \textsc{Rockstar} phase-space temporal halo finder \citep{Rockstar}; merger trees were assembled using the \textsc{Consistent Trees} code \citep{BehrooziTree}, which reconstructs missing halos in merger trees to preserve gravitational consistency across simulation timesteps.

As discussed in \S \ref{s:basic} and \S \ref{s:amodel}, our approach requires the average growth rate, $\langle\dot{M}_\mathrm{peak}\rangle$, for halos as a function of redshift and mass.  \cite{BWC12} already determined the \textit{median} mass accretion histories for halos in Bolshoi; we therefore build on this existing fit, which is expressed as a function of the $z=0$ halo mass and the scale factor.  The full calibration and fitting procedure are discussed in Appendix \ref{a:mar_calibration}.  For the halo mass function, we adopt the modified \cite{tinker-umf} form in \cite{BWC12}, which was fit to the Bolshoi halo mass function for $0<z<8$.

Bolshoi's constraints on halos weaken above $z=8$, due to its mass completeness limit.  However, extrapolating the \cite{BWC12} halo mass function fit to higher redshifts compares well to the fit in \cite{Watson13b}, which is calibrated to $z=26$.  We find maximum differences of 15--25\% between the two mass functions from $z=10$ to $z=15$; these are consistent with systematic biases from the choice of halo finder \citep{Knebe11,Knebe13,Watson13b}.  We also have compared to the average mass accretion histories in \cite{Wu13b}, finding a maximum difference of 5\% in the specific mass accretion rates at $z>10$.

\begin{figure*}
\vspace{-8ex}
\plotgrace{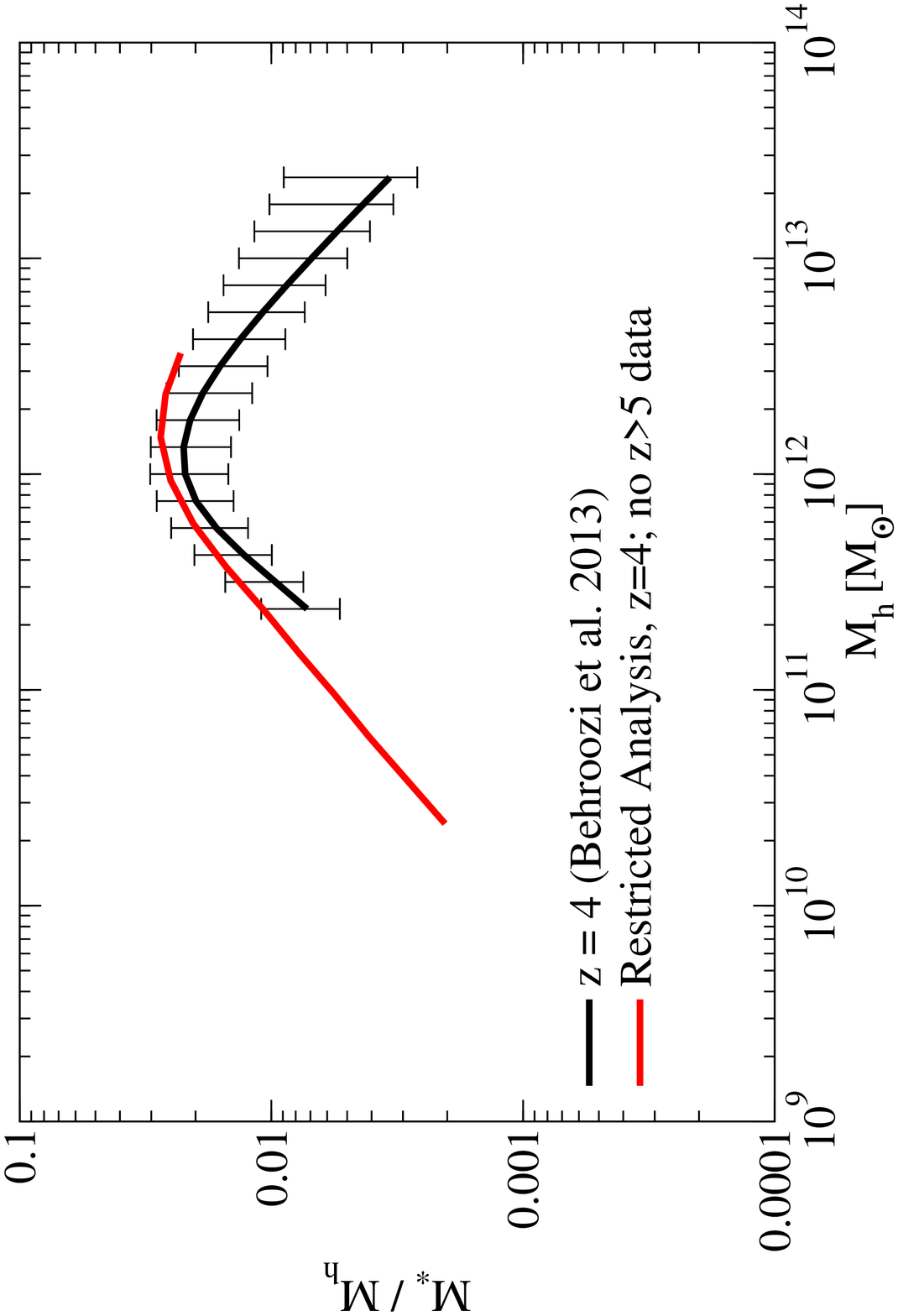}\plotgrace{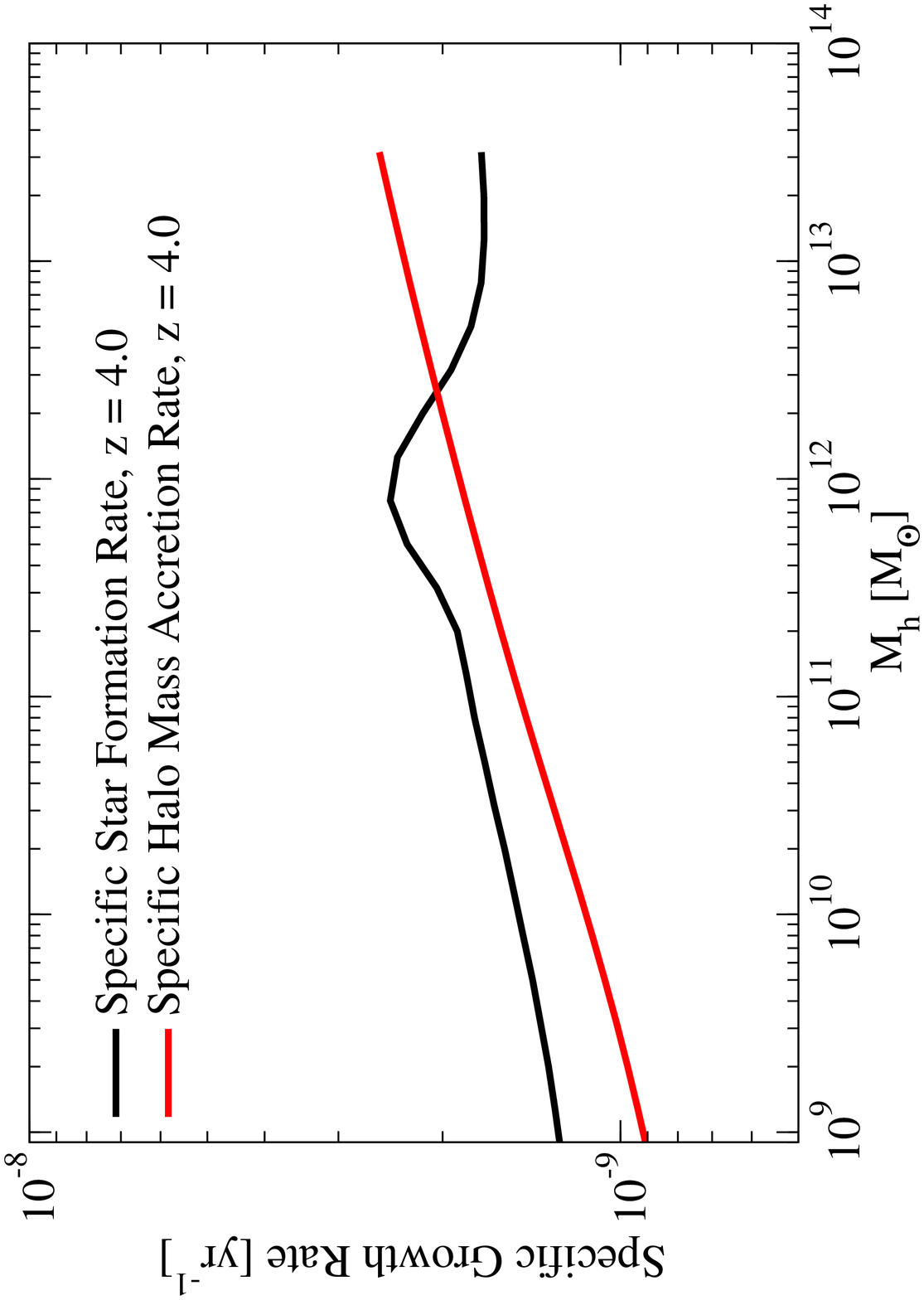}\\[-8ex]
\plotgrace{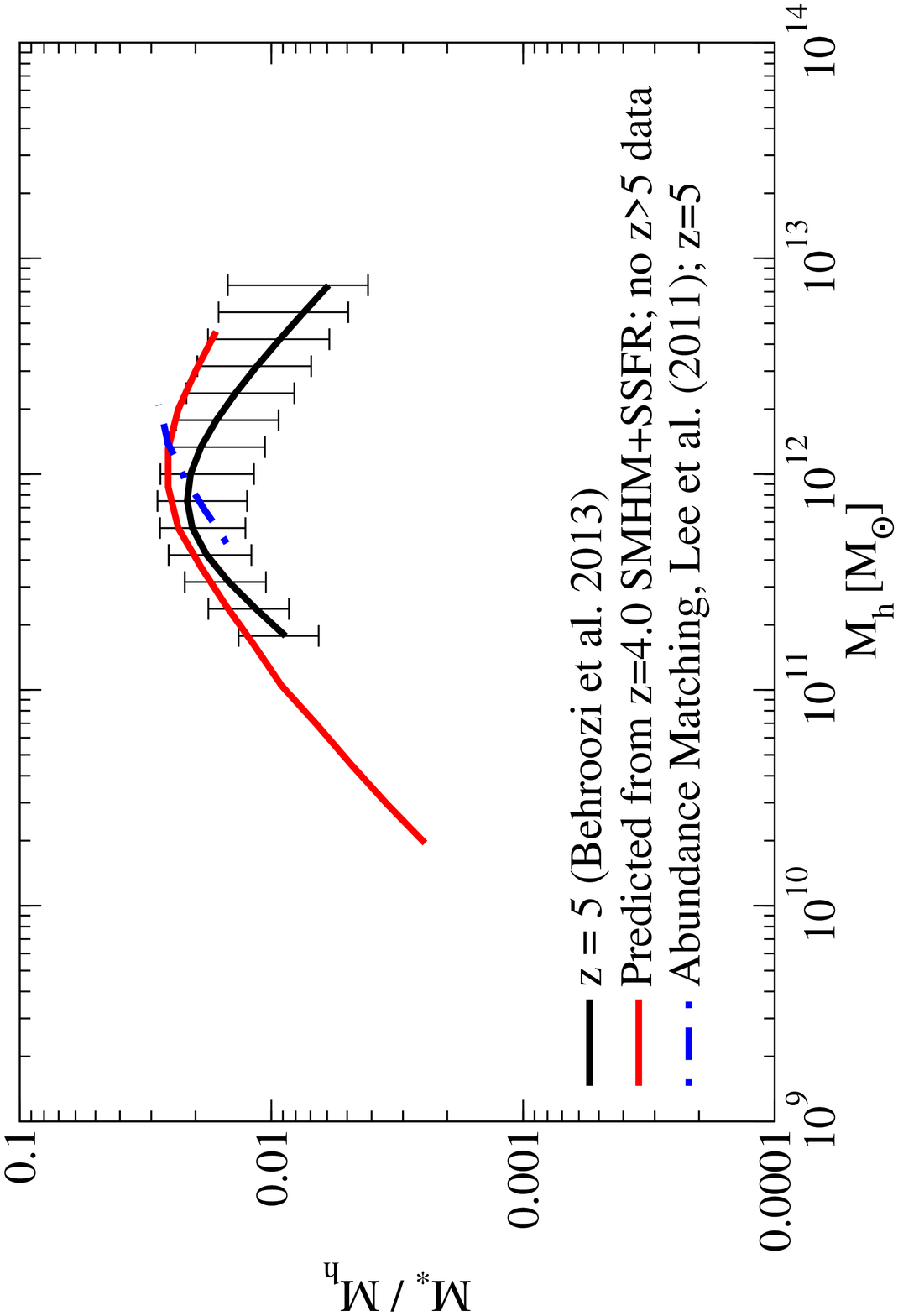}\plotgrace{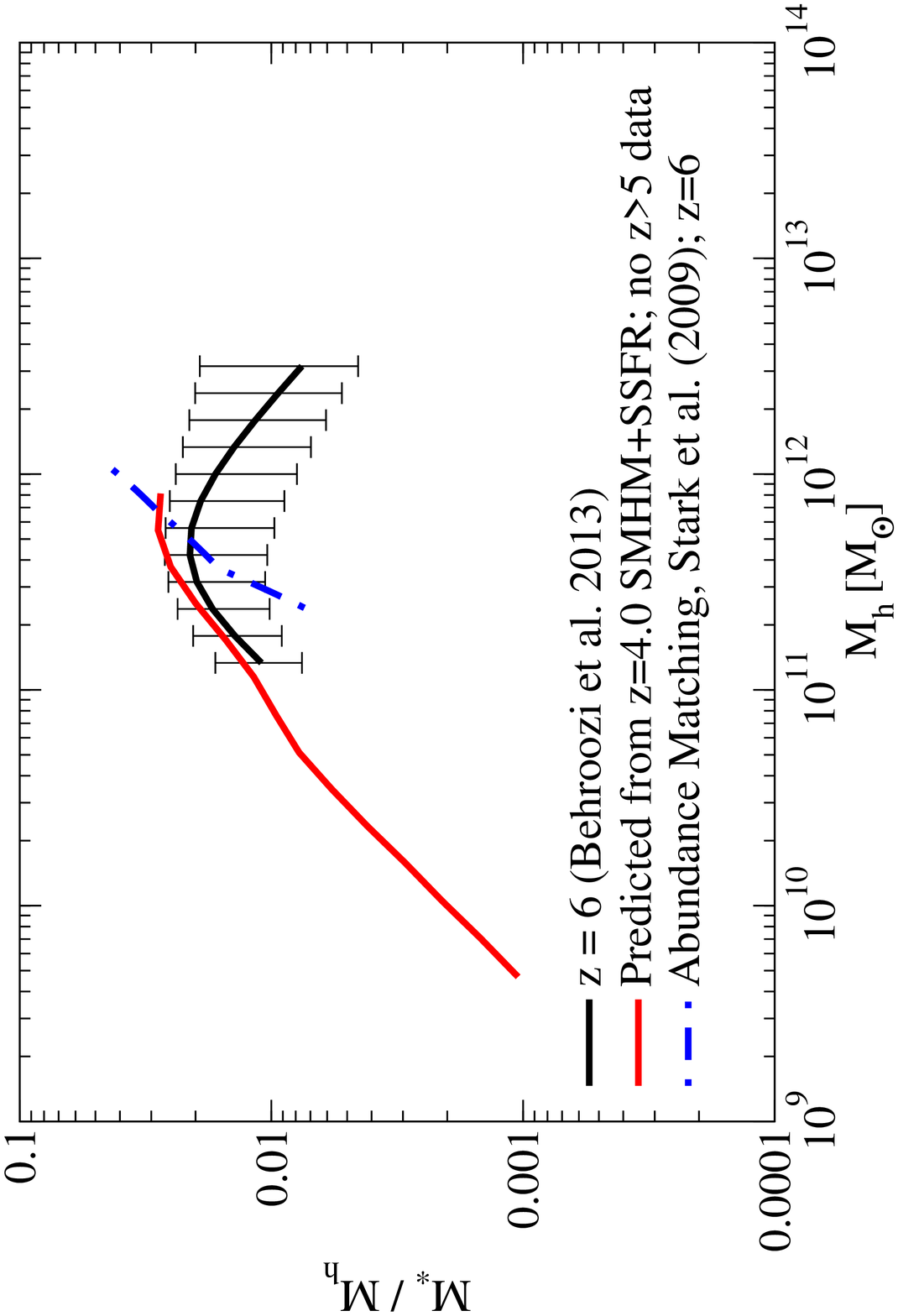}\\[-8ex]
\plotgrace{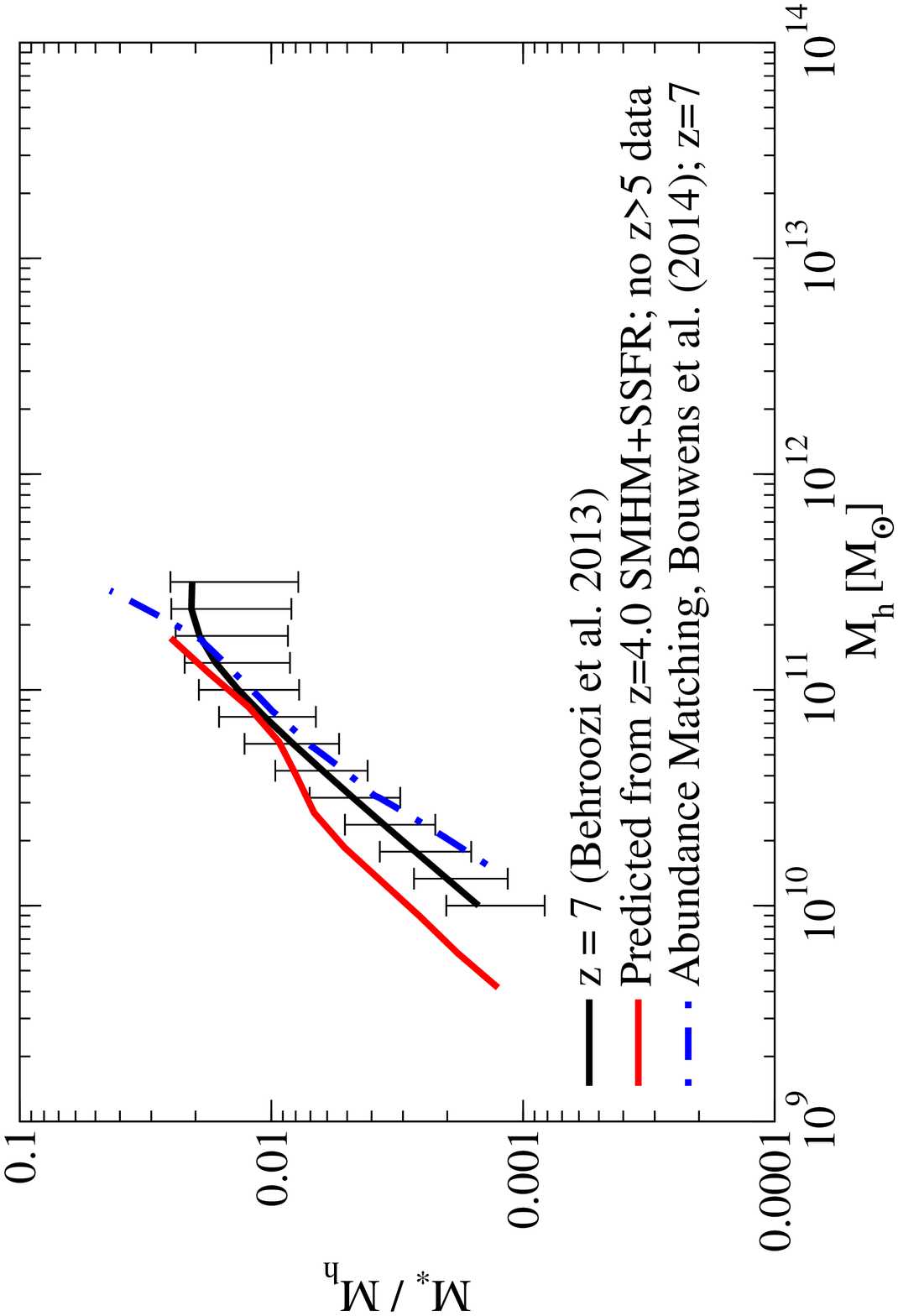}\plotgrace{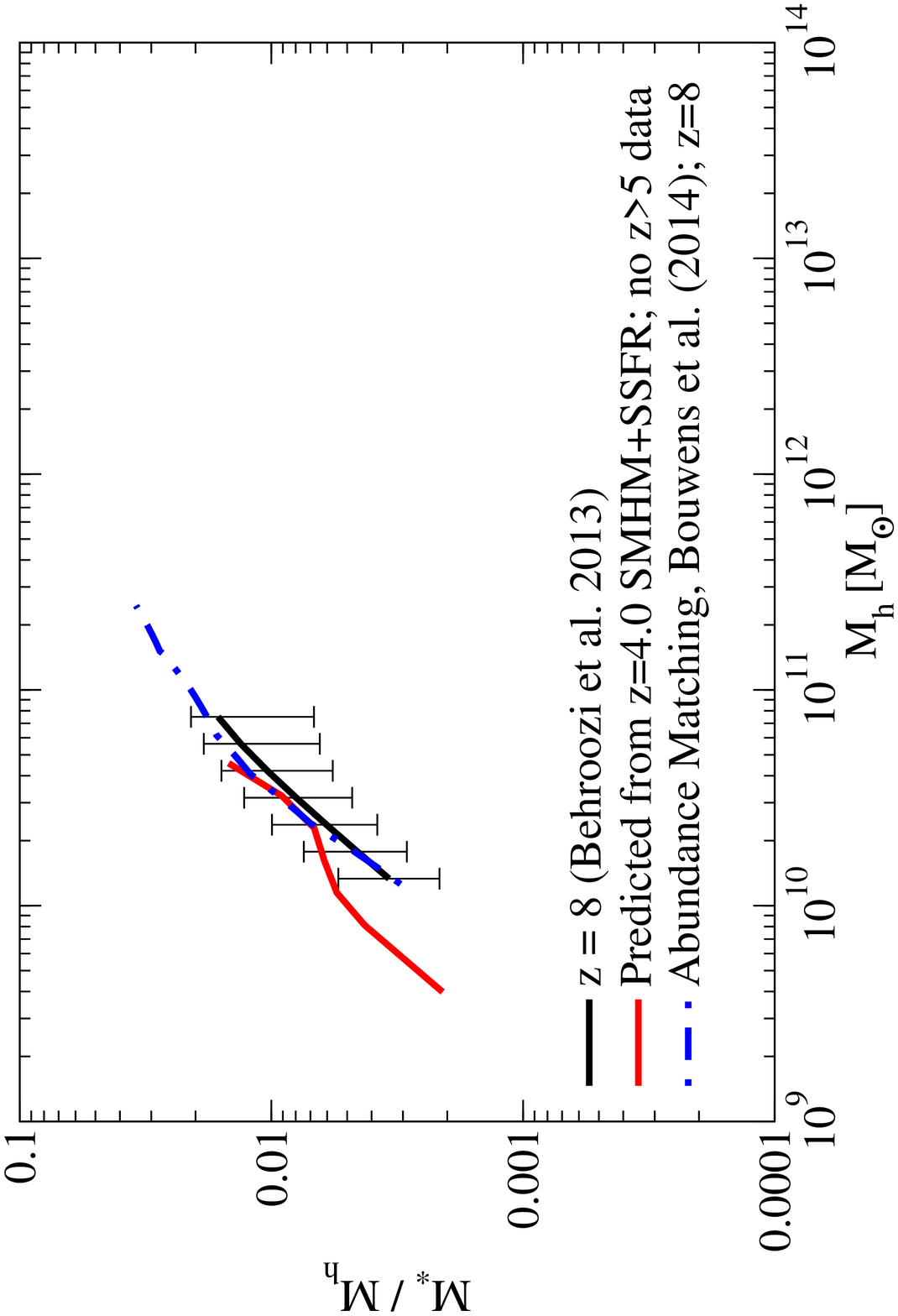}\\[-3ex]
\caption{Tests of the method in \S \ref{s:theory} for predicting high-redshift stellar mass to halo mass (SMHM) ratios.  \textbf{Top} panels: Inputs for predicting galaxy evolution (Eq.\ \ref{e:ssfr}).  These include the SMHM ratio (top-left panel), the specific star formation rate (top-right panel), and the specific halo mass accretion rate (top-right panel) at $z=4$.  The SMHM ratio and SSFRs were derived from the method in \S \ref{s:amodel}, excluding all observational constraints at $z>5$.  \textbf{Middle} and \textbf{Bottom} panels: predictions from Eqs.\ \ref{e:ssfr} and \ref{e:smhm}, compared to constraints from \cite{BWC12}.  Error bars in the \cite{BWC12} results represent 68\% confidence intervals, and are largely dominated by observational systematics.  The blue lines show nonparametric abundance matching results for $z=5$ to $z=8$, assuming zero scatter between stellar mass and halo mass; the stellar mass functions used at each redshift are denoted in the figure legends.  For the highest redshifts, stellar mass functions were derived from \cite{Bouwens14} luminosity functions using the conversion in \cite{Gonzalez10}; see \cite{BWC12} for full details.}
\label{f:pred_z4}
\end{figure*}

\begin{figure}
\vspace{-4ex}
\plotgrace{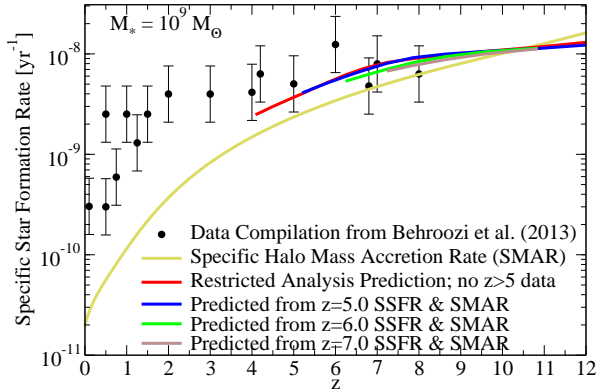}
\caption{Observed specific star formation rates (SSFRs) for $10^{9}\Msun$ galaxies compared to the specific mass accretion rates (SMARs) of their host halos; corrections for nebular emission lines are included.  See similar figure in \cite{Weinmann11}.  A high SSFR to SMAR ratio (e.g., at $z<2$) means that the fraction of incoming baryons converted into stars is currently much higher than it was for the galaxies' progenitors (see \S \ref{s:theory}).  A ratio near unity (e.g., at $z>4$) means that the baryon conversion fraction has not increased as significantly.  Also shown are extrapolations of SSFRs to $z=12$ from Eq.\ \ref{e:ssfr}, calculated from the SSFR to SMAR ratio at $z=4$, 5, 6, or 7, respectively. (See \S \ref{s:tests} and \S\ref{s:pred_ssfr}).  The lack of $10^{9}\Msun$ galaxies at $z>12$ prevents further extrapolation from being useful.}
\label{f:obs_ssfr}
\end{figure}

\begin{figure}
\vspace{-4ex}
\plotgrace{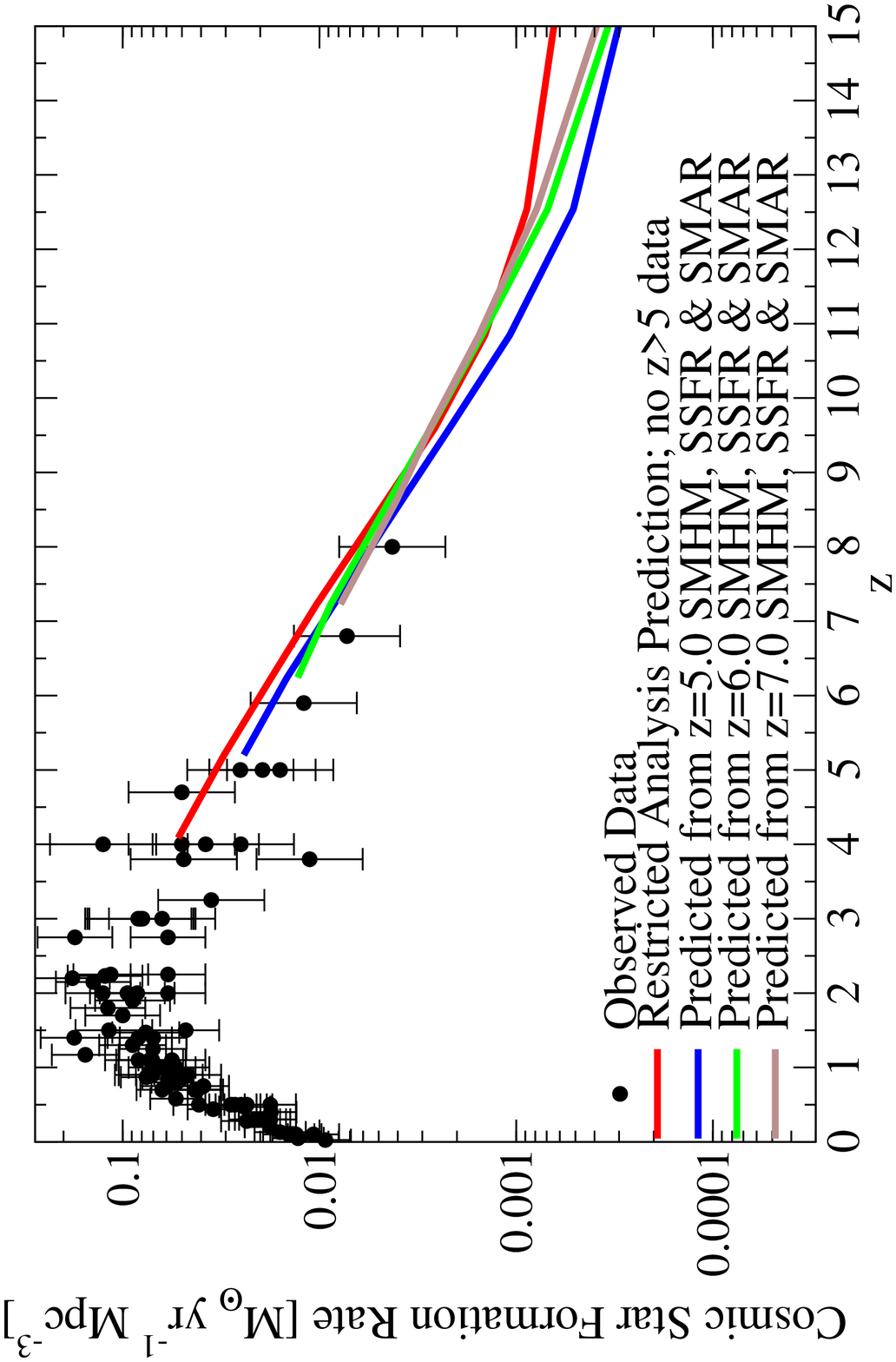}
\caption{A comparison of observed cosmic star formation rates (CSFRs; data sources from \citealt{BWC12}) to extrapolations of CSFRs to $z\sim 15$ from Eq.\ \ref{e:ssfr}, assuming that the ratio of the SSFR to the SMAR is fixed to its value at $z=4$, 5, 6, or 7, respectively. (See text).  Predicted observations assume a UV luminosity threshold of $M_{1500,AB} < -18$ for detecting individual galaxies.}
\label{f:obs_csfr}
\end{figure}

\section{Tests and Error Analyses}
\label{s:tests}

\subsection{The Stellar Mass---Halo Mass Relation}

\label{s:t_smhm}

In this section, we check how well Eqs.\ \ref{e:smhm}--\ref{e:ssfr} can be used to derive the $z>4$ stellar mass--halo mass (SMHM) relation from the $z=4$ galaxy SMHM relation, SSFRs, and halo specific mass accretion rates.  Starting at $z=4$ is motivated because we approximated galaxies' historical stellar mass---halo mass (SMHM) relationships as pure power laws (Eq.\ \ref{e:smhm}).  This approximation is most appropriate if the galaxy's history is dominated by a single feedback mode (e.g., stellar feedback) for star formation (see \S \ref{s:basic}); at $z\le 4$, AGN-mode or other feedback in massive galaxies results in rapidly declining star formation histories \citep{BWC12}.\footnote{As discussed in Appendix \ref{a:single_ssfr}, Eqs.\ \ref{e:smhm}--\ref{e:ssfr} still give correct results for galaxies with more complicated histories, although only for the galaxies' recent past.}  Note that while any starting redshift $z\ge 4$ is reasonable, we choose $z=4$ because it gives the longest redshift baseline ($\Delta z = 4$ for galaxies up to $z=8$) over which we can test the predictions of Eqs.\ \ref{e:smhm}--\ref{e:ssfr}.

Predicting the high-redshift SMHM relation is only a fair test if no high-redshift data has been used as a constraint.  To that end, we have rerun the abundance modeling framework discussed in \S \ref{s:amodel} using observational data constraints covering $0 < z \le 5$.  As noted in \S \ref{s:ssfrs} and \S \ref{s:amodel}, the inclusion of stellar mass function data at $z=5$ allows the growth of the stellar mass function to serve as a constraint on galaxies' specific star formation rates at $z=4$.  The resulting best-fit SMHM relation and SSFRs at $z=4$ are shown in comparison to the results of \cite{BWC12} (which used data constraints from $0<z\le 8$) in the top panels of Fig.\ \ref{f:pred_z4}.  The best-fitting SMHM relation of the restricted analysis is (as expected) slightly different from the best-fitting result of \cite{BWC12}, but it is still well within the observational error bars.

The quantitative predictions from assuming that galaxies' ratios of SSFRs to halo specific mass accretion rates (SMARs) remain constant (Eq.\ \ref{e:ssfr}) over $5\le z \le 8$ are shown in the middle and lower panels of Fig.\ \ref{f:pred_z4}.  These predictions are extremely consistent with constraints on the evolution of the stellar mass---halo mass relationship to the highest available redshift ($z=8$) from \cite{BWC12}.  They are also consistent with nonparametric abundance matching of stellar mass functions at those redshifts (Fig.\ \ref{f:pred_z4}), within the expected systematic observational errors.  While the shape and amplitude of the SMHM relation remain relatively constant, the halo mass corresponding to peak integrated star formation efficiency evolves from $\sim 10^{12}\Msun$ at $z=4$ to $\sim 10^{11}\Msun$ at $z=8$.

For $z=4$ galaxies, the ratio of their specific star formation rates to their specific mass accretion rates is fairly close to unity: $\alpha \sim$ 1--1.5 (Fig.\ \ref{f:pred_z4}, top-right panel).  If the SSFRs and SMARs were equal ($\alpha=1$), the relative growth rates of galaxy stellar mass and host halo mass are the same---so that the historical stellar mass--halo mass \textit{ratio} remains constant.  For $\alpha$ slightly greater than unity, this will still be approximately true, so that the historical SMHM ratios will have at most a weak dependence on their historical halo mass.  So, the galaxies forming stars at peak efficiency at $z>4$ will be the progenitors of the galaxies forming stars at peak efficiency at $z=4$.  Similarly, the halo mass at which the SMHM ratio reaches any chosen amplitude is also expected to evolve (see also \S \ref{s:pred_smhm}).

The uncertainties in these predictions grow as the redshift baseline increases.  At high redshifts, halo growth is very nearly exponential with redshift \citep{Wechsler02,McBride09,Wu13b,BWC12}; we find from the simulations in \S \ref{s:sims} that $\frac{dM_h}{dz}$ is $0.25$ dex.  From Eq.\ \ref{e:ssfr}, the errors in stellar mass predictions then increase as 
\begin{equation}
\label{e:errs}
\Delta M_\ast \approx -0.25 \epsilon \alpha \Delta z \textrm{ dex}
\end{equation}
where $\alpha\sim 1$ is the ratio of SSFRs to SMARs (see, e.g., Fig.\ \ref{f:pred_z4}), and $\epsilon$ is the fractional uncertainty in $\alpha$.  The predicted evolution and the fully-constrained evolution over $z=4$ to $z=8$ typically differ by $0.25$ dex or less (Fig.\ \ref{f:pred_z4}), which implies that $\epsilon$ is $\sim$25\% or less.

\subsection{Specific and Cosmic Star Formation Rates}

\label{s:t_ssfr}

We next calculate how Eq.\ \ref{e:ssfr} predicts that the SSFR for $10^{9}\Msun$ galaxies changes with redshift (Fig.\ \ref{f:obs_ssfr}), given SSFRs and SMARs at $z=4$ (Fig.\ \ref{f:pred_z4}).  At fixed redshift, more massive galaxies have lower specific star formation rates (SSFRs) and larger specific halo mass accretion rates (SMARs) than smaller galaxies (see, e.g., Fig.\ \ref{f:pred_z4}).  As a result, they would be expected to maintain a smaller SSFR---SMAR ratio over their growth history (Eq.\ \ref{e:ssfr}).  The more massive the galaxy is at $z=4$, the higher the redshift when it reached $10^{9}\Msun$.  Since more massive galaxies have lower SSFR---SMAR ratios (Fig.\ \ref{f:pred_z4}), we expect that the SSFR---SMAR ratio at fixed stellar mass will decline with increasing redshift.

This is exactly what is shown in Fig.\ \ref{f:obs_ssfr}.  While the specific mass accretion rate rises steadily with redshift, the declining SSFR--SMAR ratio leads to an apparent ``plateau'' \citep{Weinmann11}, which is extremely consistent with the observed specific star formation rates at high redshifts.

We also show the expected observed cosmic star formation rates in Fig.\ \ref{f:obs_csfr}, again calculated from galaxy SSFRs at $z=4$.  To calculate these, we applied the expected specific star formation rates and the expected stellar mass---halo mass relationship (\S \ref{s:t_smhm}) to the halo mass function \citep{BWC12}.  We integrated the resulting number density down to an assumed UV luminosity threshold of $M_{1500,AB} < -18$ (matching recent high-redshift observations \citep{Oesch13}), where UV luminosities were computed without including dust as described in Appendix \ref{a:lfs}.  As shown in Fig.\ \ref{f:obs_csfr}, the cosmic star formation rates expected from the SSFR to SMAR ratio of $z=4$ galaxies are extremely consistent with existing observations from $z=5$ to $z=8$.

We note that the uncertainties on specific star formation rates are not as sensitive to the redshift baseline because the errors are not cumulative.  The fractional uncertainty $\epsilon$ in the SSFR to SMAR ratio $\alpha$ translates directly into the fractional uncertainty in specific star formation rates.  As $\epsilon \lesssim 25\%$ (\S \ref{s:t_smhm}), this suggests that specific star formation rate predictions are extremely robust, as long as the assumption of a power-law historical stellar mass---halo mass relation (Eq.\ \ref{e:smhm}) holds.   Galaxy star formation rates will be the product of the SSFR and stellar mass, which will result in fractional errors of
\begin{equation}
\label{e:errs_sfr}
\frac{\Delta SFR}{SFR} = \log_{10}(1+\epsilon) - 0.25 \epsilon \alpha \Delta z \textrm{ dex}
\end{equation}
Since the total integrated stellar mass must remain constant, an error in $\alpha$ will result in redistributing when past cosmic star formation happened.  The error in Eq.\ \ref{e:errs_sfr} thus has a sign change when $\log_{10}(1+\epsilon) = 0.25 \epsilon \alpha \Delta z$.  Noting that $\log_{10}(1+\epsilon)$ is well-approximated by $\epsilon \log_{10}(e)$ for small $\epsilon$, this fractional error becomes zero for redshift baselines of $\Delta z = 4 \log_{10}(e) \alpha^{-1}$, regardless of what value $\epsilon$ actually takes(!)  For typical values of $\alpha$ ($\sim 1.25$), the total additional error from extrapolating the cosmic star formation rate is minimized for a redshift baseline of $\Delta z=1.2$.

\begin{figure}
\plotgrace{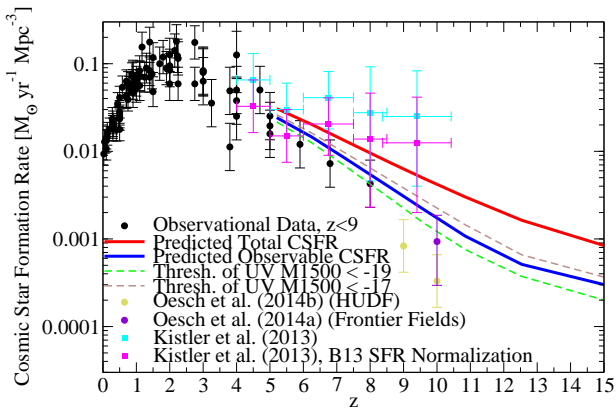}
\caption{Comparison of our predictions (using galaxy SSFRs at $z=5$) for high-redshift cosmic star formation rates to measurements from \cite{Oesch13} and \cite{Kistler13}.  The red line shows the predicted total cosmic star formation rate, and the blue line shows the predicted observed cosmic star formation rate, which includes all galaxies with UV luminosities $M_{1500,AB} < -18$.  The dashed lines show how the predictions would change if the threshold were changed to $M_{1500,AB} < -17$ or $M_{1500,AB} < -19$.  Our predictions are systematically higher than the \cite{Oesch13} results by 1-2 $\sigma$, although we are consistent with the \cite{Oesch14} results.  Our predictions for the total cosmic star formation rate agree well with \cite{Kistler13}, especially if their results are normalized to the $z=4$ cosmic star formation rate from the compilation in \cite{BWC12} (purple-colored points) instead of the older compilation in \cite{Hopkins06b} (light cyan-colored points).}
\label{f:oesch}
\end{figure}

\begin{figure*}
\plotgrace{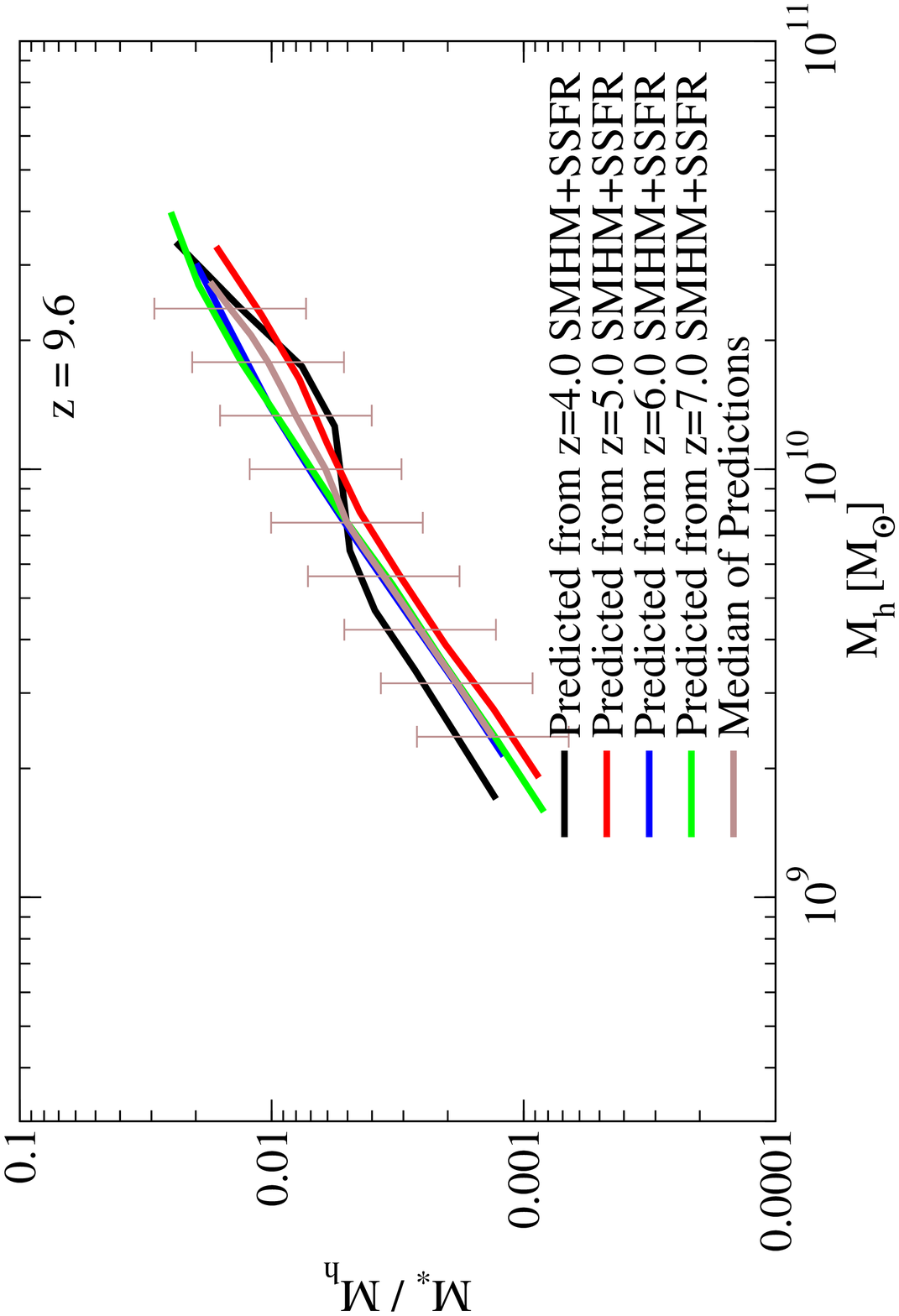}\plotgrace{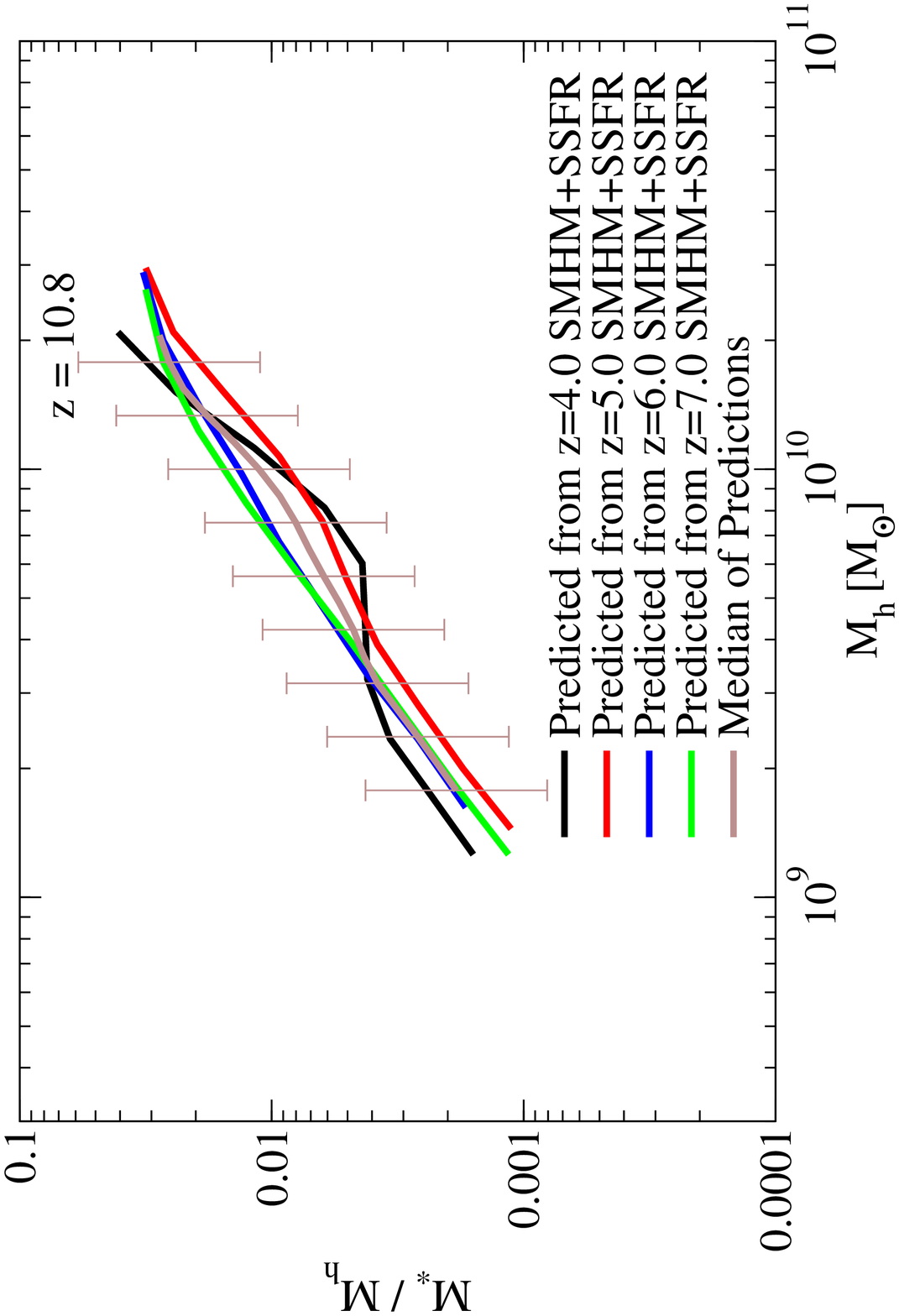}\\[-4ex]
\plotgrace{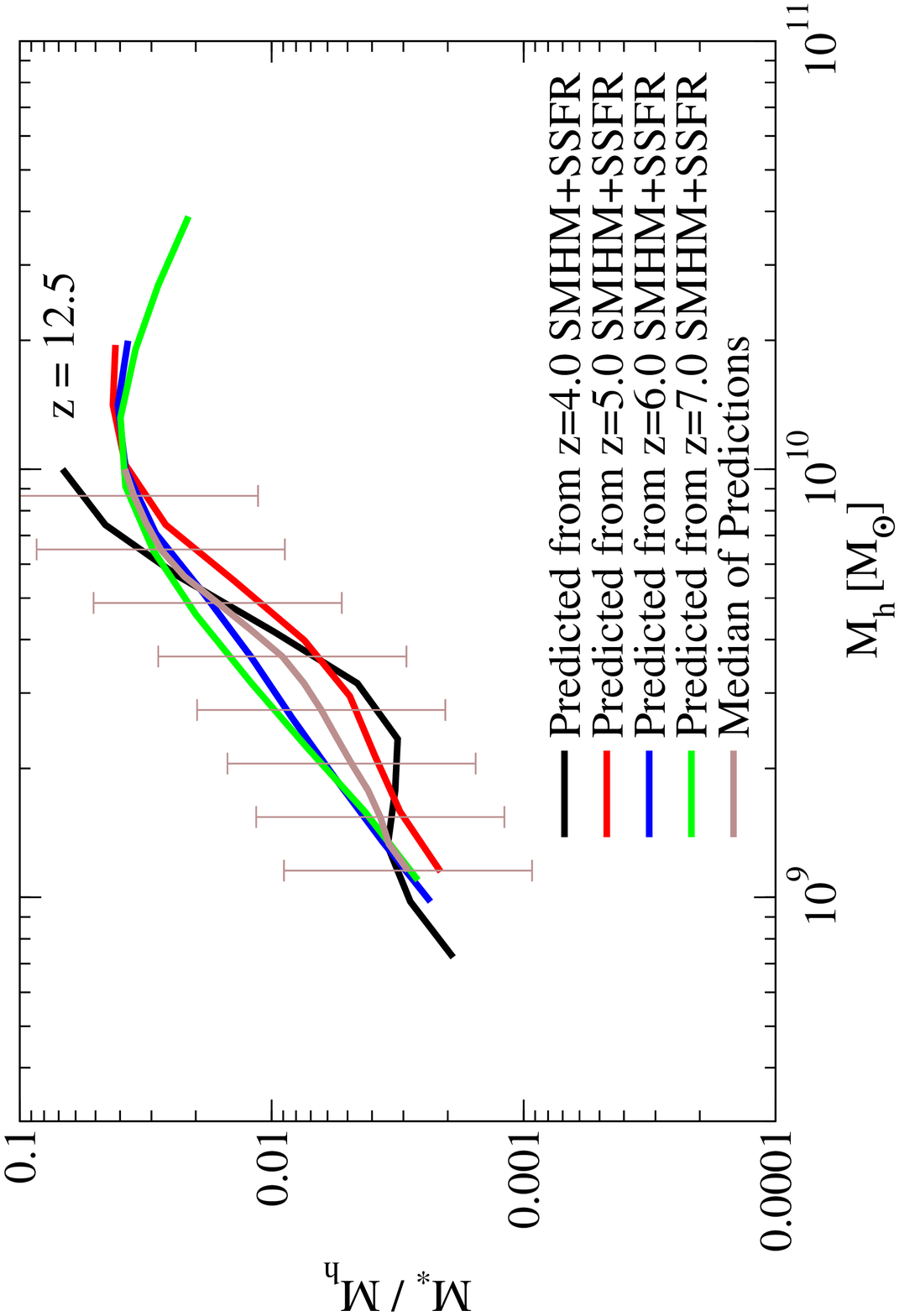}\plotgrace{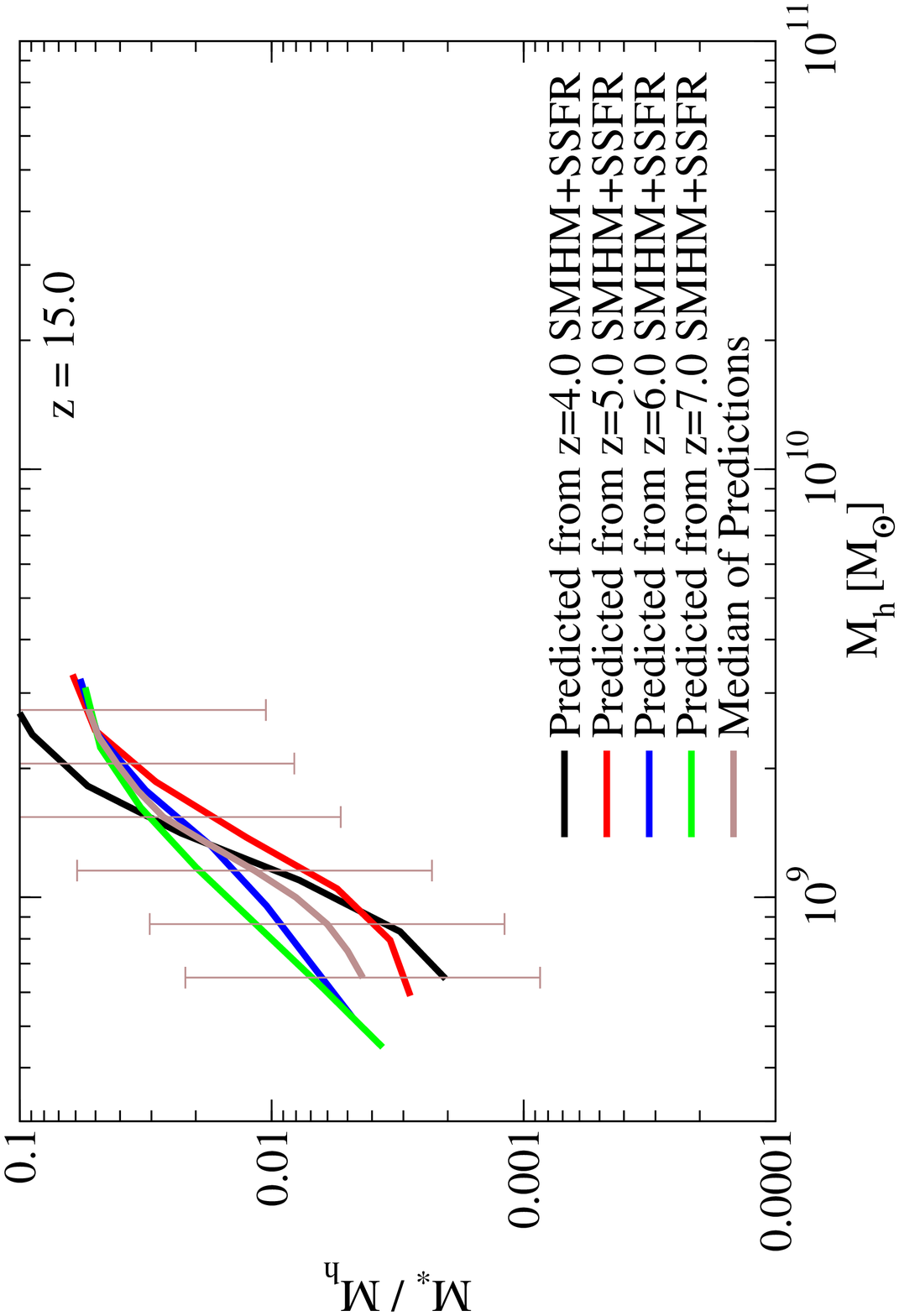}
\caption{Stellar mass---halo mass ratios for $z=9.6$ to $z=15$, predicted using Eqs.\ \ref{e:ssfr} and \ref{e:smhm}.  The black, red, blue, and green lines show predicted evolution assuming that the ratio of galaxy SSFRs to halo specific mass accretion rates is fixed at $z=4$, 5, 6, and 7, respectively.  The brown line shows the median of these predictions; the error bars give the expected 68\% uncertainties on the prediction from the error analysis in \S \ref{s:t_smhm}.}
\label{f:pred_z15}
\end{figure*}

\begin{figure}
\plotgrace{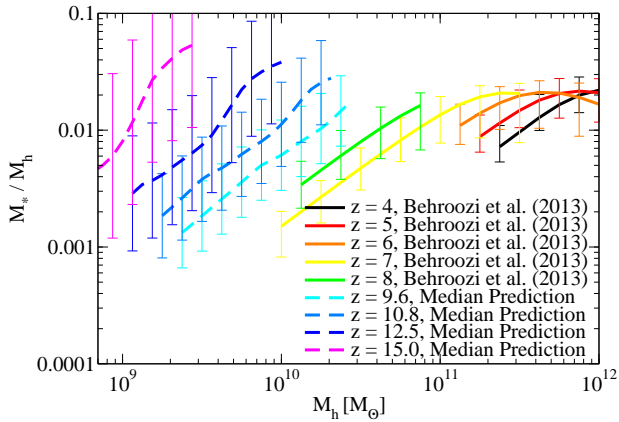}
\caption{Evolution of median prediction for the stellar mass---halo mass ratio from Fig.\ \ref{f:pred_z15} for $z=9$ to $z=15$.  Significant evolution in the efficiency at fixed halo mass is evident.}
\label{f:evolution}
\end{figure}

\begin{figure}
\plotgrace{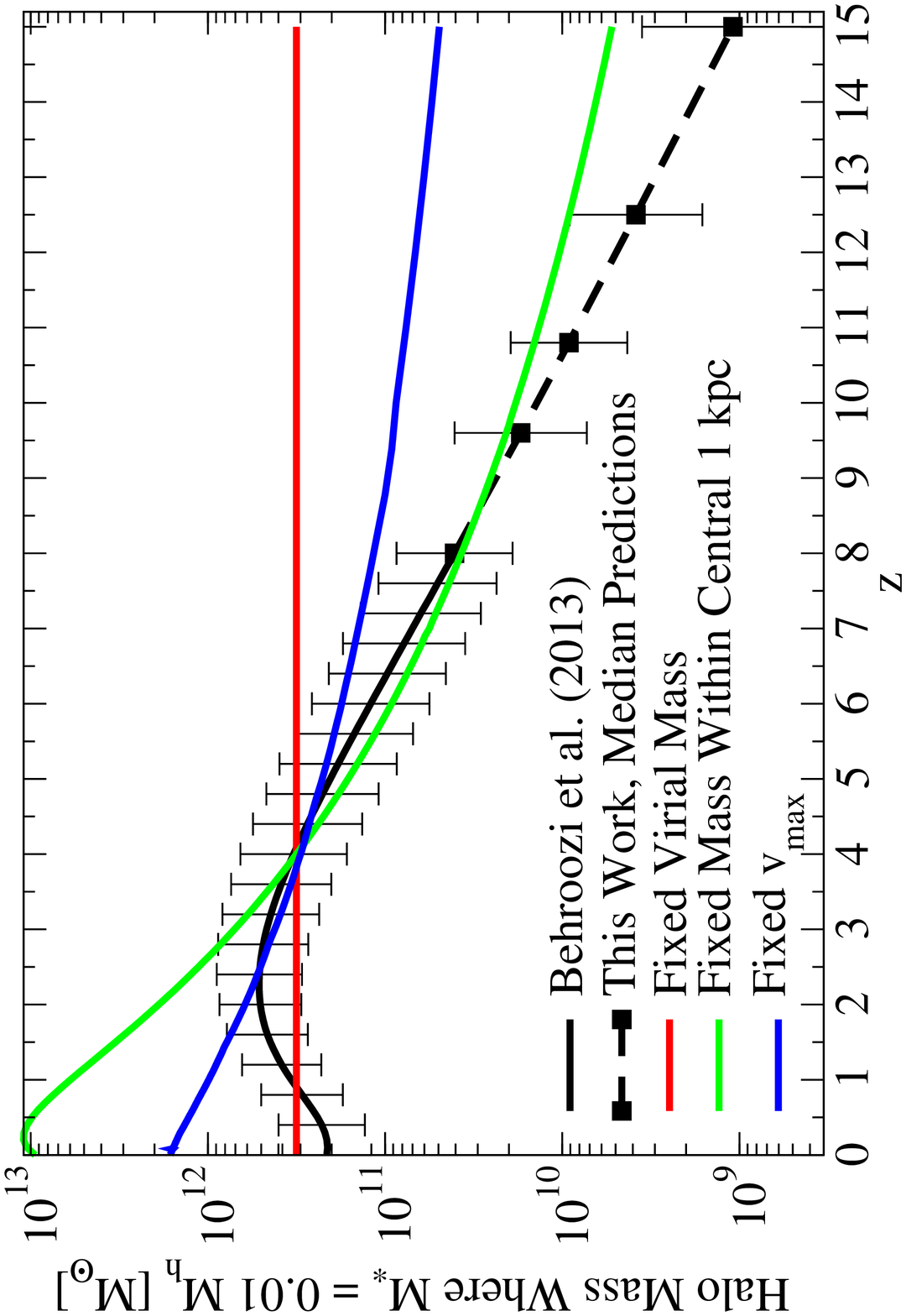}
\caption{Evolution of the halo mass ($M_h$) at which the stellar mass ($M_\ast$) first reaches one percent of $M_h$.   The black solid line shows the constraints from \cite{BWC12}; the black dots show results from this work according to Fig.\ \ref{f:evolution}.  For comparison, lines of constant mass ($M_h = 10^{11.5}\Msun$) and constant $v_\mathrm{max}$ ($v_\mathrm{max} = 175$ km s$^{-1}$) are shown; these values are chosen to match the ``one percent'' halo mass at $z=4$.  While the constant mass line describes the evolution reasonably well for $z<4$, neither constant mass nor constant $v_\mathrm{max}$ tracks the mass evolution for $z>4$.  Instead, the evolution is much closer to the average mass accretion histories of halos, suggesting that high redshift star formation efficiency may be more influenced by environmental factors (e.g., mass accretion rates) than at lower redshifts.}
\label{f:mass_ev}
\end{figure}

\section{Results}

We discuss predictions for high-redshift specific and cosmic star formation rates in \S \ref{s:pred_ssfr}, stellar masses in \S \ref{s:pred_smhm}, and sensitivity to errors in \S \ref{s:systematics}.

\label{s:results}

\subsection{Predictions for Specific and Cosmic Star Formation Rates}

\label{s:pred_ssfr}

As shown in Fig.\ \ref{f:obs_ssfr}, we expect specific star formation rates to rise slowly with redshift at fixed galaxy mass.  At $z=4$ and above, galaxy SSFRs are nearly the same magnitude as host halo specific mass accretion rates (see, e.g., Fig.\ \ref{f:pred_z4}).  As argued in \S \ref{s:basic} and \S \ref{s:t_smhm}, the ratio of SSFRs to halo specific mass accretion rates is conserved for progenitors of $z>4$ galaxy populations; hence, higher-redshift SSFRs will also be about the same magnitude as the slowly-rising halo specific mass accretion rates.  Since the choice of initial redshift is arbitrary, we show the SSFR evolution predicted from assuming constant SSFR to SMAR ratios for galaxies starting at $z=5$, $z=6$, and $z=7$ in Fig.\ \ref{f:obs_ssfr}.  Consistent with the error analysis in \S \ref{s:t_ssfr}, these predictions are all very similar to each other, as well as to the prediction from the restricted analysis in \S \ref{s:t_ssfr}.

Predictions for observed cosmic star formation rates are shown in Fig.\ \ref{f:obs_csfr}; as in \S \ref{s:t_ssfr}, we require galaxies to have a threshold UV luminosity of $M_{1500,AB} < -18$.  These star formation rates show a smooth decline from $z=4$ to $z=15$.  As with SSFRs, we show predictions from assuming constant SSFR to SMAR ratios (taken from \citealt{BWC12}) for galaxies starting at $z=5$, $z=6$, and $z=7$, as well as from the restricted analysis in \ref{s:t_ssfr}.

As noted in \S \ref{s:intro}, \cite{Oesch13} found an apparent sharp steepening in the CSFR evolution beyond $z=8$.  In our model, we do not find this steepening to occur.  Intuitively, this is because galaxy stellar mass doubling times at $z=7$ and $z=8$ are on the order of 100--200 Myrs (see Fig.\ \ref{f:obs_ssfr}).  Hence, if galaxy SSFRs at $z=7$ and $z=8$ are correct, there should still be significant star formation ongoing at $z=9$ and $z=10$.  If the \cite{Oesch13} results turn out to be correct, on the other hand, specific star formation rates at $z=7$ and $z=8$ would have to be a factor of 3 higher than current measurements \citep[i.e., beyond existing corrections for nebular emission;][]{Labbe12}.  The tension between the \cite{Oesch13} measurements and our predictions is at the 1.5--2 sigma level (Fig.\ \ref{f:oesch}).  We find that the discrepancy is too large to be explained by cosmic variance uncertainties \citep{Trenti08}, or by underestimating the effective UV luminosity threshold by a factor of 2.5 (Fig.\ \ref{f:oesch}).  Instead, these results may indicate that the UV--SFR conversion in \cite{Oesch13} underestimates the star formation rate for young stellar populations at lower metallicities \citep{Castellano14,Madau14}.  Alternately, the contamination exclusion criteria in \cite{Oesch13} may be too restrictive; e.g., removing point sources to exclude contamination from nearby stars may also affect compact galaxies at very high redshifts.  We note that high-redshift measurements are rapidly evolving; as this manuscript was being revised, \cite{Oesch14} found a doubly-imaged $z\sim10$ candidate in the Frontier Fields, which may suggest a somewhat higher cosmic star formation rate (Fig.\ \ref{f:oesch}).  

We can also predict the total (observed $+$ unobserved) CSFR; galaxies too faint to be observed at a given redshift can be constrained by their descendants at a later redshift.  A UV luminosity threshold of $M_{1500,AB}<-18$ is expected to miss nearly half of the total star formation rate at $z=8$ and miss two-thirds by $z=12$.   Long gamma-ray bursts at high redshifts can also constrain the total CSFR.  Although significant calibration uncertainties remain for how the long gamma-ray burst rate depends on metallicity and redshift,\footnote{Future work modeling the host galaxy demographics of long gamma-ray bursts (analogous to the short gamma-ray burst modeling in \citealt{BehrooziGRB}) will help improve existing uncertainties on the redshift and metallicity dependence.} our predictions are generally within the one-sigma error bars of current constraints \citep{Kistler09,Kistler13}.  We caution that \cite{Kistler09,Kistler13} normalized the ratio of the cosmic gamma ray burst rate to the CSFR using the CSFR calibration in \cite{Hopkins06b}.  As discussed in \cite{BWC12}, more recent measurements of the CSFR have been systematically lower than the \cite{Hopkins06b} calibration by 0.3 dex; this new normalization would lower the GRB-derived star formation rates by at least the same factor.  As shown in Fig.\ \ref{f:oesch}, lowering the \cite{Kistler13} results by 0.3 dex results in excellent agreement with our predictions for the total CSFR.

\subsection{Stellar Mass---Halo Mass Ratios}

\label{s:pred_smhm}

Given the ratios of galaxy SSFRs (from \citealt{BWC12}) to halo specific mass accretion rates (SMARs) at $z=4$, 5, 6, and 7 (from \S \ref{s:sims}), we have used Eqs.\ \ref{e:smhm}--\ref{e:ssfr} to predict stellar mass---halo mass (SMHM) ratios from $z=9.6$ to $z=15$ (Fig.\ \ref{f:pred_z15}).  As noted in \S \ref{s:t_smhm}, typical uncertainties in the stellar mass evolution are $\sim 0.25$ dex for a redshift baseline of $\Delta z = 4$ and $\sim 0.5$ dex for a redshift baseline of $\Delta z = 8$; these add in quadrature with existing 0.25 dex systematic uncertainties in the stellar mass normalization at $z>4$ \citep{BWC12,CurtisLake13}.  We note that reionization may lead to non-power law historical stellar mass --- halo mass ratios for halo masses below $M_h = 10^9\Msun$ \citep{Gnedin00,Noh14}.  However, several authors find that gas may cool in $10^{8}\Msun$ halos prior to reionization \citep{Gnedin00,Noh14,Wise14}, so that predictions for $M_h < 10^{9}\Msun$ at very high redshifts (e.g., $z>10$) may still be reasonable.

The median predicted evolution of the SMHM ratio is shown in Fig.\ \ref{f:evolution}.  We find that the redshift evolution seen in the SMHM ratio from $z=4$ to $z=8$ is expected to continue to very high redshifts.  More quantitatively, we can calculate (as a function of redshift) the halo mass at which the stellar mass is one percent of the halo mass.  The evolution of this halo mass with redshift is shown in Fig.\ \ref{f:mass_ev}.  At $z<4$, this mass hardly evolves, which suggests that halo mass determines galaxy star formation efficiency at $z<4$ \citep[see also][]{Behroozi13}.  At higher redshifts, this mass evolves continuously from $10^{11.5}\Msun$ at $z=4$ to $10^{9.3}\Msun$ at $z=15$---i.e., over two orders of magnitude across this redshift range.  This evolution cannot be explained by appealing to an evolving halo mass definition.  E.g., it outpaces even that expected for constant $\vmax$ (the maximum of $\sqrt{\frac{GM(<R)}{R}}$ over the halo profile), which is a direct physical measure of the halo's potential well depth.  The evolution at $z>4$ may match better with a fixed central density (e.g., density within the central 1 kpc, as shown in Fig.\ \ref{f:mass_ev}), which would correspond to a fixed cooling rate.  However, galaxy formation is a complex process, and we cannot rule out contributions from other sources.

\subsection{Sensitivity to Systematic Errors}

\label{s:systematics}

\subsubsection{Stellar Mass Functions}

\label{s:smf_errs}

Results in this paper depend mostly on the growth of the stellar mass function over redshifts from $z=8$ to $z=4$, especially since this growth largely constrains galaxy specific star formation rates (\S \ref{s:amodel}).  We therefore address several uncertainties in high-redshift stellar mass functions, including $\pm 0.25$ dex systematic biases in recovering stellar masses \citep{BWC12,CurtisLake13}, sample variance from small survey volumes ($\pm 0.1$ dex for the $z>7$ stellar mass functions; \citealt{Trenti08}), and low-redshift interlopers at the massive end of the stellar mass function.

We cannot directly resolve interloper contamination, so we avoid presenting results from the bright end of the stellar mass function where such contamination is most an issue (e.g., galaxy stellar masses $>10^{9.5}\Msun$ at $z=8$).  The specific star formation rates we calculate as a function of halo mass are robust to the other two issues.  Sample variance uncertainties are reduced because we use a smooth function to model the redshift evolution of the stellar mass function (\S \ref{s:amodel}).  Currently-known uncertainties in computing stellar masses (e.g., assumed stellar population histories, dust modeling, uncertainties in stellar population synthesis) do not result in systematic biases that change rapidly with redshift or stellar mass.  Therefore, when comparing two stellar mass functions at two nearby redshifts, systematic biases will result in similar offsets to the stellar masses in both stellar mass functions.  The \textit{relative} growth of stellar masses---i.e., specific star formation rates---computed from these stellar mass functions will therefore be largely unaffected.  While the absolute galaxy stellar masses will be affected by these biases, the inferred halo masses depend only on galaxy cumulative number densities as a function of stellar mass.  So, as long as the systematic uncertainties do not destroy the overall rank-ordering of galaxies (i.e., if there is a strong rank correlation between the true galaxy stellar masses and the observationally calculated galaxy stellar masses), the inferred halo masses will not be affected.  As a result, the specific star formation rates as a function of \textit{halo mass} are robust.

The systematic biases in absolute stellar masses do affect the stellar mass --- halo mass relationships presented in this paper.  We have included these uncertainties in the error bars in Figs.\ \ref{f:pred_z4}, \ref{f:pred_z15}, \ref{f:evolution}, and \ref{f:mass_ev}; for high-redshift predictions, we have also included uncertainties from Eq.\ \ref{e:errs}.  Yet, the conclusion that the characteristic halo mass for the stellar mass---halo mass relation changes from $z=8$ to $z=4$ is extremely robust to these uncertainties.  As shown in Figs.\ \ref{f:evolution} and \ref{f:mass_ev}, the halo mass for a given SMHM ratio is significantly lower at $z=8$ than it is at $z=4$.  Considering the slope of the SMHM relation, the stellar mass for a given halo mass is a full decade higher at $z=8$ than it is at $z=4$.  This means that, at fixed cumulative number density, galaxy stellar masses are one dex higher at $z=8$ than would be predicted by keeping the $z=8$ SMHM relationship the same as at $z=4$.  This level of evolution cannot be explained by known systematic biases, which are only of order $\pm 0.25$ dex.

We note that the stellar mass functions we have used for $z=7$ and $z=8$ were obtained by converting UV luminosity functions \citep{Bouwens14} to stellar mass functions using \cite{Gonzalez10}.  The \cite{Gonzalez10} conversion applied to luminosity functions at $z=4$ and $z=5$ results in mass functions systematically too low compared to other results at $z=4$ and $z=5$ \citep[e.g.,][]{Stark09,Marchesini10}, however, it is in agreement with \cite{Stark09} by $z=6$.  This may be because galaxies become systematically less dusty \citep{Bouwens13} and younger (Fig.\ \ref{f:obs_ssfr}) at high redshifts, and so have higher luminosity to mass ratios.  Nebular emission lines in high-redshift galaxies can bias stellar mass estimations \citep{Stark12,deBarros12}; however, using the \cite{Gonzalez10} conversion at $z\ge 6$ gives consistent results with these recent studies (D.\ Stark, priv.\ comm.).  Due to these concerns, we have generated UV luminosity functions from our assumed star formation histories and find that they agree directly with the \cite{Bouwens14} luminosity functions in Appendix \ref{a:lfs}.  Finally, we have checked that new stellar mass functions from the CANDELS survey at $z>6$ are within the error bars of the converted stellar mass functions that we use (Duncan et al., in prep.; Grazian et al., in prep.).

\begin{figure}
\plotgrace{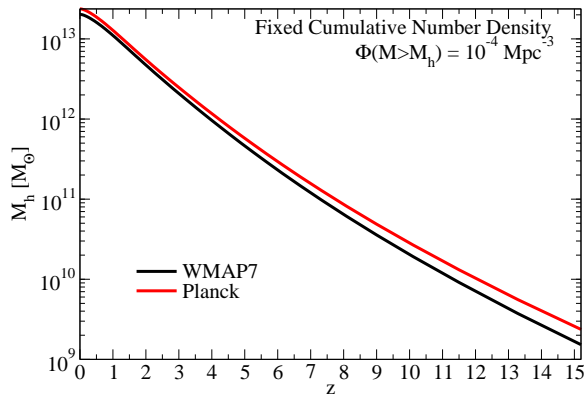}
\caption{Halo masses at a fixed cumulative number density of $10^{-4}$ Mpc$^{-3}$ for the WMAP7 and Planck cosmologies, using the mass function fit from \cite{BWC12}.  Uniform systematic offsets in halo mass do not affect the main conclusions of this paper.  The changes with redshift amount to a 0.01 dex (2\%) difference in mass accretion rates (Appendix \ref{a:mar_calibration}).  See \S \ref{s:cosmology_errs} for discussion.}
\label{f:b_vs_bp}
\end{figure}

\subsubsection{Cosmology}

\label{s:cosmology_errs}

The recent Planck cosmology results \citep{Planck} suggest that $\Omega_M$ may be higher and $h$ may be lower than the values used in this paper (see, however, \citealt{Spergel13}).  We find that the differential comoving volume as a function of redshift (affecting stellar mass functions and cosmic star formation rates) decreases by only 6--7\% for $z>4$ when we adopt $\Omega_M=0.307$ and $h=0.68$.  The increased $\Omega_M$ does result in an increase to the high-redshift halo mass function; at fixed cumulative number density, halo masses are $\sim 0.08$ dex larger at $z=4$ and $\sim 0.19$ dex larger at $z=15$ (Fig.\ \ref{f:b_vs_bp}).  This corresponds to an average change in mass accretion rates of 0.01 dex (2\%), which we explicitly verify with halo merger trees in Appendix \ref{a:mar_calibration}.

A 2\% change in halo mass accretion rates is within the existing uncertainties for the mass accretion rate fitting formula that we use (Appendix \ref{a:mar_calibration}); we therefore do not expect this to change predictions for specific or cosmic star formation rates.  The main effects on the stellar mass---halo mass relation would be systematic increases of 0.08 to 0.19 dex in halo masses at fixed galaxy mass.  For the redshift range from $z=4$ to $z=8$, the $\sim 1$ dex change in characteristic halo mass (Fig.\ \ref{f:mass_ev}) would be reduced by $0.04$ dex if we had used the Planck cosmology.  Naturally, uniform systematic offsets in halo masses do not affect our conclusions about the \textit{evolution} of the characteristic halo mass.

\section{Discussion}

\label{s:discussion}

As we have shown in \S \ref{s:theory} and \S \ref{s:tests}, the ratio of galaxies' specific star formation rates (SSFRs) to their host halo specific mass accretion rates (SMARs) sets the evolution of the galaxy progenitors' stellar mass--halo mass (SMHM) relations.  Knowledge of two of these three quantities allows one to solve for the third, as discussed in \S \ref{s:results}.  In this section, we discuss several consequences of this three-way relationship, as well as implications of our predicted high-redshift galaxy evolution.  We consider the specific star formation rate ``plateau'' in \S \ref{s:d_ssfr}, reionization in \S \ref{s:d_reion}, expectations for JWST in \S \ref{s:d_jwst}, 
abundance matching in \S \ref{s:d_clf_am}, and comparisons to other work in \S \ref{s:d_comp}.

\subsection{High-Redshift Star Formation and the Changing Characteristic Mass Scale of Galaxy Formation}

\label{s:d_ssfr}

As shown in Figs.\ \ref{f:evolution} and \ref{f:mass_ev}, we find that the evolution of the SMHM relation can be very simply described.  At $z<4$, the SMHM ratio depends much more on halo mass than on redshift \citep{Behroozi13}.  At $z>4$, however, the characteristic halo mass evolves strongly with redshift.  More quantitatively, the halo mass at which $M_\ast$ is 1\% of $M_h$ evolves as
\begin{equation}
M_{1\%} \sim \left\{\begin{array}{l l} 10^{11.5}\Msun & \textrm{if }z<4\\ 10^{11.5-0.23(z-4)}\Msun & \textrm{if }z\ge 4\end{array}\right.
\end{equation}
This evolution at high redshift is close to the average accretion rate of halos (Fig.\ \ref{f:mass_ev}), so the SMHM ratio is approximately unchanging as a function of cumulative number density or of peak height, $\nu$.

As discussed in \S \ref{s:systematics} and as shown in Fig.\ \ref{f:mass_ev}, this mass evolution is much larger than can be explained by known systematic biases.  We note that this mass evolution is not only consistent with flattening in specific star formation rates at high redshifts \citep{Weinmann11}, but also that they are equivalent statements.  For a galaxy population with a given stellar mass and redshift, the ratio of the galaxies' SSFRs to their host halos' specific mass accretion rates is equal to the power-law slopes of their historical SMHM relations (Eqs.\ \ref{e:smhm}--\ref{e:ssfr}).  If these historical slopes are not completely parallel to the slope of the SMHM relation at the starting redshift, then galaxies will evolve off their original SMHM relationship---that is to say, the SMHM relationship as a whole must evolve with redshift.  As a corollary, evolution in the ratio of galaxy SSFRs to their host halo specific halo mass accretion rates (e.g., Fig.\ \ref{f:obs_ssfr}) \textit{must} be accompanied by evolution in the SMHM relationship: as the ratio changes, the evolutionary trajectories of galaxies will change, and no single SMHM relationship can be parallel to all galaxy trajectories.

Flattening of galaxy specific star formation rates at high redshifts implies a decrease in the ratio of galaxy SSFRs to their host halo specific mass accretion rates.  In Fig.\ \ref{f:obs_ssfr}, the ratio of galaxy SSFRs to host halo mass accretion rates is consistent (within observational errors) with the slope of the SMHM relation at fixed redshift, at least for $z<4$.  However, the ratio at higher redshifts is too low to be consistent.  From Eq.\ \ref{e:ssfr_smar}, if this ratio falls below the slope of the SMHM relationship at fixed redshift, and if the SMHM relationship has a positive slope, then the SMHM relationship will evolve to higher efficiencies at higher redshifts.

So, the growth rate of the stellar mass function and the flattening in observed specific star formation rates both provide evidence that halos with masses below $10^{12}\Msun$ formed stars more efficiently at very high redshifts than they did for $0<z<4$.  Part of the reason for this may be a change in the gravitational potential well depth at fixed halo mass (Fig.\ \ref{f:mass_ev}, \S \ref{s:pred_smhm}).  This is not likely the only explanation, as it would require evolution in the characteristic halo mass over $0<z<4$ (Fig.\ \ref{f:mass_ev}); feedback efficiency at high redshift may also be reduced.  However, other factors like increased central density at high redshifts (Fig.\ \ref{f:mass_ev}) may decrease cooling times enough to explain the evolution.

\subsection{Reionization}

\label{s:d_reion}

\begin{figure}
\vspace{-5ex}
\plotgrace{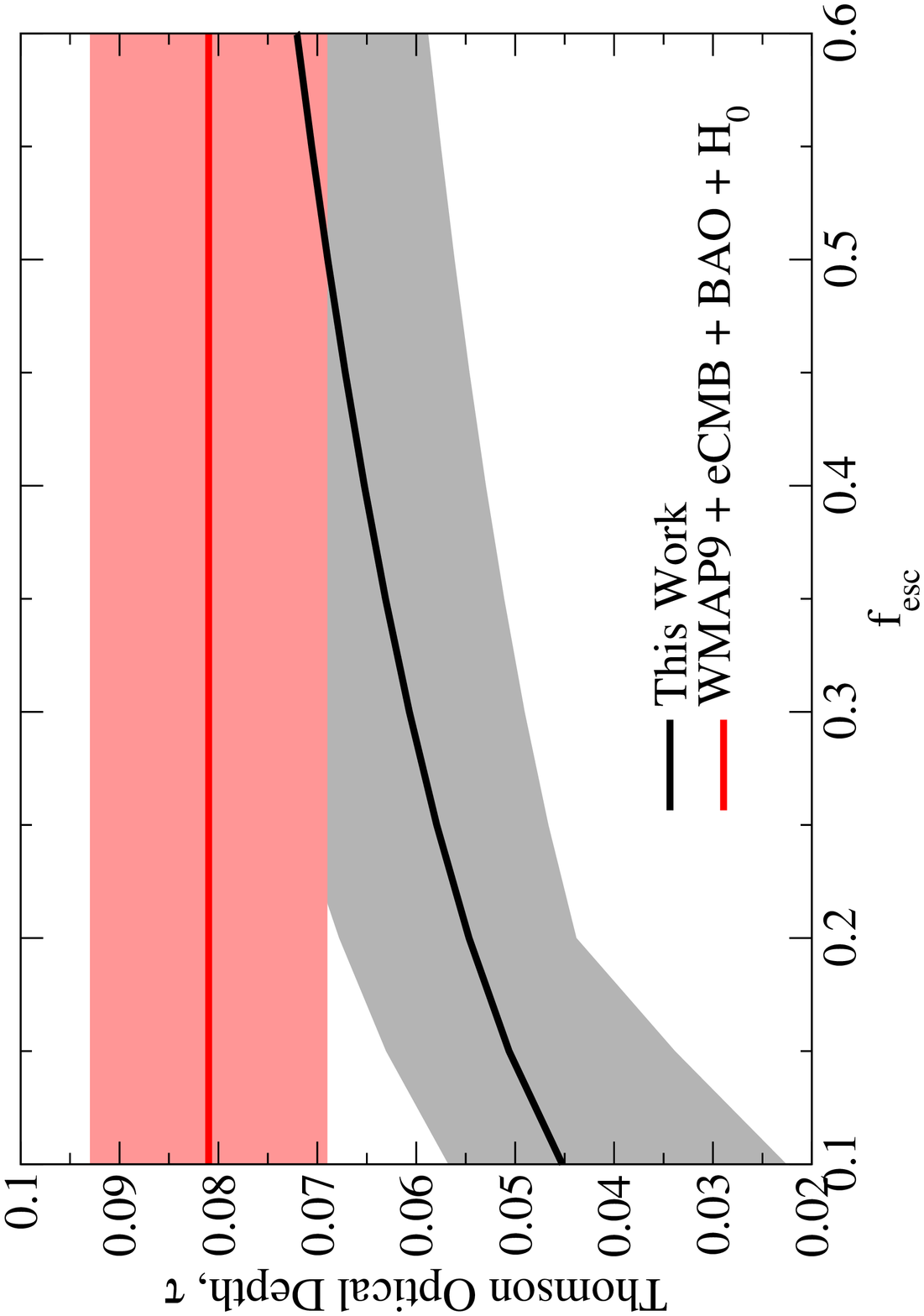}
\caption{Optical depth as a function of ionizing photon escape fraction.  The grey shaded region corresponds to the best-fit WMAP9 68\% limits from \cite{WMAP9}.  The red shaded region correspond to 68\% limits for our cosmic star formation rate estimates.  The errors come largely from $\pm 0.3$ dex systematic uncertainties in the total cosmic star formation rate.}
\label{f:optical_depth}
\plotgrace{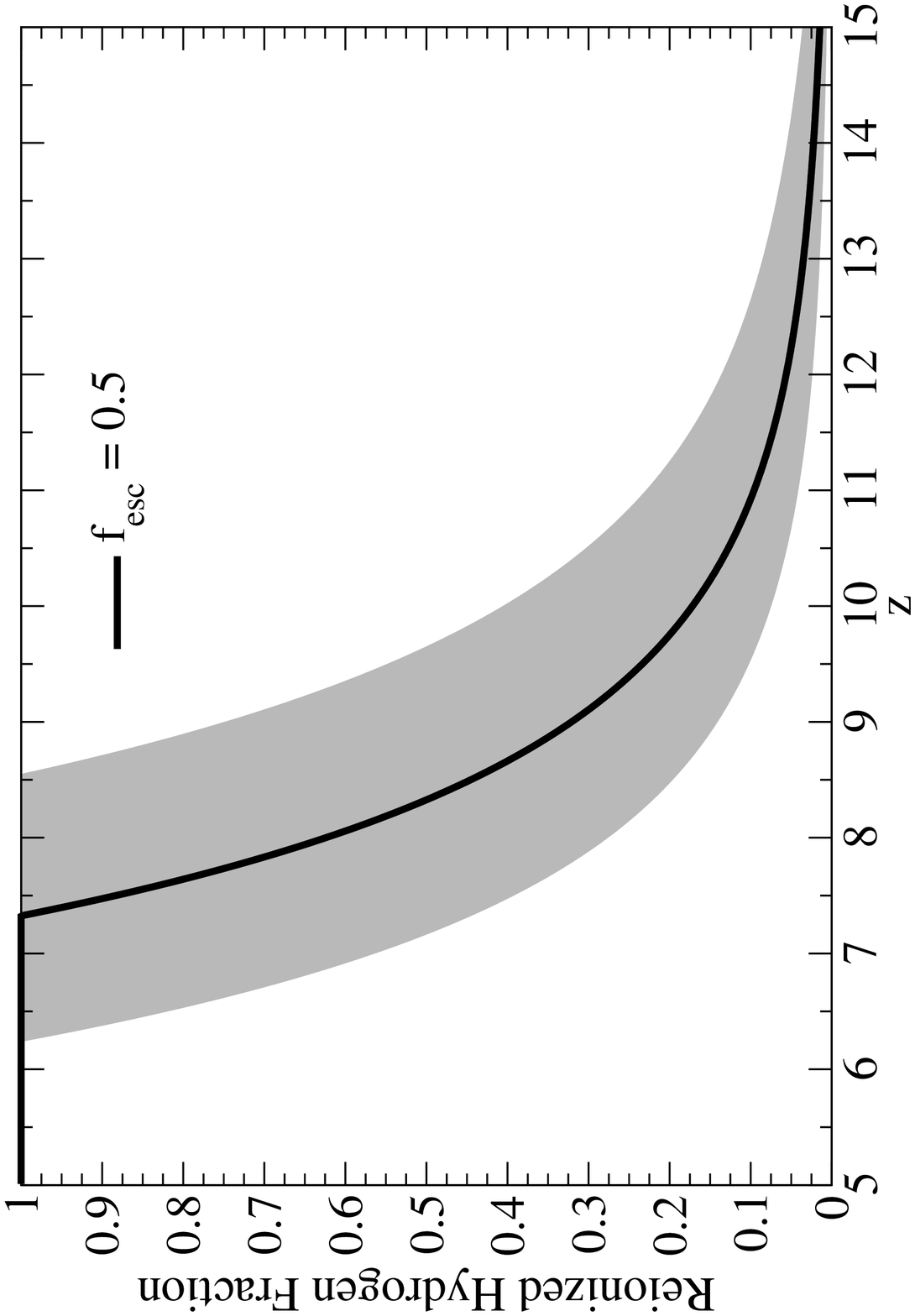}
\caption{The reionization history of the universe assuming a star-formation weighted average $f_\mathrm{esc}$ of $0.5$.  As with Fig.\ \ref{f:optical_depth}, the 68\% uncertainties (grey shaded region) come largely from systematic uncertainties in the cosmic star formation rate.}
\label{f:re_hist}
\plotgrace{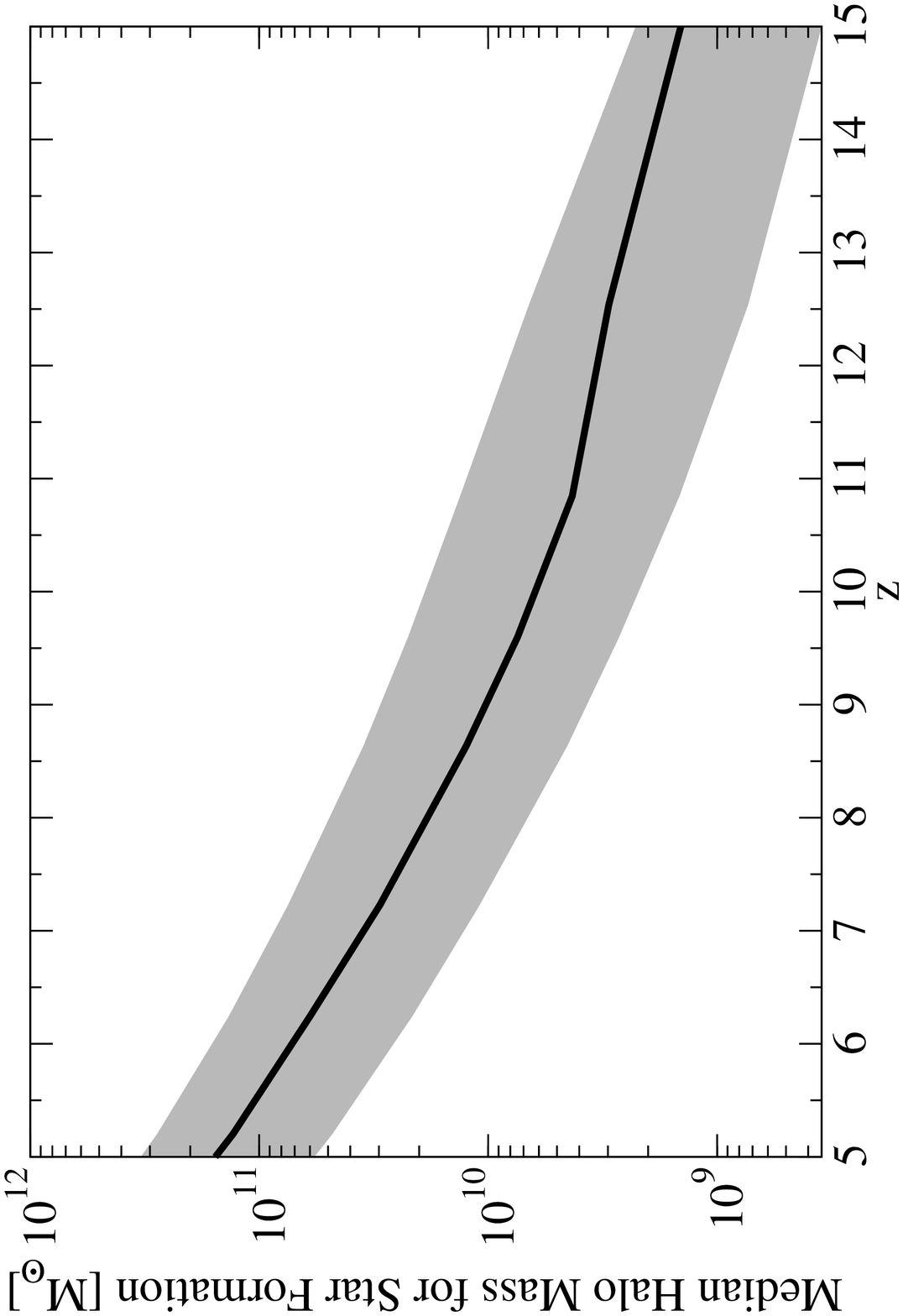}
\caption{The halo mass below which 50\% of star formation occurs as a function of redshift; i.e., the median halo mass for star formation.  The errors are dominated by uncertainties on the faint-end slope of the star formation rate function.}
\label{f:median_hm}
\end{figure}

Knowing the cosmic star formation history, it is possible to calculate the hydrogen reionization history of the universe.  We use the minimal reionization model of \cite{Haardt12}:
\begin{equation}
\frac{dQ}{dt} = \frac{\dot{n}_\mathrm{ion}}{n_\mathrm{H+He}} - \frac{Q}{t_\mathrm{rec}}
\end{equation}
where $Q$ is the reionized fraction, $\dot{n}_\mathrm{ion}$ is the creation rate density of ionizing photons, $n_\mathrm{H+He}$ is the number density of hydrogen and helium atoms, and $t_\mathrm{rec}$ is the recombination time.  In this model, single ionization of helium is assumed to happen at the same time as hydrogen; we assume that the helium mass fraction is $Y=0.24$.  Ionizing photon production is given by
\begin{equation}
\dot{n}_\mathrm{ion} = N_\gamma f_\mathrm{esc} CSFR
\end{equation}
where $N_\gamma$ is the number of ionizing photons per unit mass of stars, $f_\mathrm{esc}$ is the average escape fraction from galaxies, and $CSFR$ is the cosmic star formation rate.  For a low-metallicity Salpeter IMF, $N_\gamma$ would be $5\times 10^{60}\Msun^{-1}$ \citep{Alvarez12}.  Since the Chabrier IMF that we use has fewer low-mass stars but the same number of high-mass (ionizing) stars, we take $N_\gamma$ equal to $7.65\times 10^{60} \Msun^{-1}$; this corresponds to 6400 photons per baryon in stars.  While inferences at $z<5$ suggest escape fractions less than $10\%$, the escape fraction at high redshifts is very uncertain \citep{Hayes11}.  Simulations suggest that the escape fractions can be very large \citep{Wise09,Wise14}; current models typically adopt either a constant $f_\mathrm{esc}$ or an $f_\mathrm{esc}$ which rises rapidly with redshift \citep{Haardt12}.  In this paper, we use a constant $f_{esc}$ to simplify the interpretation of our results; we note that this would roughly correspond to the star-formation weighted average $f_\mathrm{esc}$ for models in which $f_\mathrm{esc}$ evolves with time.  To explore the range of uncertainties, we allow $f_\mathrm{esc}$ to vary from 0.1 to 0.6.

The recombination time is given by
\begin{equation}
t_\mathrm{rec} = (\alpha_B n_\mathrm{H+He} C)^{-1}
\end{equation}
where we take $\alpha_B$ to be the case B recombination coefficient at 10,000K for hydrogen ($2.79\times 10^{-79}$ Mpc$^3$ yr$^{-1}$) and $C$ to be the clumping factor.  The determination of $C$ is also uncertain, with estimates ranging from $C=2$ to $C=4$ at high redshifts \citep{Haardt12}; we take $C=3$ as our fiducial value and allow $C$ to vary from 2 to 4 for modeling uncertainties.

For any reionization history, we can also calculate an optical depth $\tau$ for Thomson scattering, given by
\begin{equation}
\tau = \sigma_T \int_0^{t_\mathrm{now}}  (Q n_\mathrm{H+He} + Q_\mathrm{HeIII} n_\mathrm{He})c \mathrm{d} t
\end{equation}
where $\sigma_T$ is the Thomson cross-section of an electron ($6.986\times 10^{-74}$ Mpc$^2$), $Q_\mathrm{HeIII}$ is the fraction of helium that has been doubly ionized, and $n_\mathrm{He}$ is the number density of helium.  Since the double ionization of helium happens much later than single ionization, it changes $\tau$ by only $\sim 0.001$; we therefore fix the redshift of double ionization to $z=3.5$.

The largest uncertainty for $\tau$ is the escape fraction, followed by uncertainties in the normalization of the cosmic star formation rate (Fig.\ \ref{f:optical_depth}), followed by subdominant uncertainties in the clumping factor.  
  If the cosmic star formation rate at $z>4$ was currently underestimated by a factor of 2, an average escape fraction of $0.2$ would be consistent with the most recent WMAP results at the $1$--$\sigma$ level.  However, if current high-redshift CSFRs are correct, $f_\mathrm{esc}$ would need to average at least $0.5$ during reionization.  These escape fractions agree with other models which match the optical depth (e.g., \citealt{Haardt12,Alvarez12,Bouwens12,Robertson13}) and are plausible for certain types of galaxies \citep{Wardlow13,Dijkstra14}, but are higher than typically seen in low-redshift observations \citep{Hayes11}.

We also show the expected reionization history of the universe in Fig.\ \ref{f:re_hist} for $f_\mathrm{esc}=0.5$.  As with the optical depth, the normalization of the cosmic star formation rate is a substantial uncertainty.  For this escape fraction, 68\% confidence limits on the redshift of half-reionization are $7<z<10$.  Absent independent constraints, the escape fraction is completely degenerate with the normalization of the cosmic star formation history in these models.  This makes it difficult to use $\tau$ or the reionization history as a constraint on galaxy formation (or vice versa).

As noted in \cite{Alvarez12} and \cite{Wise14}, an escape fraction which varies strongly with halo mass would result in an escape fraction which varies strongly with redshift (as assumed in \citealt{Haardt12}).  This is because the median halo mass for star formation (i.e., the halo mass below which 50\% of star formation occurs) is a strong function of redshift (Fig.\ \ref{f:median_hm}).  From Fig.\ \ref{f:median_hm} and Fig.\ \ref{f:evolution}, we expect that the galaxies which would reionize the universe at $z>8$ are typically below $10^{8}\Msun$ and their host halos are typically below $10^{10}\Msun$. While escape fractions for these low-mass galaxies can be observed locally, we note that the nature of star formation in their high-redshift counterparts will likely be very different, as evidenced by the very different stellar mass---halo mass ratios (Fig.\ \ref{f:evolution}).

\begin{figure}
\plotgrace{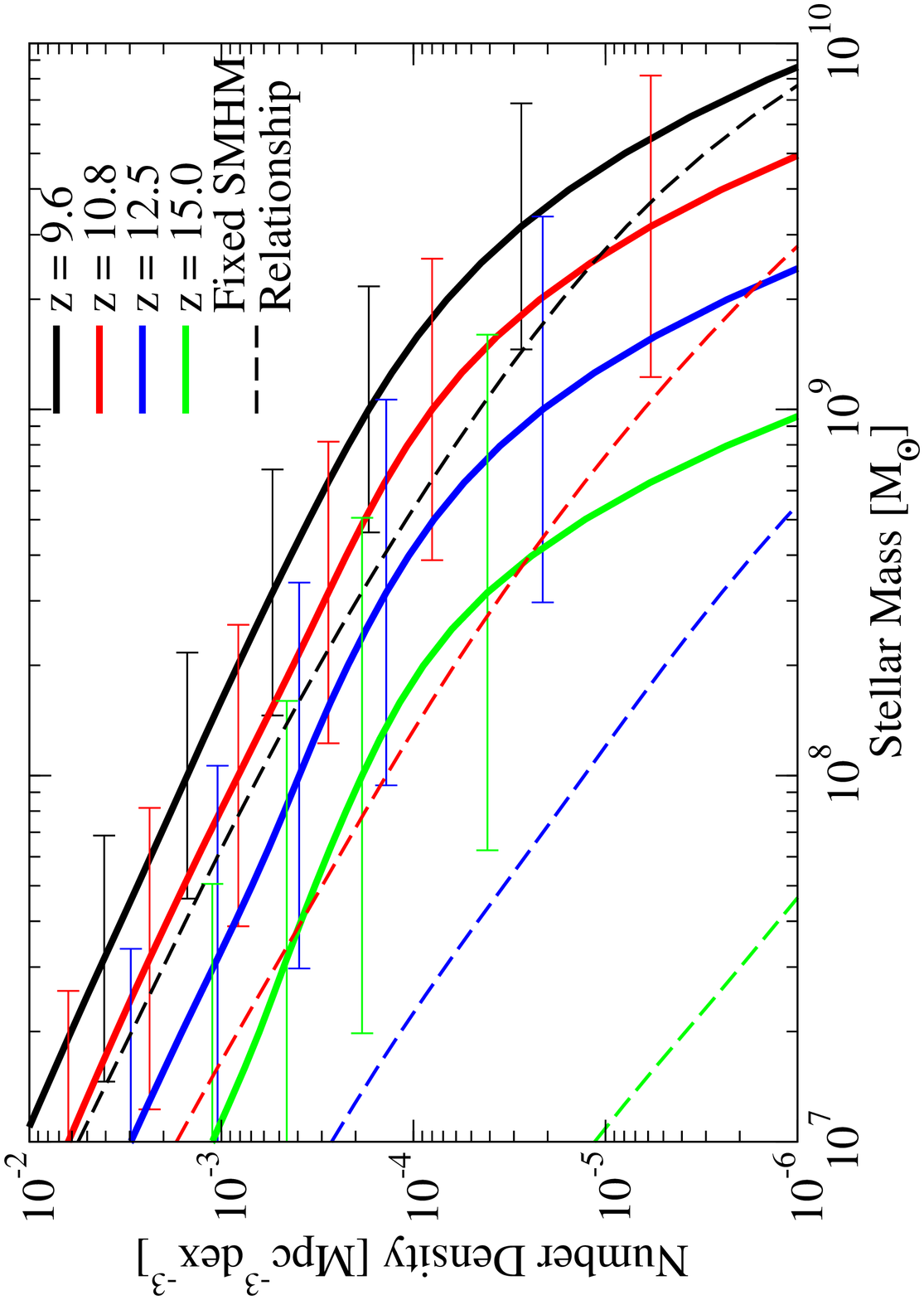}
\caption{Predictions for the evolution of the galaxy stellar mass function to $z=15$ (solid lines).  Error bars show uncertainties in the evolution of stellar masses, from Eq.\ \ref{e:errs}.  Dashed lines show how the stellar mass functions would evolve if the stellar mass--halo mass relationship instead remained fixed to its $z=8$ value.  This latter case would result in over 2 orders of magnitude fewer galaxies observed by JWST at $z=15$.}
\label{f:smf_highz}
\plotgrace{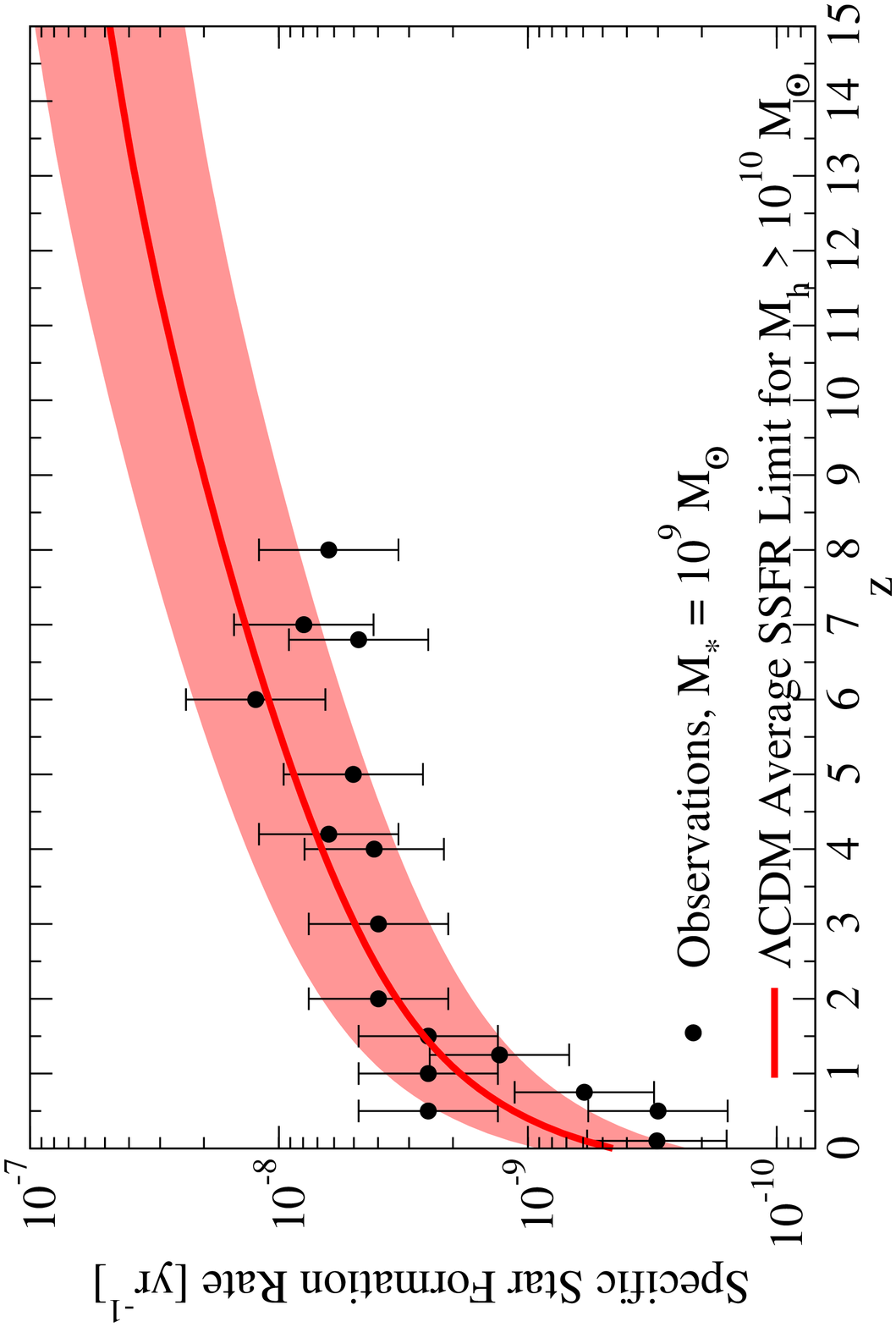}\\[-7ex]
\plotgrace{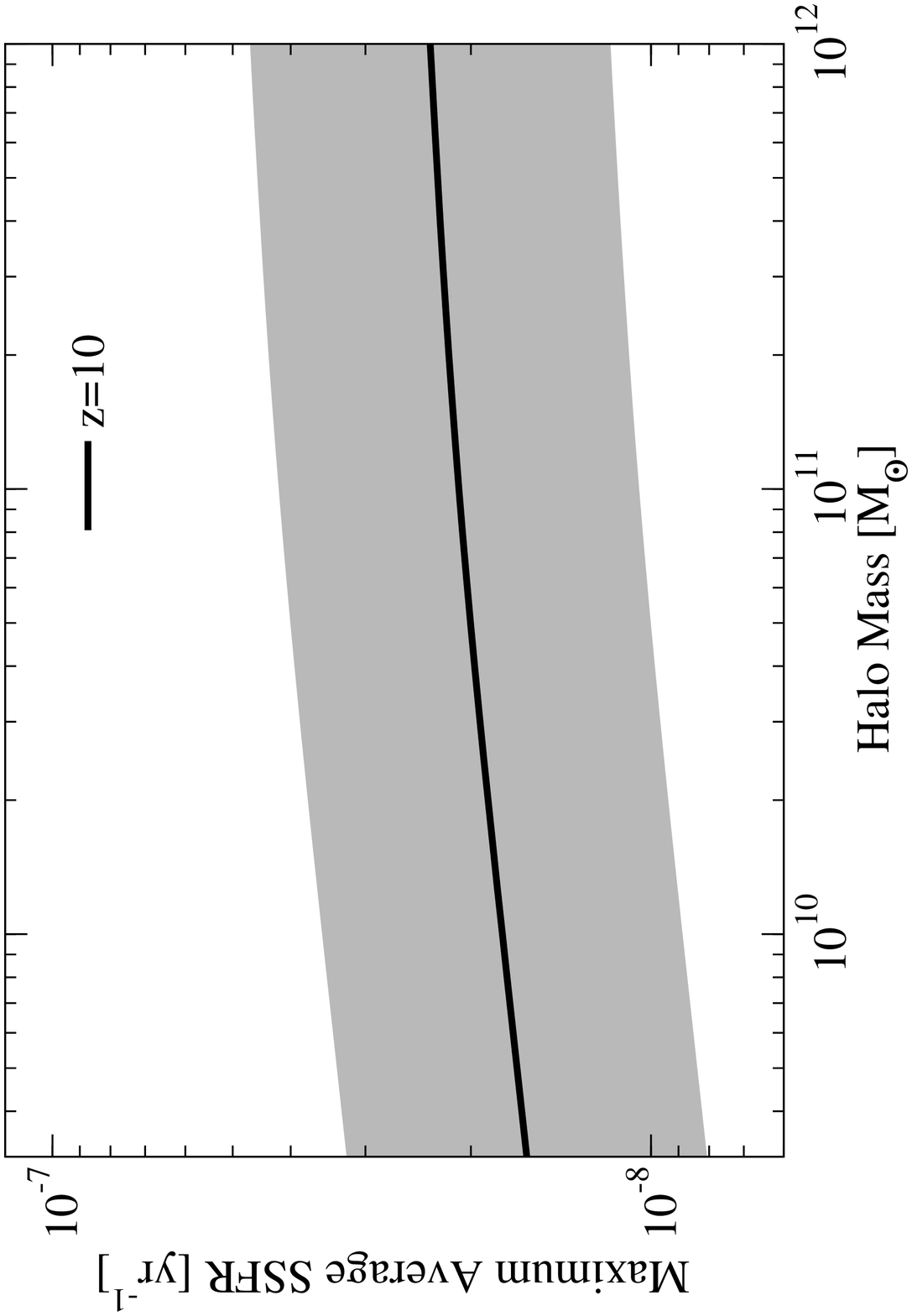}
\caption{\textbf{Top} panel: the maximum possible population-averaged specific star formation rates for galaxies with host halo masses $M_h>10^{10}\Msun$, as a function of redshift.  Error bars show the allowable uncertainties from \textit{all} unknown galaxy formation physics.  Individual galaxies can temporarily exceed these limits, but the averaged population specific star formation rates are fundamentally limited by specific halo mass growth rates in $\Lambda$CDM. (See \S \ref{s:d_jwst}).  \textbf{Bottom} panel: The maximum attainable averaged specific star formation rates at $z=10$, as a function of host halo mass.  Other redshifts also show a very weak dependence on halo mass.}
\label{f:max_ssfr}
\end{figure}

\subsection{Expectations for JWST}

\label{s:d_jwst}

If low-mass halos continue to become more efficient at $z>8$ (Fig.\ \ref{f:evolution}), JWST should be able to see many $M_\ast > 10^8 \Msun$ galaxies out to at least $z=15$ (Fig.\ \ref{f:smf_highz}; see also Appendix \ref{a:lfs} for luminosity functions).  By comparison, if the stellar mass---halo mass relation were fixed at $z=8$ and did not evolve to $z=15$, the expected number density of galaxies at $z=15$ would drop by over two orders of magnitude (Fig.\ \ref{f:smf_highz}).  We expect that JWST will rule out at least one of these two scenarios very shortly after launch.  Due to the rapid buildup of the halo mass function from $z=15$ to $z=8$, the number density of galaxies observed with JWST will be very constraining for galaxy formation models.

We find that the expected faint-end slopes for the stellar mass functions are -1.85 to -1.95 from $z=9$ to $z=15$.  These are steep, suggesting that deeper JWST pointings will result in many more observed galaxies.  However, none of the slopes are significantly steeper than the faint-end slopes we expect at $z=8$ (-1.8), suggesting that the evolution in faint-end slopes observed over $4<z<8$ \citep{Bouwens12} does not continue to higher redshifts.

We note that Eq.\ \ref{e:ssfr} also constrains the \textit{maximum} attainable specific star formation rates for galaxy populations as a function of redshift.  From Eq.\ \ref{e:ssfr}, we have:
\begin{equation}
SSFR = \alpha SMAR
\end{equation}
So, maximum SSFR would require maximizing the specific halo mass accretion rates (SMARs) as well as the power-law slope ($\alpha$) of the historical SMHM relation.  The maximum value of $\alpha$ is limited by available mechanisms to regulate star formation.  For halos which are just large enough to accrete reionized gas ($\sim 10^{9}\Msun$; \citealt{Gnedin00}) and form their first stars, the specific star formation rates could be effectively infinite.  At larger masses (e.g., $>10^{10}\Msun$), the most important mechanisms become stellar winds and supernovae from massive stars \citep{Agertz13}.  Typical values of $\alpha$ for $10^{10}\Msun$ and larger halos range from $\alpha\approx 2$ at $z=0$ to $\alpha \approx 1$ at $z>4$.  It is difficult to imagine that $\alpha$ could be much higher than $2$ because the energy and momentum requirements for leaving a halo both scale as a low power of the halo mass.  E.g., the energy needed to escape from a halo scales as halo mass to the two-thirds power, and the escape velocity scales as halo mass to the one-third power.  That said, we consider values from $\alpha=1$ to $\alpha = 4$ to cover any exotic future feedback mechanisms.

Note that \textit{all} variation from galaxy physics is contained in the slope $\alpha$, and the range $\alpha=1$ to $\alpha=4$ constitutes only a $\pm0.3$dex uncertainty in the maximum specific star formation rates for halos larger than $10^{10}\Msun$.  The uncertainty in specific halo mass accretion rates is negligible in comparison, as those are directly measurable from dark matter simulations (Appendix \ref{a:mar_calibration}).  Hence, the maximum average SSFRs for a mass-selected galaxy population are limited to be within a factor of 1--4 of the specific mass accretion rates at all redshifts, as shown in the top panel of Fig.\ \ref{f:max_ssfr}.

These constraints are consistent with existing measurements of galaxy SSFRs (Fig.\ \ref{f:max_ssfr}).  They also place strong constraints on what typical stellar populations will look like at higher redshifts.  E.g., at $z=10$, averaged specific star formation rates are limited to no more than $10^{-7.3}$ yr$^{-1}$.  If a survey returned higher average specific star formation rates for galaxies in a chosen stellar mass bin, it would most likely indicate that the survey was biased towards selecting highly star-forming galaxies, or that the analysis pipeline was giving biased stellar masses or star formation rates.\footnote{Of course, it could also indicate that $\Lambda$CDM is wrong---though this avenue seems an unlikely way that this could be proven.}

\subsection{The ``Too Big to Fail'' Problem and Satellite Galaxy Abundance Matching}

\label{s:d_clf_am}

Redshift evolution in the stellar mass---halo mass relation at $z>4$ can strongly affect the stellar mass for satellite galaxies. For example, the scatter in satellite stellar masses at fixed peak halo mass could be very large, depending on the distribution of satellite formation times.  This is especially true for the satellite halo mass range expected to host the brightest dwarf satellites of the Milky Way ($\sim10^9$--$10^{10}\Msun$).  \cite{BK12} identified several Milky Way satellites which were much too bright compared to $z=0$ stellar mass--halo mass relation.  Conversely, \cite{BK12} found that not all of the most massive satellite halos of the Milky Way could host the brightest satellite galaxies (i.e., the ``too big to fail'' problem).  While several papers have shown that the observations may be explained by a combination of cyclic feedback and tidal interactions with the disk of the Milky Way \citep{Governato12,Zolotov12,Teyssier13,Pontzen14}, it remains possible that some of the satellites were formed and accreted very early in the history of the universe.  In this case, Figs.\ \ref{f:evolution} and \ref{f:mass_ev} would suggest that these early-formed satellites should have a higher stellar mass than later-formed satellites at fixed circular velocity.

We briefly note that abundance matching methods typically match galaxies to halos using the SMHM relation at a single redshift.  For satellites, this creates some consistency issues, because it is often not clear whether it is better to use the SMHM relation at the time of the satellite's accretion or peak halo mass/peak maximum circular velocity \citep{Behroozi10,Yang11,Moster12}.  At $z<2$, clustering and conditional luminosity functions indicate that satellite galaxy masses are best set by using the SMHM relation at a single redshift \citep{Reddick12,Watson13}; physically, this indicates continued star formation in a fraction of satellites after accretion \citep{Wetzel12}.  However, strong evolution in the SMHM relation at high redshifts suggests that abundance matching using the SMHM relationship at the time of accretion or peak halo properties will better capture satellite galaxy masses at $z>4$.

\subsection{Comparison with Previous Semi-Empirical Models}

\label{s:d_comp}

Several previous models, such as \cite{Trenti10} and \cite{Wyithe13}, have assumed that high-redshift star formation occurs in bursty episodes.  Under this interpretation, many galaxies at $z>4$ are unobserved because they are temporarily not forming stars.  This assumption is in part based on rest-frame UV angular clustering measurements at $z>4$, such as \cite{Lee09}.  We note, however, that modeling angular clustering measurements can be extremely challenging.  The \cite{Lee09} conclusions are largely based on the discrepancy between the 1-halo autocorrelation term between their model and the observed data.  However, the 1-halo term is dominated by satellites, which tend to have lower star formation rates than central galaxies.  As a result, the \cite{Lee09} results largely constrain the duty cycle of \textit{satellites}; if only 30\% of satellites are star-forming at $z=4$, this would still allow 90\% of all galaxies to be star-forming, since the satellites make up a small fraction at these redshifts.  Additional issues generic to all angular correlation functions are appropriately propagating uncertainties in the redshift distribution (which are difficult to determine for faint galaxies with UV-only detections) and understanding how the sample selection function (e.g., color-color cuts) affects the redshift-dependent average bias of the selected galaxies.  We note that \cite{conroy:06} finds no duty cycle to be necessary when matching angular correlation functions at $z=4$ and $z=5$, further suggesting that it is very difficult to use angular correlation functions as robust constraints on galaxy duty cycles.

Separately, \cite{Wyithe13} argued for lower duty cycles so that the growth of the stellar mass density matches the cosmic star formation rate; however, previous discrepancies have largely been resolved \citep{Reddy09,Bernardi10,Moster12,BWC12}.  Also, the concern in \cite{Wyithe13} that $z>4$ galaxies cannot have formed stars at their observed rates for an entire Hubble time is resolved by strong evidence for individual galaxies having rising star formation rates \citep{Papovich11,BehrooziNumberD,BWC12,Moster12,Lee14}, as well as the overall rise in the total cosmic star formation rate with time.

Because of problems with angular correlation functions, duty cycles at $z>4$ may not be fully determined until future deep rest-frame optical observations \citep{Wyithe13}.  In our approach, we chose not to assume low duty cycles due to several heuristic arguments.  Most importantly, UV light continues to be emitted by stellar populations up to 100 Myr old \citep{Conroy09,Castellano14,Madau14}.  At $z=8$, this is comparable to specific halo mass accretion rates---so even if a major starburst expels all the gas from a galaxy, an equal amount of replacement gas will have re-accreted by the time the first stellar population fades in the UV.  At $z=4$, halo specific mass accretion rates are somewhat lower ($2-3\times 10^{-9}$ yr$^{-1}$), but the typical halo still grows by $20-25\%$ in 100 Myr.    These high accretion rates make it difficult for stellar feedback to completely quench star formation on long timescales.  Second, high quenched fractions at $z=4$ are inconsistent with $K$-band selected stellar mass functions at $z<4$, because it would imply a discontinuous jump between LBG-selected (at $z>4$) and $K$-selected (at $z<4$) stellar mass functions.  Instead, the growth of LBG-selected stellar mass functions into $K$-selected mass functions is extremely consistent with measured specific star formation rates \citep{BWC12,Moster12}.\footnote{This argument also excludes dust from hiding many galaxies at $z>4$.} Third, the quenched fractions measured in $K$-selected samples at $2<z<4$ are very low \citep{Brammer11,Muzzin13} except at the very highest stellar masses, which is inconsistent with a large quenched fraction at $z>4$---unless the quenched galaxies at $z>4$ all return to being star forming at $z<4$.  Fourth, regardless of the feedback prescription, the observational duty cycle of galaxies above $10^{9}\Msun$ is unity in hydrodynamical simulations (G.\ Snyder, P.\ Hopkins, priv.\ comm.; \citealt{Jaacks12b,Wise14}).

Despite the differences with our approach, both \cite{Trenti10} and \cite{Wyithe13} also find that galaxy star formation must be more efficient at high redshifts---in their language, that galaxies have higher duty cycles.  Their conclusion is driven by the same evidence as ours; namely, that if the same stellar mass--halo mass or star formation---halo mass relationship for $z=4$ halos were applied to $z=8$ halos, the resulting stellar mass, SFR, and luminosity functions would be an order of magnitude less than observational constraints (\S \ref{s:smf_errs}).  However, we note that the \cite{Trenti10} extension to high redshifts ($z>8$) is very different than ours, because their duty cycle is nearly 100\% at $z=8$ and has no more room to grow.  Hence, the \cite{Trenti10} predictions are very similar to the case in Fig.\ \ref{f:smf_highz}, where the stellar mass---halo mass relationship is assumed to be fixed for $z>8$.  As noted in \S \ref{s:d_jwst}, JWST should be able to very quickly distinguish these two scenarios.

We note finally that \cite{Tacchella13} took a very different approach, which was to model halo star formation histories as a combination of a burst and a longer-duration star formation mode.  While this model provides an excellent match to luminosity functions, star formation histories of halos are not self-consistent across redshifts (S.\ Tacchella, priv.\ comm.).  For example, the star-formation history of $z=5$ halos at $z>6$ in the model is not consistent with the star-formation histories of $z=6$ halos at $z>6$; so if applied to merger trees, the model would require retroactive modification of galaxy star formation histories.  It is nonetheless still interesting mathematically to consider the resulting redshift evolution; the \cite{Tacchella13} model also does not predict any break in the slope of the cosmic star formation rate for $z>8$.

\section{Conclusions}
\label{s:conclusions}
We have presented a simple theoretical basis for predicting galaxy stellar mass, specific star formation rate, and host halo mass evolution at high redshifts (\S \ref{s:basic}).  We reach the following conclusions:
\begin{enumerate}
\item The ratio of galaxies' specific star formation rates to their host halos' specific mass accretion rates can be used to predict galaxies' evolution in the stellar mass---halo mass plane with redshift.  (\S \ref{s:basic}, \S \ref{s:t_smhm}, \S \ref{s:t_ssfr}, Appendix \ref{a:single_ssfr}).
\item Tests of this predictive model with galaxy properties at $z=4$ successfully match observational constraints on the stellar mass---halo mass relation, specific star formation rates, and cosmic star formation rates for $z=5$ to $z=8$ (\S \ref{s:t_smhm}, \S \ref{s:t_ssfr}).
\item Galaxy specific star formation rates are expected to increase only mildly at higher redshifts (\S \ref{s:pred_ssfr}).
\item Cosmic star formation rates are not expected to rapidly fall at $z>8.5$, in contrast to recent \textit{HST} measurements, but consistent with estimates from long gamma-ray burst rates (\S \ref{s:pred_ssfr}).
\item Halos smaller than $10^{12}\Msun$ are expected to be increasingly efficient at star formation at $z>4$ (\S \ref{s:pred_smhm}), which is strongly linked with the ``plateau'' in galaxy SSFRs at fixed stellar mass (\S \ref{s:d_ssfr}).
\item This evolution in efficiency is too strong to allow star formation efficiency to be set only by gravitational potential well depth at $z>4$ (\S \ref{s:pred_smhm}); at $z>4$, halos maintain similar stellar mass---halo mass ratios along their progenitor histories.
\item Matching WMAP constraints on the optical depth with our model requires high average escape fractions of $f_\mathrm{esc}>0.2$ during reionization (\S \ref{s:d_reion}).
\item JWST should see many $M_\ast > 10^{8}\Msun$ galaxies over the entire redshift range $8<z<15$ (\S \ref{s:d_jwst}).
\end{enumerate}

\acknowledgements
PB was supported by a Giacconi Fellowship through the Space Telescope Science Institute, which is operated by the Association of Universities for Research in Astronomy, Incorporated, under NASA contract NAS5-26555.  We thank Marcelo Alvarez, John Beacom, Richard Ellis, Steve Finkelstein, Harry Ferguson, Phil Hopkins, Juna Kollmeier, Avi Loeb, Cameron McBride, Casey Papovich, Eliot Quataert, Greg Snyder, John Wise, Risa Wechsler, Andrew Wetzel, and Ann Zabludoff for very useful discussions.  We also give special thanks to Matt Becker for running the \textit{Lb125} simulation, to Charlie Conroy for adding JWST filters to FSPS, to St\'ephane Charlot and Gustavo Bruzual for providing verification data for our luminosity calculations, and to the anonymous referee for many helpful suggestions.


\bibliography{master_bib}

\appendix

\begin{figure*}
\vspace{-4ex}
\plotminigrace{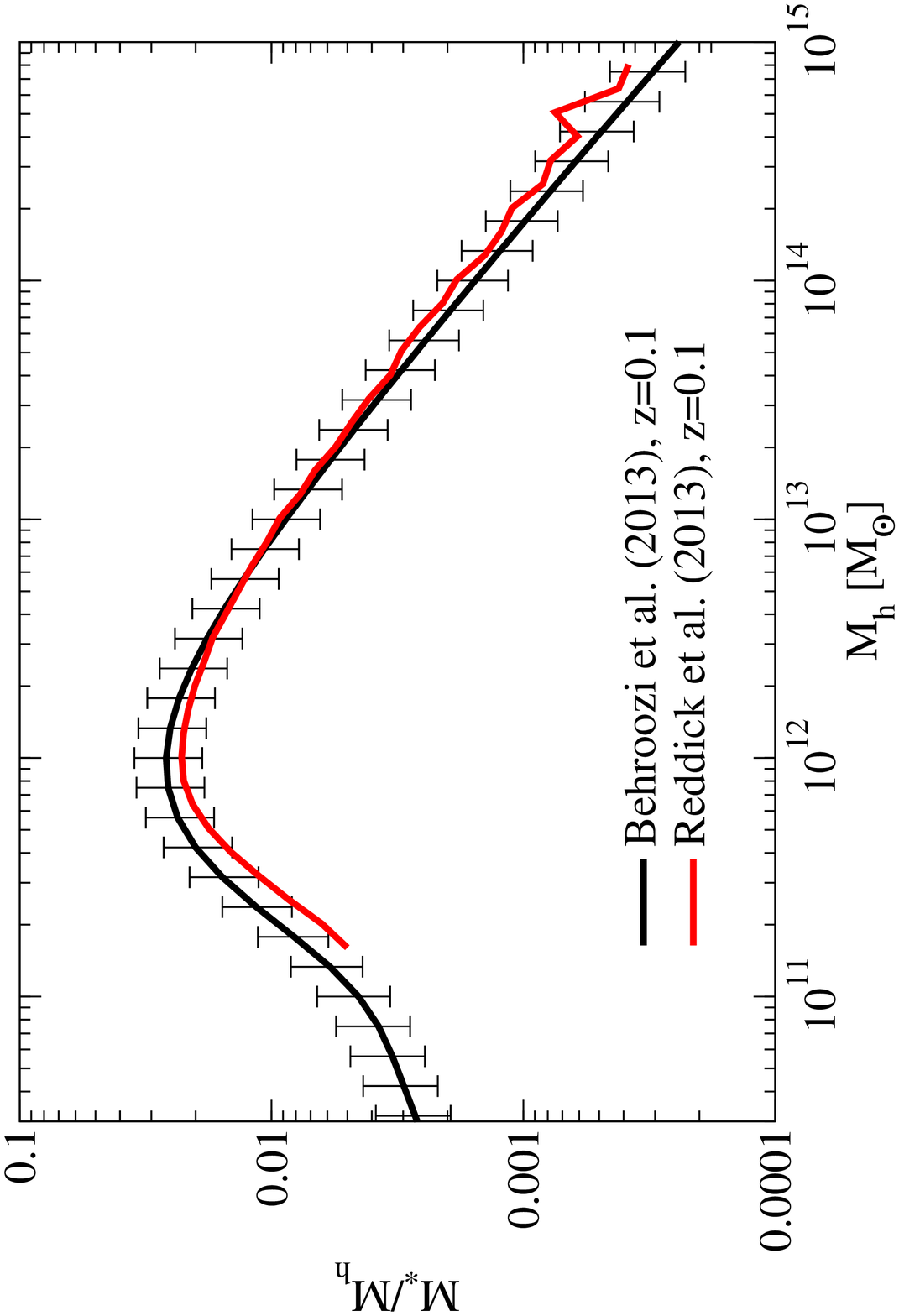}\plotminigrace{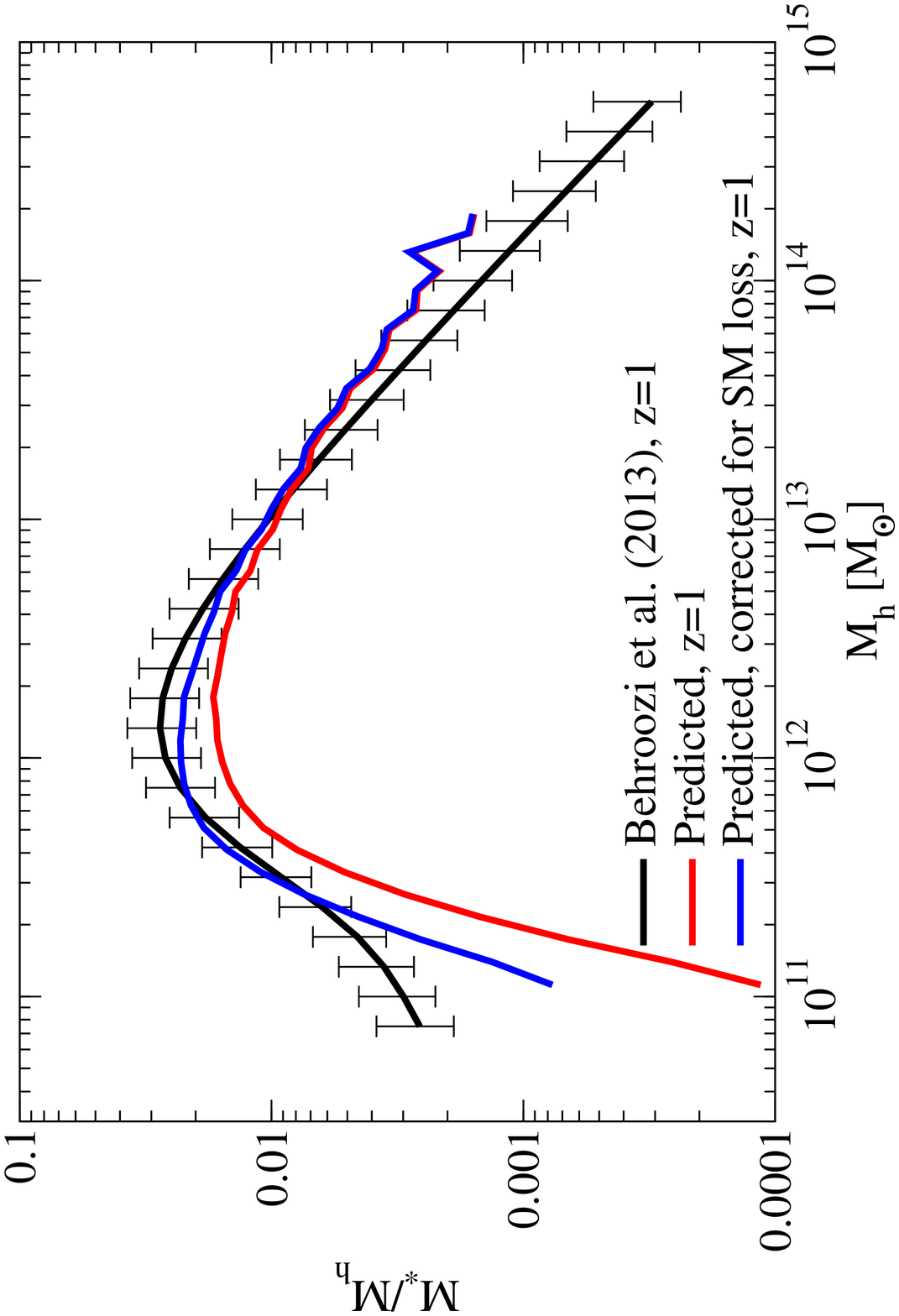}\\[-4ex]
\plotminigrace{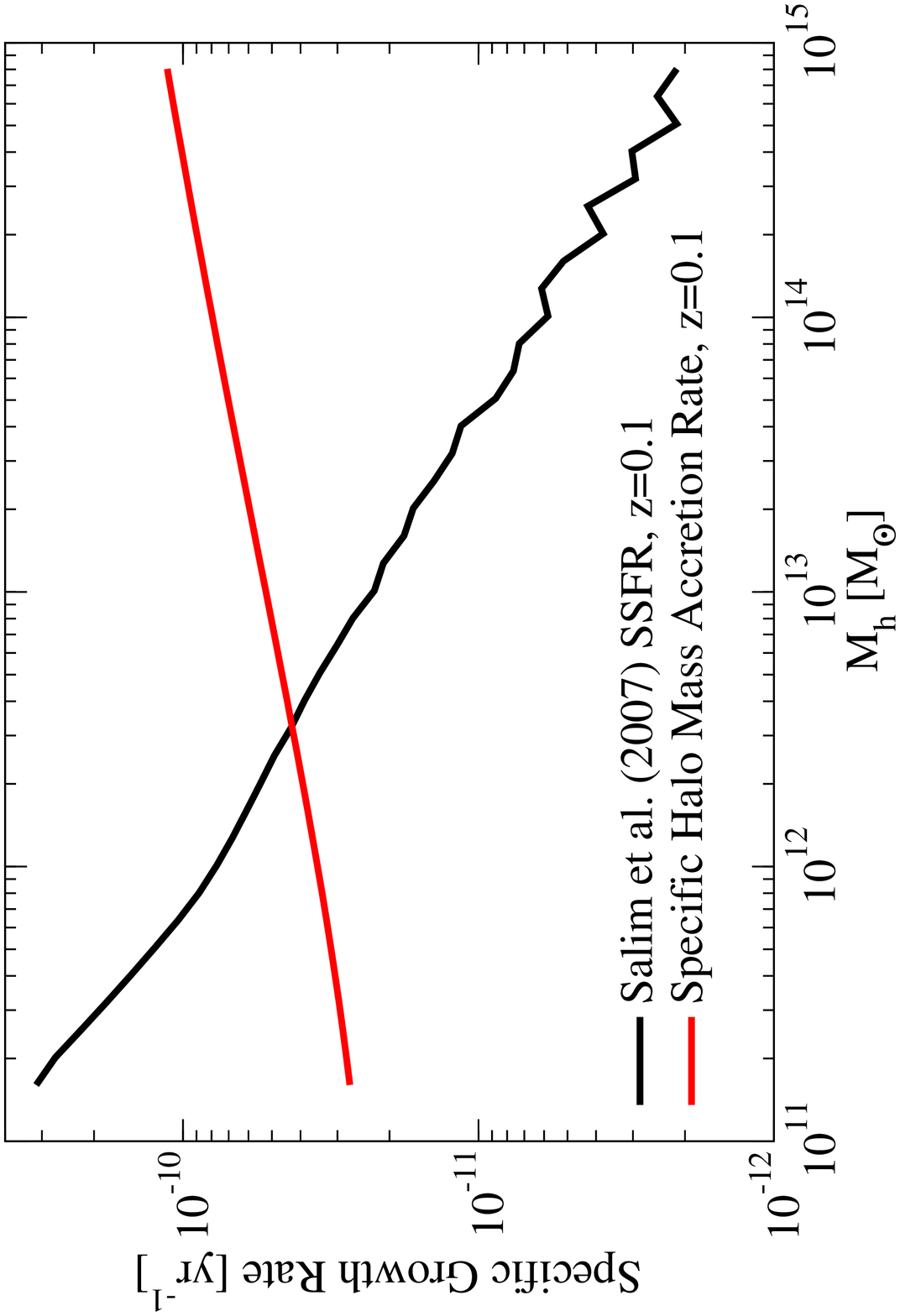}\plotminigrace{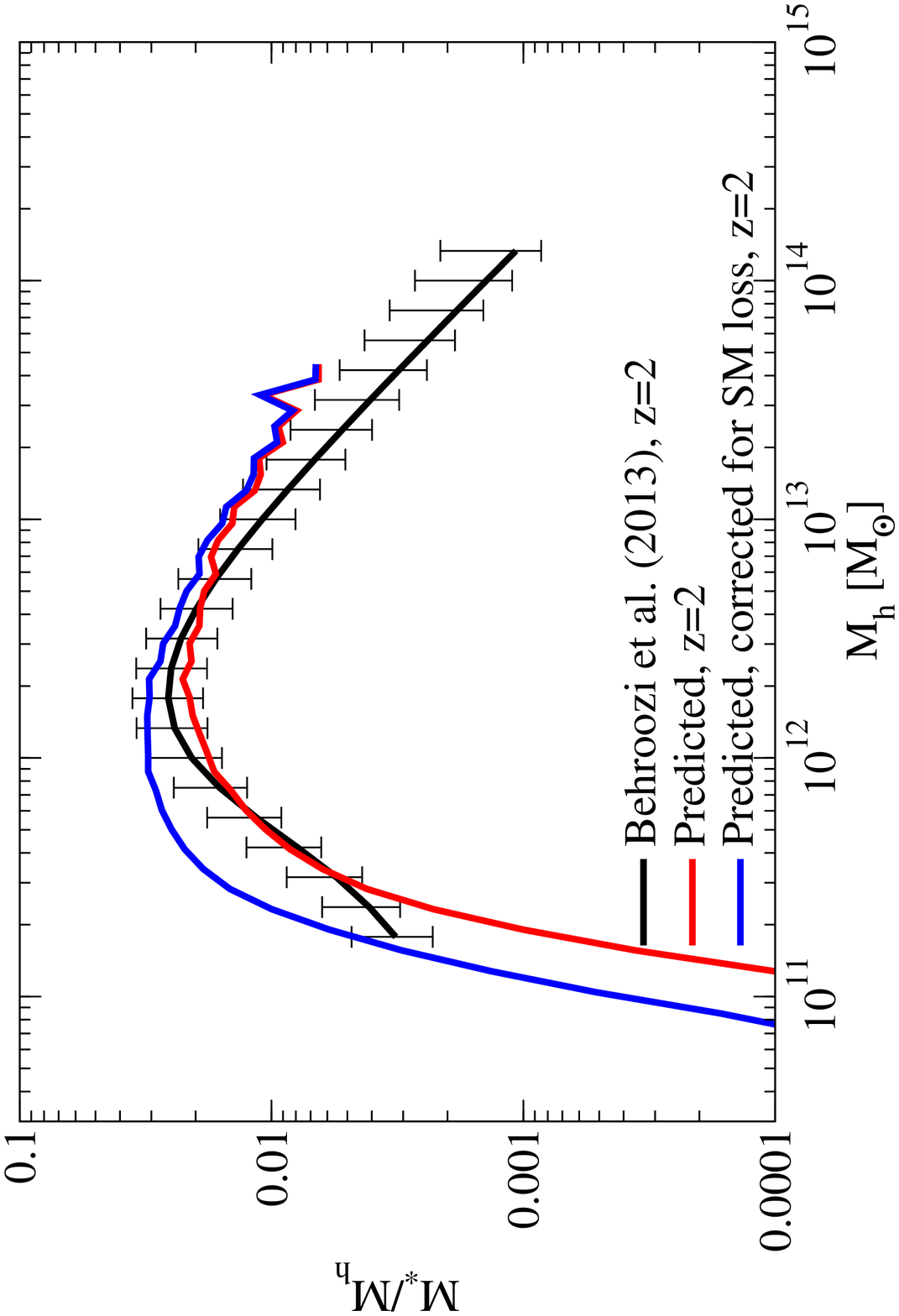}\\[-4ex]
\plotminigrace{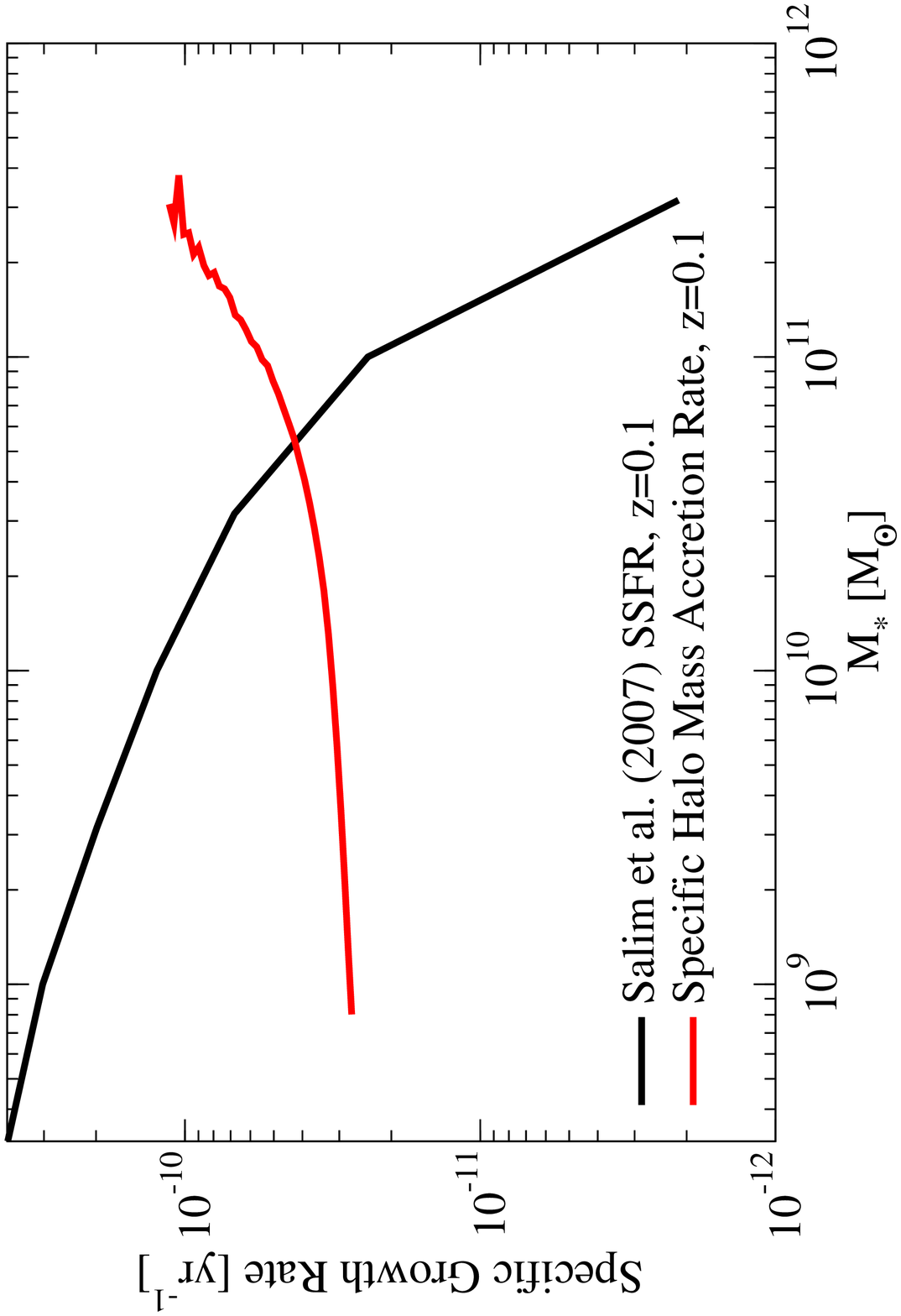}\plotminigrace{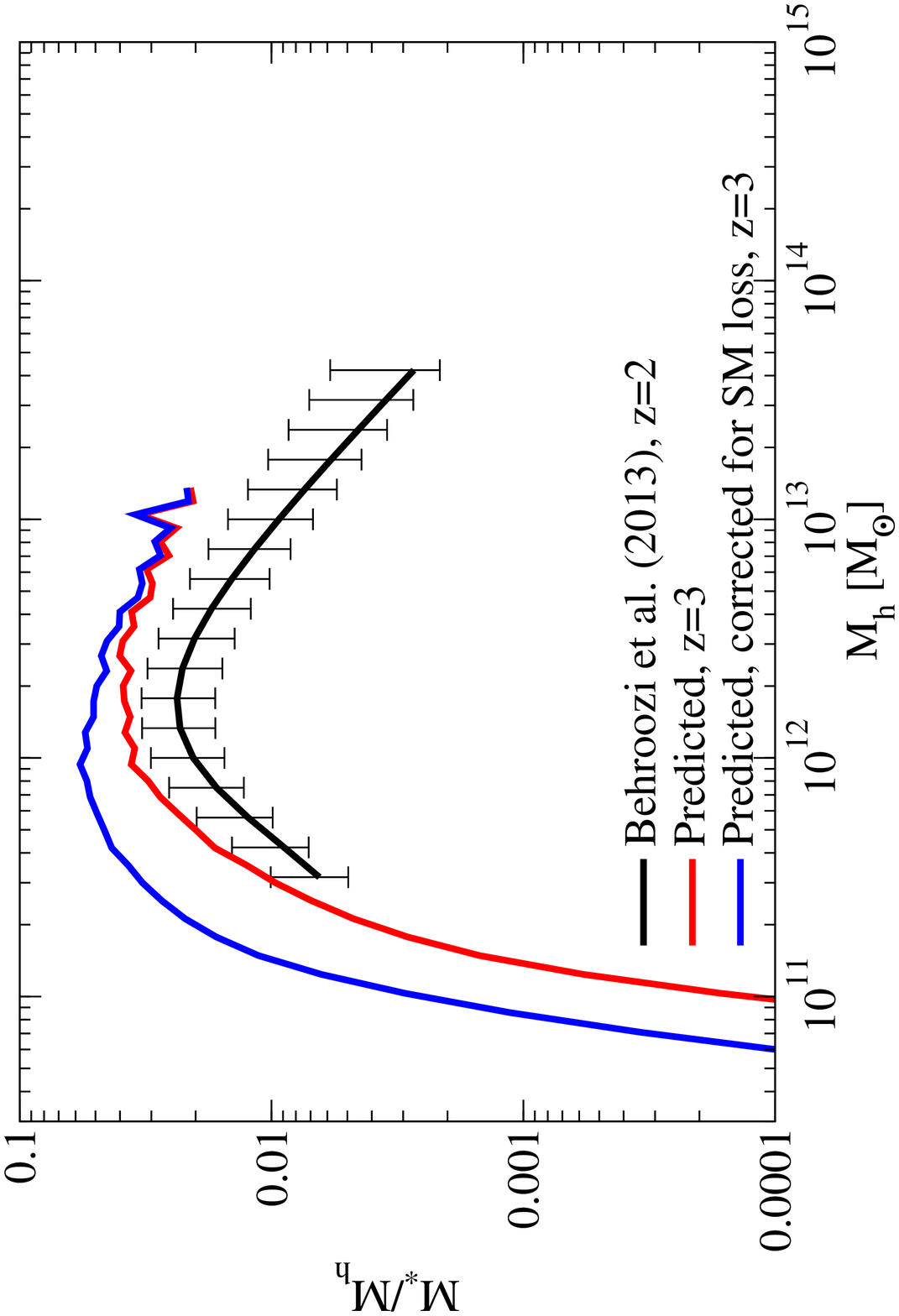}
\caption{Test of the methodology in \S \ref{s:basic} at low redshifts.  The \textbf{top-left} panel shows the stellar mass--halo mass (SMHM) ratio for $z=0.1$ derived in \cite{Reddick12} by pure abundance matching (i.e., considering a single redshift only), compared to the SMHM ratio in \cite{BWC12}, derived by abundance modeling (\S \ref{s:amodel}).  The \textbf{middle-left} panel shows the specific star formation rates from \cite{Salim07} as a function of halo mass (using \citealt{Reddick12} to convert between stellar mass and halo mass) along with specific halo mass accretion rates as a function of halo mass (Appendix \ref{a:mar_calibration}).  The \textbf{bottom-left} panel shows the same data, except this time plotted as a function of stellar mass.  The \textbf{top-right} panel compares the \cite{BWC12} constraints on the SMHM ratio at $z=1$ (black line) to predictions for the $z=1$ SMHM ratio using the $z=0.1$ \cite{Reddick12} SMHM ratio and the \cite{Salim07} SSFRs with the methodology in \S \ref{s:basic} (red line).  The blue line shows the predictions if the \cite{Salim07} SSFRs are not corrected for passive stellar mass loss (\S \ref{s:effects}).  The \textbf{middle-right} and \textbf{bottom-right} panels show the analogous comparisons at $z=2$ and $z=3$.  As explained in the text, the assumption of a power-law historical SMHM relation for galaxies at $z=0.1$ breaks down at $z=2$ and above.}
\label{f:pred_z0.1}
\end{figure*}

\section{Tests of the Methodology at $z=0.1$}

\label{a:single_ssfr}

Because we used abundance modeling to calculate stellar mass---halo mass (SMHM) relations as well as specific star formation rates (\S \ref{s:amodel}) throughout this paper, it may seem that some of the simplicity of \S \ref{s:basic} has been lost.  Here, we show that the method still works without the extra machinery of \S \ref{s:amodel}, albeit with some cost to accuracy.

In the left panels of Fig.\ \ref{f:pred_z0.1}, we show the SMHM ratios derived in \cite{Reddick12}, along with specific star formation rates from \cite{Salim07}.  \cite{Reddick12} derived their results from abundance matching the \cite{Moustakas12} stellar mass function at $z=0.1$ to halos in the Bolshoi simulation (\S \ref{s:sims}).  The bottom two panels on the left of Fig.\ \ref{f:pred_z0.1} also show the specific halo mass accretion rates from Appendix \ref{a:mar_calibration}.  These panels show the stark contrast between the formation of galaxies and halos at low redshifts: galaxies in low-mass halos are forming stars much more rapidly than their halos are growing, and the opposite is true for galaxies in high-mass halos.

In the right panels of Fig.\ \ref{f:pred_z0.1}, we show predictions for the SMHM ratios at $z=1,$ 2, and 3 from using the \cite{Reddick12} and \cite{Salim07} data directly in Eqs.\ \ref{e:ssfr} and \ref{e:smhm}.  We note that the stellar mass estimators used in \cite{Moustakas12} and \cite{Salim07} are different, each using SED fitting with their own stellar population history models and dust models.  We also note that the \cite{Salim07} SSFRs are the ratio of current SFR to currently remaining stellar mass, whereas Eq.\ \ref{e:ssfr} requires the ratio of current SFR to the total stellar mass ever formed.  For a 13 Gyr-old population, the ratio of currently remaining to total formed stellar mass is 54\% (Eq.\ \ref{e:sm_loss}), so we also show predictions if we multiply the \cite{Salim07} SSFRs by this fraction.  These predictions therefore bracket the range of allowable ages for stellar populations.  We make no attempt to correct for stellar mass gained in mergers (\S \ref{s:effects}).  \cite{BWC12} find that the merger contribution is negligible for $M_h < 10^{12}\Msun$ halos, but can be important for $M_h > 10^{13.5}\Msun$ halos at $z<2$.

Nonetheless, this predictive range is in remarkable agreement with constraints from \cite{BWC12} at $z=1$ and $z=2$.  We find it astonishing that the physical properties of galaxies three-quarters of the way back to the Big Bang are predictable using observations of practically our own backyard.  The agreement at $z>2$ is not so good, and it is worth understanding why the relationship breaks down at higher redshifts.  In fact, at low redshifts, a simple pure power-law star formation efficiency history like that assumed in Eq.\ \ref{e:smhm} is only a reasonable description for low-mass galaxies---i.e., galaxies with masses less than $10^{10.5}\Msun$ in halos with masses less than $10^{12}\Msun$.  Predictions for high-mass galaxies work because their specific star formation rates are low.  As long as the predicted specific star formation rates are less than the inverse Hubble time, huge relative errors no longer matter: the buildup of the galaxy stellar mass from $z=2$ to $z=0$ will be only small fraction of the total (see also Eq.\ \ref{e:errs}).  However, at $z>2$, many of the massive galaxies today were forming most of their stars \citep{BWC12}, and the growth of galaxies relative to their halos was much faster (e.g., Fig.\ \ref{f:pred_z4}).  Getting the magnitude of the specific star formation rates correct therefore becomes more important, and the simple extrapolation in Eq.\ \ref{e:smhm} breaks down.

\begin{figure}
\begin{center}
\plotminigrace{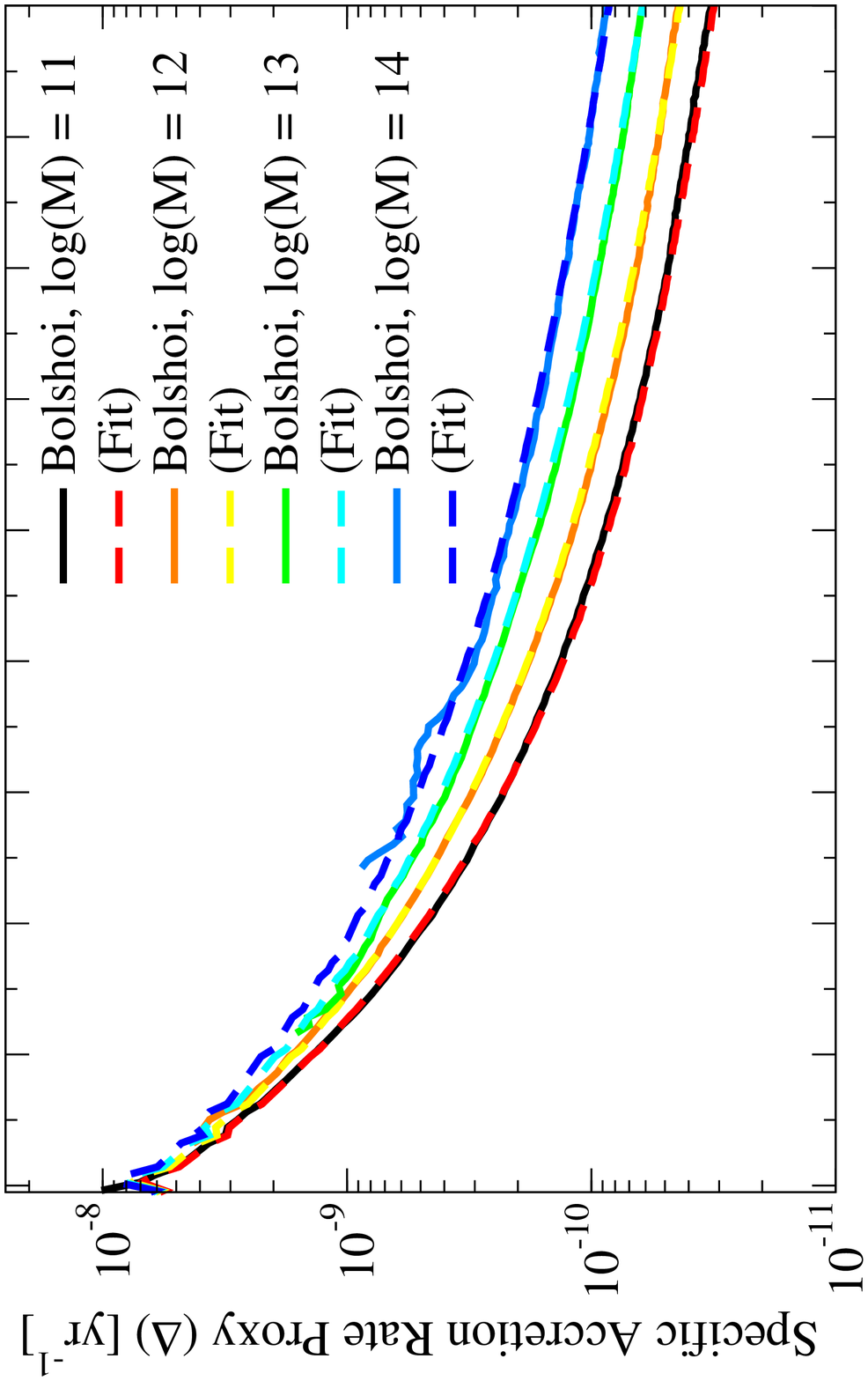}\plotminigrace{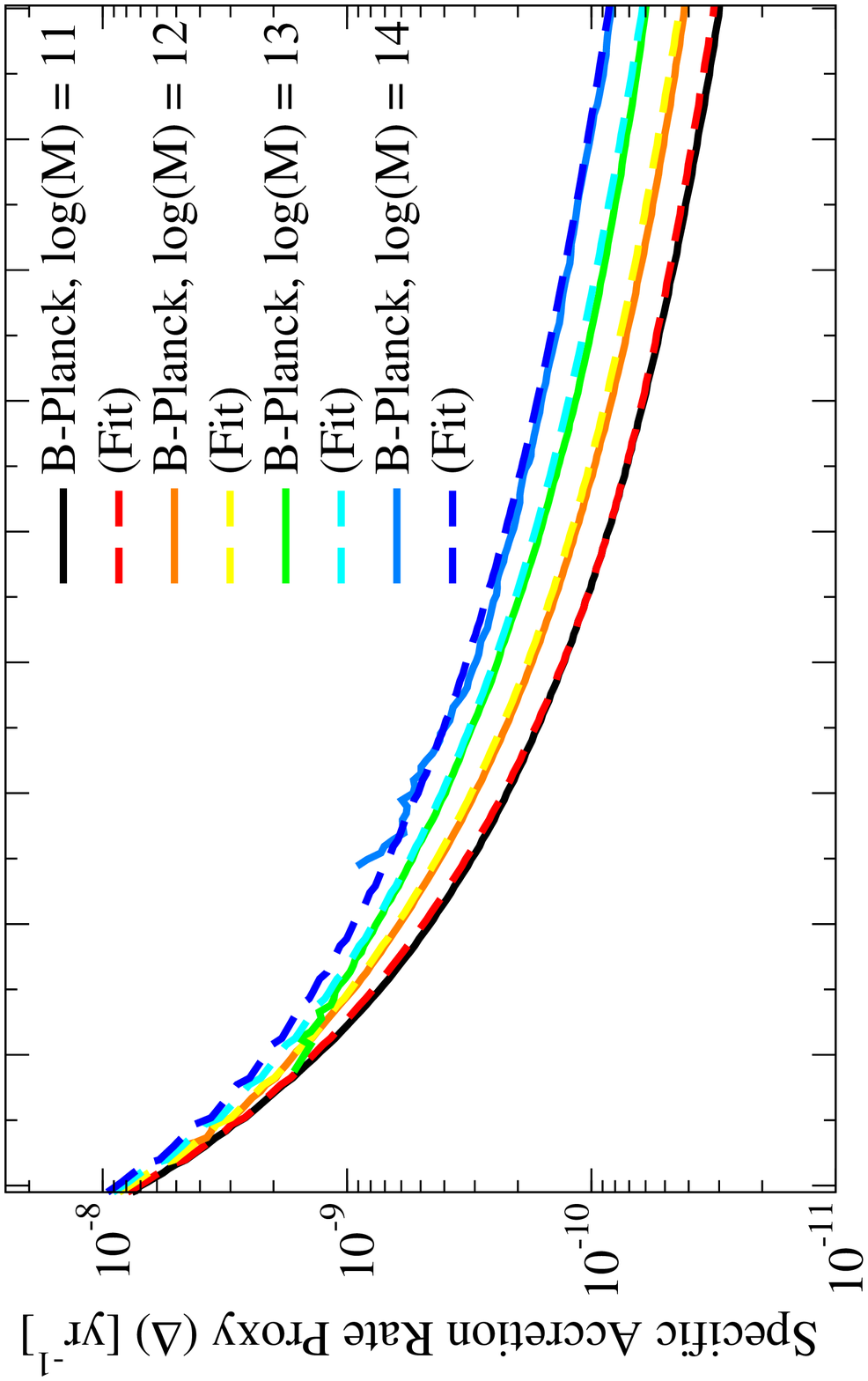}\\[-17ex]
\plotminigrace{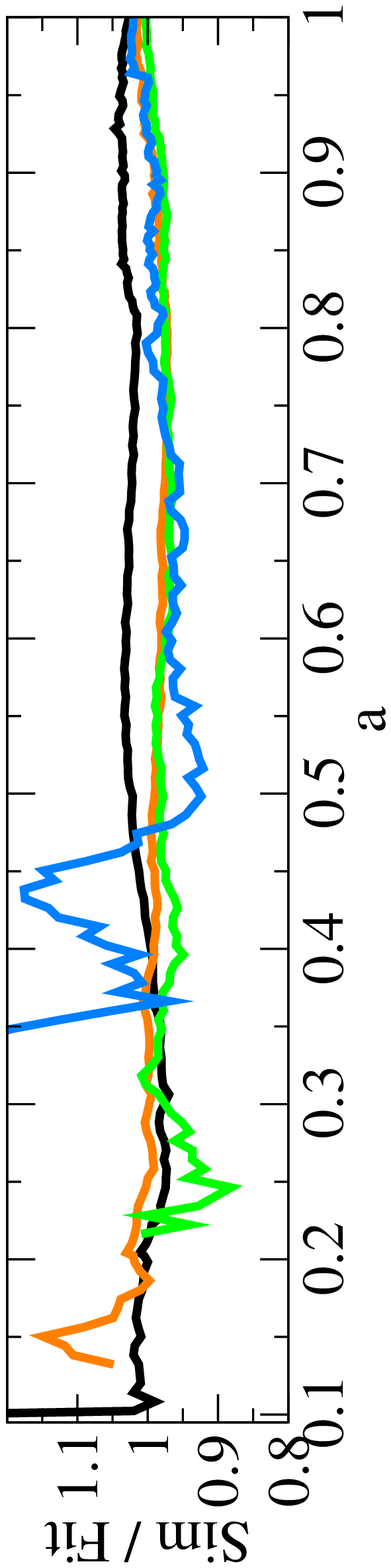}\plotminigrace{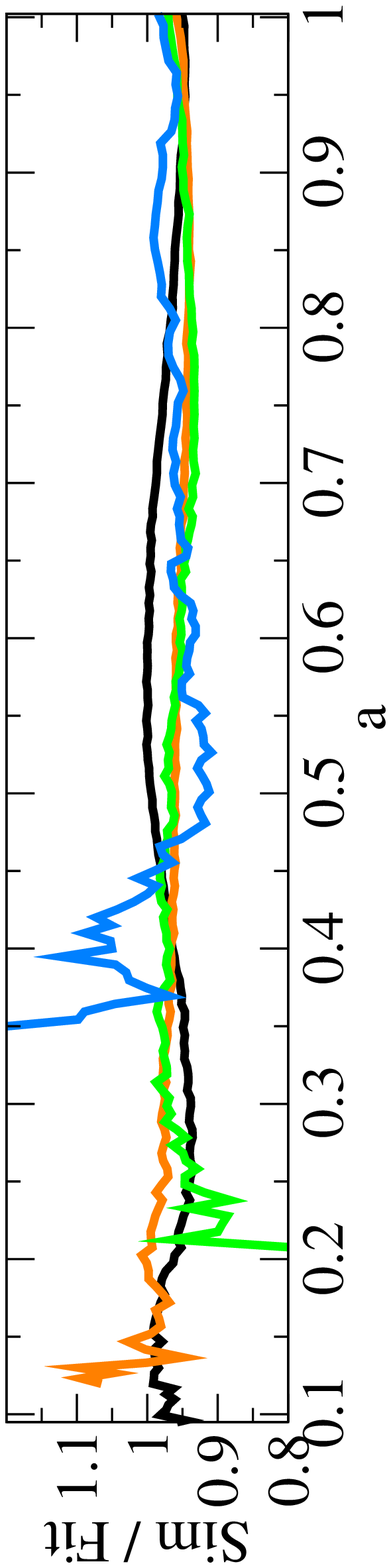}\\[-35ex]
\plotminigrace{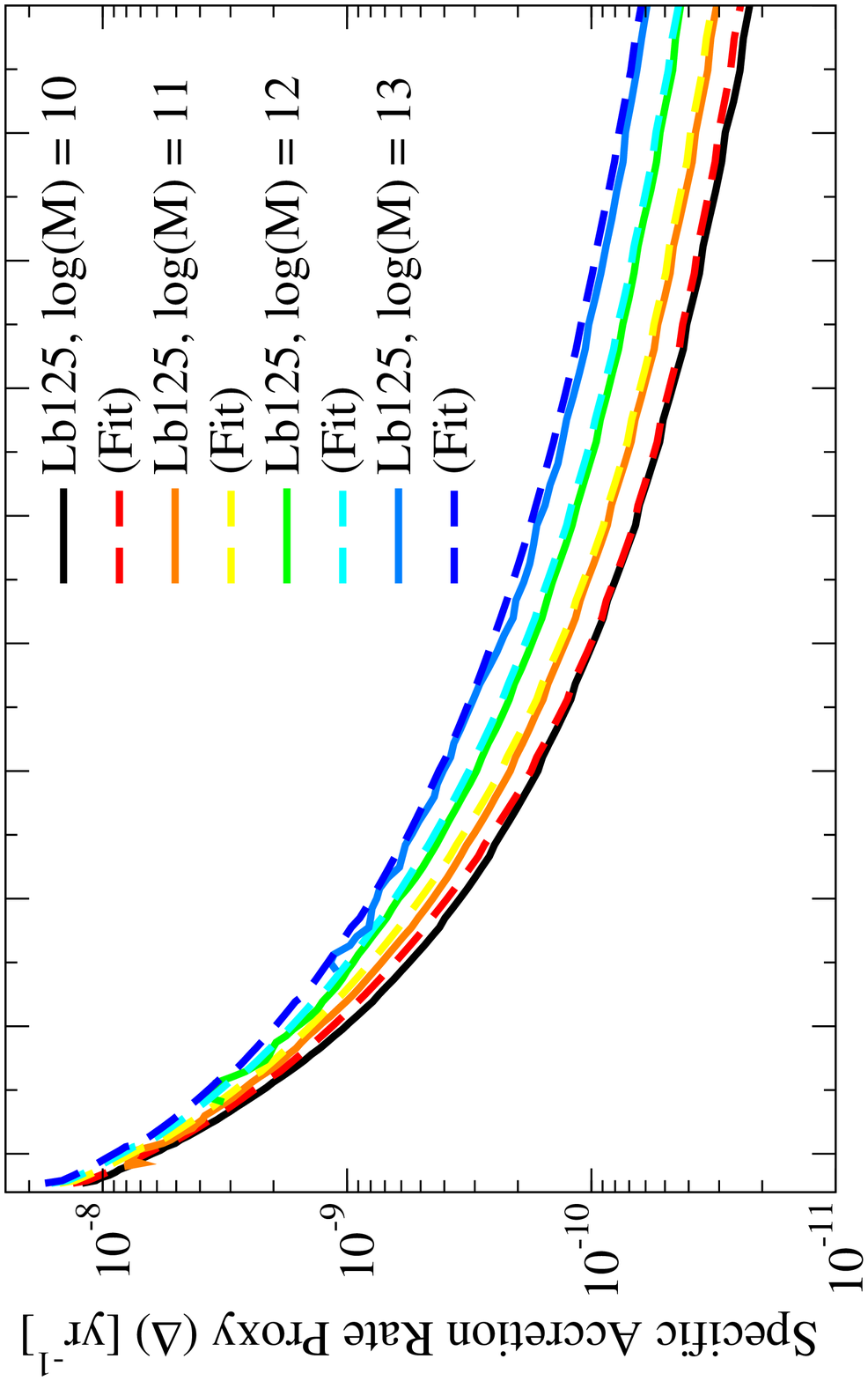}\\[-17ex]
\plotminigrace{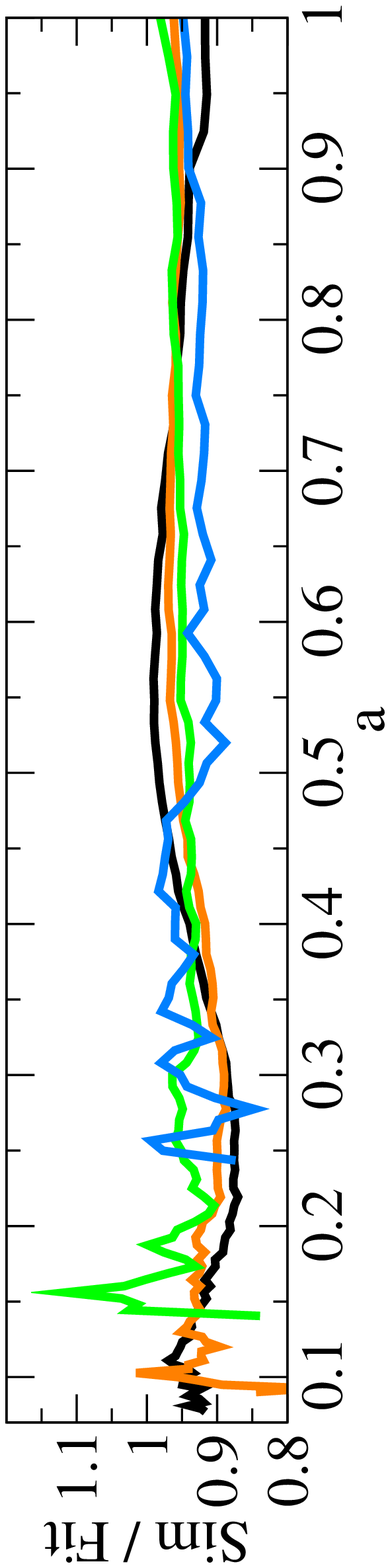}\\[-30ex]
\end{center}
\vspace{-35ex}
\caption{\textbf{Left} panel: specific mass accretion rates as a function of halo mass and scale factor in the \textit{Bolshoi} simulation (measured using the $\Delta$ proxy, Eq.\ \ref{e:macc_fit}) compared to the $\Delta$ values from Eq.\ \ref{e:macc_fit}.  The lower-left panel shows the ratio of the values measured in the simulation to the fit; typical deviations are 2--3\%.  \textbf{Right} panel: same, for the \textit{Bolshoi-Planck} simulation, which uses a cosmology similar to the \textit{Planck} best-fit values \citep{Planck}.  \textbf{Bottom} panel: same, for the \textit{Lb125} simulation, which uses a cosmology in-between the \textit{Bolshoi} and \textit{Bolshoi-Planck} cosmologies.}
\label{f:mass_calib}
\end{figure}

\section{Calibrating Mass Accretion Rates}

\label{a:mar_calibration}

We calculate halo mass accretion rates from the growth of halos between discrete simulation timesteps.  As usual, if the timesteps are spaced too closely, halo growth will be small relative to measurement errors (e.g., Poisson noise and halo finding algorithm instabilities).  If the timesteps are too far apart, the halo growth rate will change over the time interval, and the inferred rates will be biased.  To work around this problem, we adopt a parametrized fitting function for the average halo growth rate.  In this way, the inferred halo growth between any two redshifts can be checked for consistency with direct measurements from simulations.

The choice of fitting function is important, since we are modeling halo growth over large redshift ranges (e.g., $z=8$ to $z=15$).  Common fitting forms (e.g., \citealt{Wechsler02,McBride09,Wu13b}) approximate the early growth of halos as exponential in redshift.  While not a bad approximation at a single halo mass, halos at different masses will have different exponential growth rates.  Since larger halos form later, their growth profiles are better fit by steeper exponential growth rates \citep{Wechsler02,McBride09}; extrapolating this faster growth to high redshifts results in the progenitor mass histories of high-mass halos unphysically falling below the progenitor mass histories of lower-mass halos.

\cite{BWC12} avoid this problem by fitting the growth history of halos at a single mass ($10^{13}\Msun$) with a standard form, and then fitting mass growth histories for other halo masses relative to the trajectory for $10^{13}\Msun$ halos.  This results in an accurate, although very complicated, fit for mass growth histories, which we reproduce here:
\begin{eqnarray}
M_\mathrm{med}(M_0,z) &=& M_{13}(z) 10^{f(M_0,z)}\\
M_{13}(z) &=& 10^{13.276} (1+z)^{3.00} (1+\frac{z}{2})^{-6.11}\exp(-0.503z)\Msun\\
f(M_0,z) & = & \log_{10}\left(\frac{M_0}{M_{13}(0)}\right)\frac{g(M_0,1)}{g(M_0,\frac{1}{1+z})}\\
g(M_0,a) & = & 1 + \exp(-4.651(a-a_0(M_0))\\
a_0(M_0) & = & 0.205 - \log_{10}\left[\left(\frac{10^{9.649}\Msun}{M_0}\right)^{0.18} + 1\right]
\end{eqnarray}
where $M(M_0,z)$ gives the median virial mass \citep{mvir_conv} for progenitors of halos with mass $M_0$ at $z=0$.

Since we are interested in average accretion rates, instead of the growth of the median halo mass, we cannot use the values directly from this function.  However, rather than re-fitting all the parameters, we consider simple modifications to the time derivative of this function:
\begin{equation}
\label{e:macc_form}
\langle \dot{M}_\mathrm{peak} \rangle(M_\mathrm{peak}, a) = \dot{M}_\mathrm{med}(M_\mathrm{0}(M_\mathrm{peak}, a),a) F(M_\mathrm{peak}, a)
\end{equation}
where $M_\mathrm{0}(M_\mathrm{peak}, a)$ is the inverse function of $M_\mathrm{med}(M_0, a)$; i.e., $M_\mathrm{med}(M_\mathrm{0}(M_\mathrm{peak}, a),a)  = M_\mathrm{peak}$.

To compare with simulations, we have measured the following proxy for specific mass accretion rate for every halo in Bolshoi for every timestep redshift $z$:
\begin{equation}
\label{e:delta_m}
\Delta = \frac{M_\mathrm{peak}(z) - M_\mathrm{peak}(z+0.5)}{M_\mathrm{peak}(z) [ t(z) - t(z+0.5) ]}
\end{equation}
where $t(z)$ is the time since the Big Bang at redshift $z$.  Calculating mass growth over a redshift delta of $0.5$ ensures a typical halo growth of $0.125$ dex, which is well beyond the typical 1\%--5\% mass fluctuations expected for halo finder mass recovery \citep{BehrooziTree}.  When a timestep at exactly $z+0.5$ was not available, we used the closest timestep available.  Bolshoi has 180 timesteps, which are spaced between 40-90 Myr apart, so this granularity becomes important at $z>4$.

For each trial parameter set in Eq.\ \ref{e:macc_fit}, we calculated the expected values for $\Delta$ as a function of halo mass and redshift.  We find that the following formula for average mass accretion rates is the best match to the Bolshoi simulation:
\begin{equation}
\label{e:macc_fit}
\langle \dot{M}_\mathrm{peak} \rangle(M_\mathrm{peak}, a) = \dot{M}_\mathrm{med}(M_\mathrm{0}(M_\mathrm{peak}, a),a) [0.22(a-0.4)^2+0.85]\frac{\left(\frac{M_\mathrm{peak}}{10^{12}\Msun}\right)^{0.05a}}{\left(1.0 + \frac{10^{11-0.5a}\Msun}{M_\mathrm{peak}}\right)^{0.04+0.1a}}
\end{equation}
We show a comparison between this formula's expected values for $\Delta$ and the actual values in Bolshoi as a function of halo mass and redshift in the left panel of Fig.\ \ref{f:mass_calib}.  The fitting-formula's typical deviations are 2--3\% from the measured values.

We have also checked this formula against two newer simulations. \textit{Bolshoi-Planck} uses a cosmology ($\Omega_M = 0.307$, $\Omega_\Lambda = 0.693$, $h=0.68$, $\sigma_8 = 0.823$, $n_s = 0.96$) similar to the \textit{Planck} best-fit cosmology \citep{Planck}.  This simulation used the same box size (250 Mpc $h^{-1}$), particle number ($2048^3$), force resolution ($1$ kpc $h^{-1}$), and simulation code (\textsc{art}) as \textit{Bolshoi}.  As noted in \S \ref{s:cosmology_errs}, this cosmology's specific mass accretion rates are expected to be about 2\% different from that of the \textit{Bolshoi} simulation.  This expectation is borne out in Fig.\ \ref{f:mass_calib}, right panel, where the measured deviations from the fitting formula in Eq.\ \ref{e:macc_fit} are 4--5\%.  We also used the \textit{Lb125} simulation to constrain the low-mass and high-redshift accretion rate behavior that is not as well sampled by \textit{Bolshoi}.  \textit{Lb125} uses a cosmology in between the \textit{Bolshoi} and \textit{Bolshoi-Planck} cosmologies ($\Omega_M = 0.286$, $\Omega_\Lambda = 0.714$, $h=0.7$, $\sigma_8 = 0.82$, $n_s = 0.96$).  This simulation used a smaller box size (125 Mpc $h^{-1}$), with the same particle number ($2048^3$) as \textit{Bolshoi} but with slightly higher force resolution (Plummer-equivalent of $0.5$ kpc $h^{-1}$) and using the \textsc{gadget} simulation code \citep{Springel05}.  The smaller box size (and consequent lack of larger-scale modes) reduces average halo accretion rates at fixed halo mass to 90-95\% of those in \textit{Bolshoi}.

\section{Luminosity Functions}

\label{a:lfs}

\begin{figure}
\begin{center}
\plotminigrace{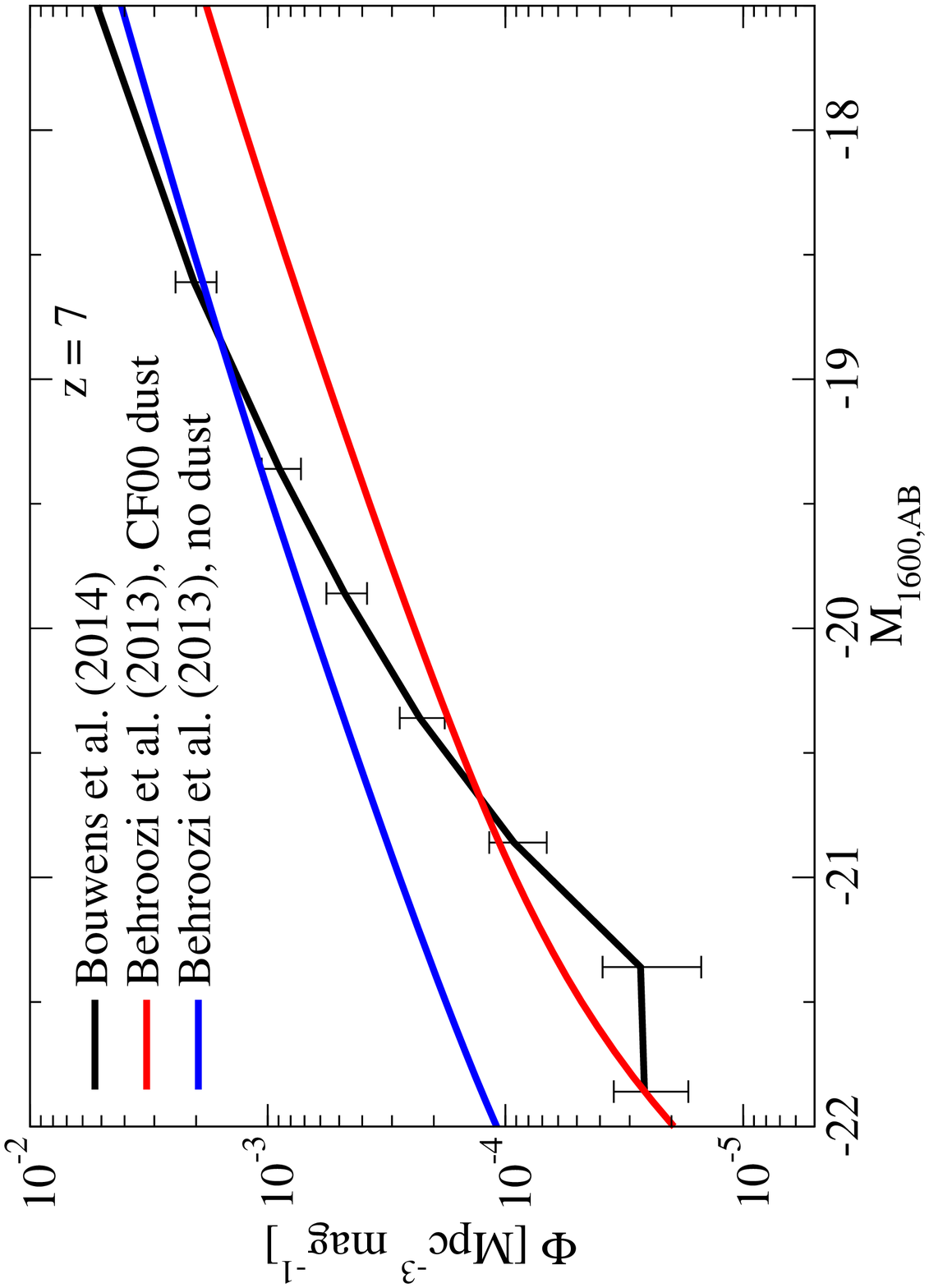}\plotminigrace{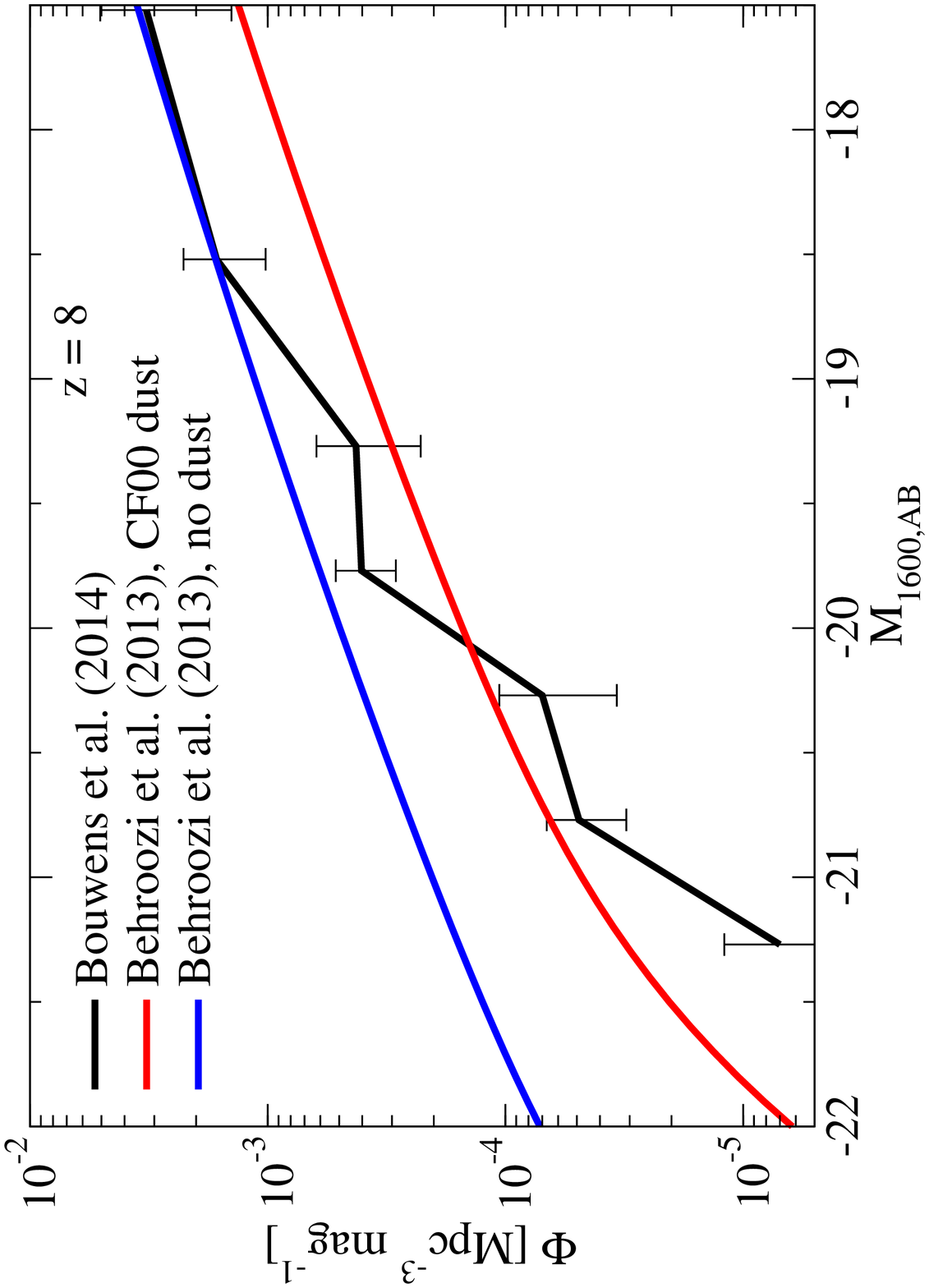}\\[-5ex]
\plotminigrace{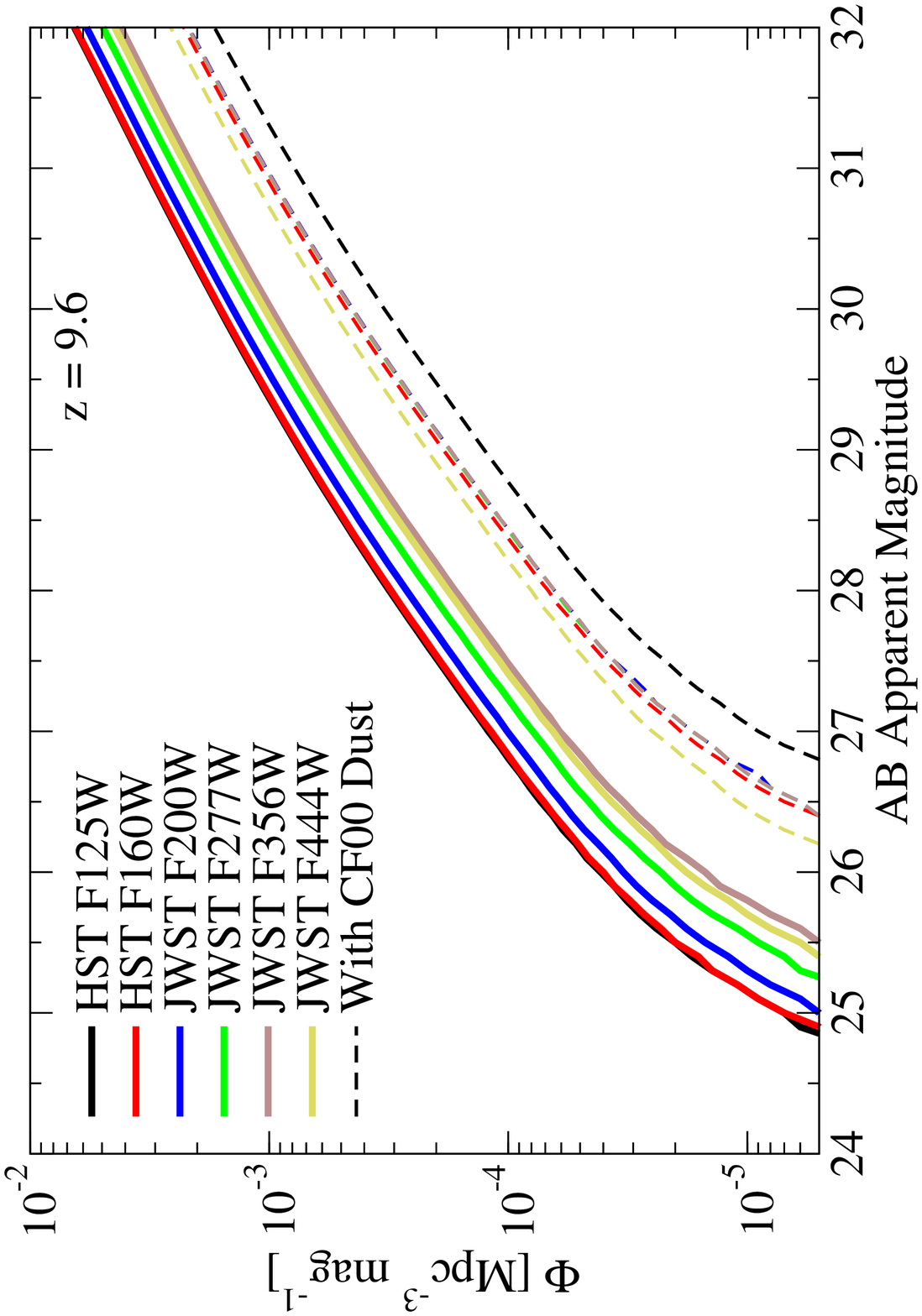}\plotminigrace{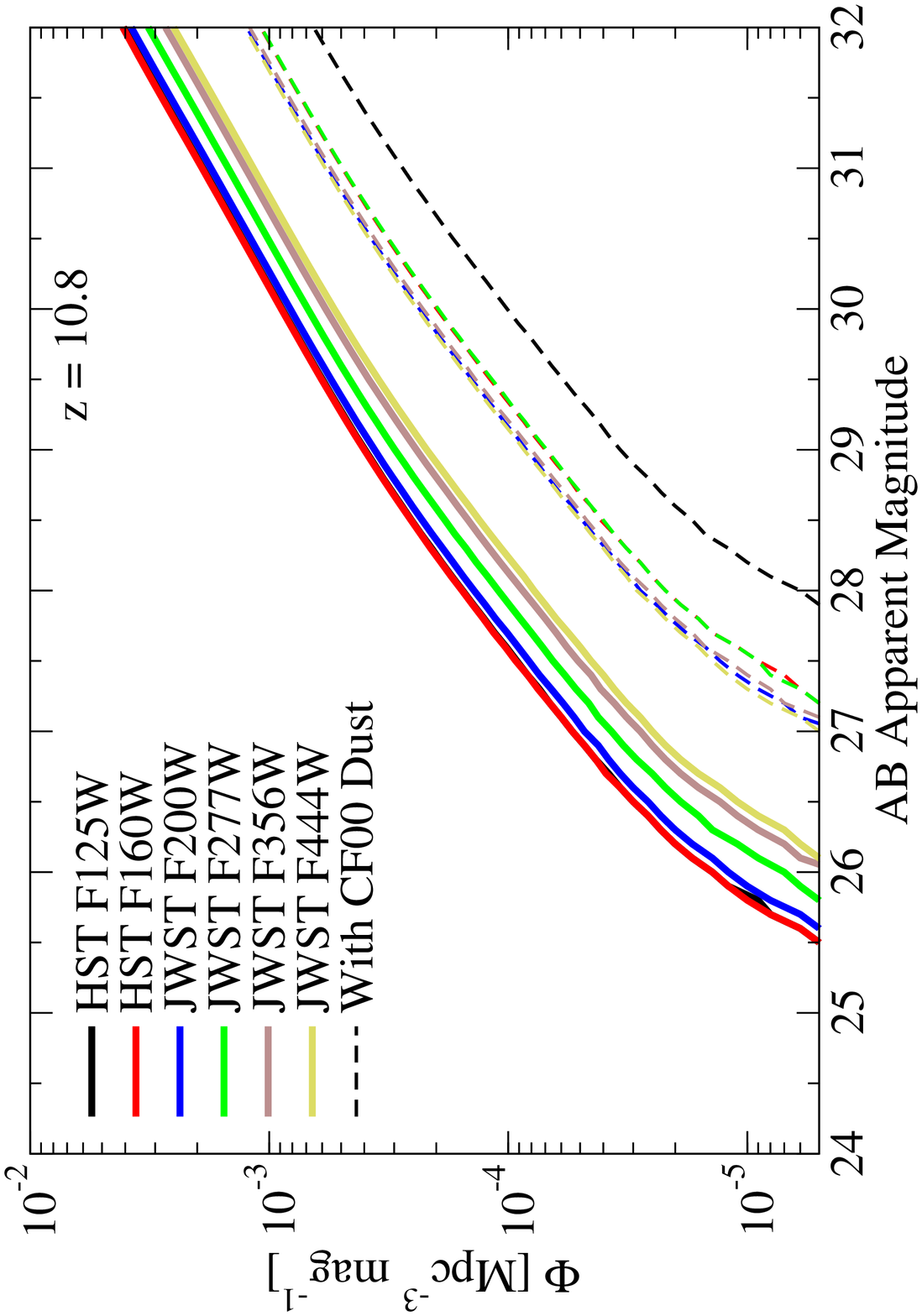}\\[-5ex]
\plotminigrace{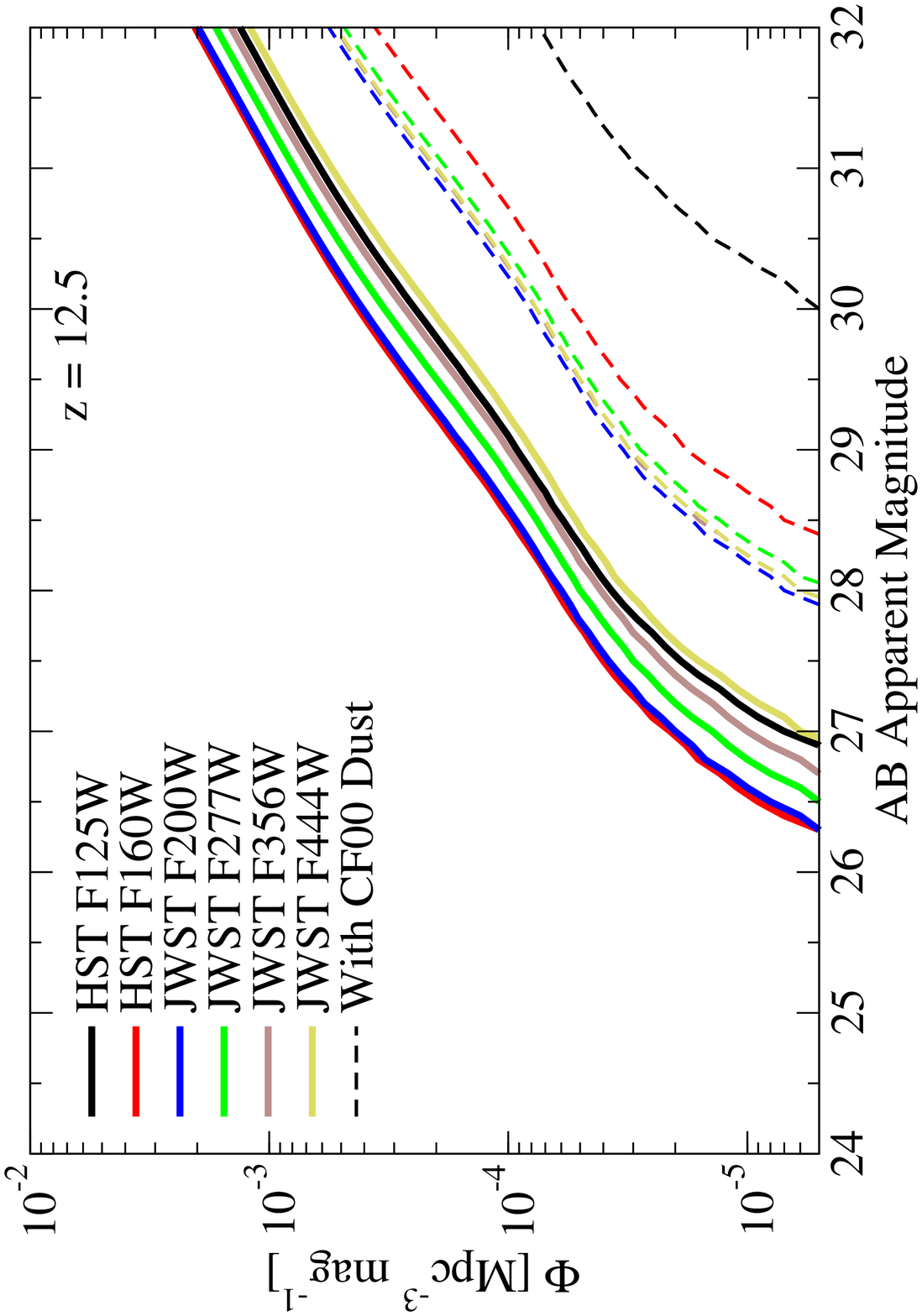}\plotminigrace{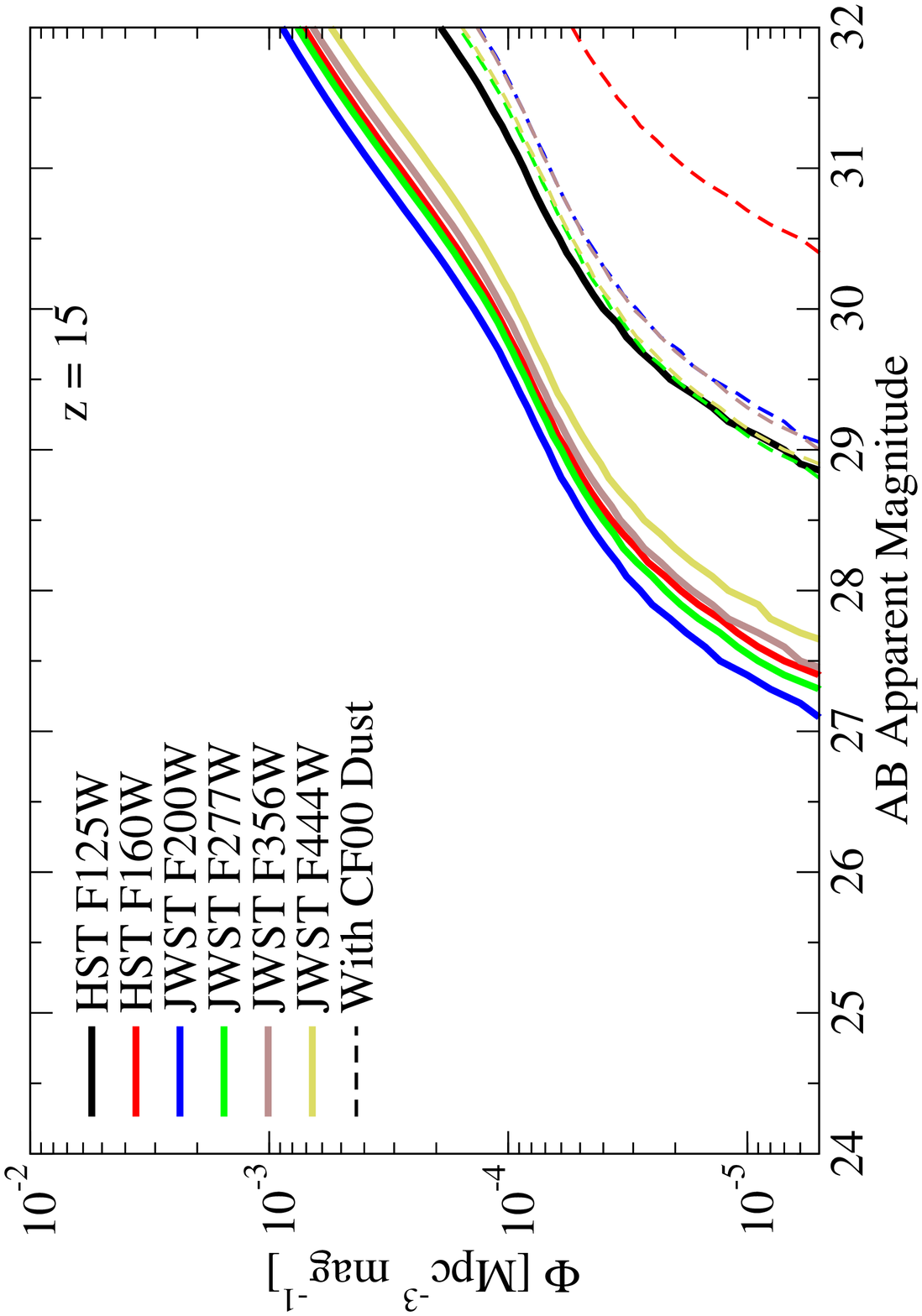}\\[-5ex]
\end{center}
\caption{Luminosity functions at $z\ge 7$.  \textbf{Top} panels: predicted UV luminosity functions at $z=7$ and $z=8$, compared to those in \cite{Bouwens14}.  Dust is a significant systematic uncertainty, and variation in dust with luminosity is the main reason why observed UV luminosity functions have steeper faint-end slopes than the corresponding stellar mass functions.  We show luminosity functions both with and without the addition of a \cite{cf-00} dust model (``CF00''). \textbf{Middle} and \textbf{bottom} panels: predicted luminosity functions in several HST WFC3 (F125W and F160W) and JWST NIRCAM (F200W--F444W) filters, both without (\textit{solid lines}) and with (\textit{dashed lines}) a \cite{cf-00} dust model.  Even at $z=15$, the Lyman break does not enter any of the JWST filters shown here.  Error bars are not shown for clarity; systematic errors from the uncertainties in stellar mass growth histories are 0.75, 0.9, 1.25, and 1.75 mag at $z=9.6,$ 10.8, 12.5, and 15, respectively.}
\label{f:lfs}
\end{figure}

Since we use converted UV luminosity functions to constrain the growth of stellar mass at $z\ge 7$, it is important to check that the stellar mass growth we infer is self-consistent with the observed UV luminosity functions.  We follow the same procedure for generating luminosities as in \cite{BehrooziGRB}; specifically, we generate UV luminosity functions using the FSPS stellar population synthesis code \citep{conroy-09,Conroy10}, and assume that the metallicity of stars produced follows the extrapolated gas-phase metallicity as a function of stellar mass and redshift from \cite{Maiolino08}.  Dust represents a significant uncertainty, however, as the only practical method for inferring the dust content of individual galaxies at $z>7$ is to use the UV slope $\beta$ \citep{Bouwens13}.  We therefore show all results both with and without the effects of a \cite{cf-00} dust model.

UV luminosity functions at $z=7$ and $z=8$ generated from the best-fit stellar growth models in \cite{BWC12} are shown in the top panels of Fig.\ \ref{f:lfs}.  As suggested by the evolution of $\beta$ with luminosity, more luminous galaxies are likely dustier than fainter galaxies \citep{Bouwens13}.  This seems in qualitative agreement with our generated luminosity functions, as the observations lie closer to the models including dust for bright galaxies, and are closer to the models without dust for faint galaxies.

We have also generated apparent luminosity functions in several HST WFC3 (F125W and F160W) and JWST NIRCAM (F200W--F444W) filters from our predictions in \S \ref{s:results}, which appear in the bottom panels of Fig.\ \ref{f:lfs}.  If the dust-luminosity relation extends to higher redshifts, it is likely that the true luminosity functions will be closer to the models including dust for bright galaxies, and closer to the models without dust for faint galaxies.  However, we caution that the systematic errors from uncertainties in stellar mass growth histories are significant; these are 0.75, 0.9, 1.25, and 1.75 mag at $z=9.6,$ 10.8, 12.5, and 15, respectively.

\end{document}